# Astronomical arguments in Newton's Chronology[1]

Yaël Nazé, FNRS Research Associate
Département AGO, Université de Liège, Allée du 6 Août 17, Bât B5C, B4000-Liège, Belgique

*Abstract: In his Chronology, Newton uses astronomical "evidence" to support its extreme rejuvenation of ancient times. These elements, having a scientific varnish, provide some credibility to the work. They have been fiercely debated for a century, with a gradual undermining of Newton's assumptions. However, this has not dented the prestige of the English scientist.*

« la partie qui me reste à examiner est ce que l'on a fait passer pour la plus importante et la plus ingénieuse des découvertes dont on a prétendu que son Livre étoit rempli » (Fréret, *Défense*, p415)

## 1. Introduction

About Newton, people are often reminded of *Principia* and *Opticks*, major works considered as founders, with few others, of modern science. But the thoughts of the English scientist were not limited to these two subjects. There is also what some consider as his 'dark side', notably his advanced historical studies. Through ages, some have argued that these have been a hobby, an idiosyncrasy of old age, or a consequence of a nerve problem that occurred in 1693 (as advocated by Laplace and Biot). However, Newton's correspondence, manuscripts, and library clearly testify that it was in no way the passing interest of a sick mind[2]. Chronology was his major concern during his last years but Newton did spend time on the subject throughout his life. He even tried various ways of interpretation, sometimes changing opinion on one point or another, before developing the *Chronology* which was published posthumously.

With this work, Newton follows the tradition of his time. Indeed, chronology was at the time a key issue, hotly debated. Newton ensures that "history without chronology is confused"[3], and no historian today can prove him wrong on this point. In his time, two main theories existed: one placing Creation 4000 years before our era, and another one placing it several centuries before. The existence of the ancient Egyptian and Chinese civilizations were however a big problem, since writings suggested them to be very old. In this context, Newton firmly believed in the accuracy of Scriptures and seniority (implying superiority) of Jewish civilization. The *Chronology* was a way to prove this supremacy while shifting Apocalypse away to a more distant future. This could not be done, however, without difficulty.

First, Newton takes some freedom with his famous "Hypotheses non fingo." He blames other chronologists for the uniqueness of their proofs, their absence of neutrality when they search for evidence, and their indiscriminate use of ancient texts. However, he does not seem to remember his own principles for his own work and content himself with a selective choice of sources and a minimal skepticism when his thesis is supported by a given text. A good example is the attribution of Eudoxus celestial sphere to Chiron, on the basis of a single verse of poetry, at most...

Newton also supports a radically short chronology – even shorter than the shortest solution of the time – by subtracting five centuries to the classic timeline. His main arguments are based on the average duration of a reign, and on astronomy. These are not, by themselves, original arguments[4], but Newton had to be original in that area as in others ... and the use of the precession is here the "act of sheer genius"[5].

---

[1] La version française de cet article est accessible via arXiv:1212.4943
[2] Emerson mentioned that it is the 'work of his whole lifetime' in his "Defence of the Chronology" (appendix to "A short comment on Sir I. Newton Principia" (1760, p150); see also "Newton Historian" by F.E. Manuel (Cambridge Univ. Press, 1693).
[3] New College MSS II fol 72, quoted by F.E. Manuel, p37.

This article aims to analyze the astronomical evidence provided by Newton and the debates that followed. After a brief introduction recalling the context of the book and the debates it started, it continues with the examination of each astronomical point advanced by Newton (calendar and duration the tropical year, precession as a tool for dating events), but also by other debaters.

This article was written before the publication of "Newton and the Origin of civilization" (J.Z. Buchwald & M. Feingold, 2013). However, since it is published after, references to this book have been added. The book focuses on Newton and discusses several issues of the *Chronology* (precession, reign duration, Egyptian empire). This short article considers only the astronomy, that used by Newton (precession but also calendar problems) as well as the astronomical elements presented by others during the debate surrounding the *Chronology*; It is therefore complementary to Buchwald & Feingold's book.

## 2. *Publication and reactions to « Chronology »*

Newton wrote extensively throughout his life about chronology and world history. However, as often happened, he was reluctant to publish. It finally occurred through a somewhat fantastic intrigue. It all starts with the *Abrégé de la Chronologie de Mr le Chevalier Isaac Newton* (hereafter 'Abrégé'). This was a short note initially made for the Princess of Wales, but then brought to France by Abbé Conti, who had had a long discussion on chronological issues with the English scientist. Although it was not supposed to be publicly released, Conti showed the note to several people, so that the *Abrégé* was eventually published in France in 1725, without the authorization of Newton but with critical observations of Nicolas Fréret (hereafter 'Observations')[6]. Newton replied sharply to this event by a short acid note in Philosophical Transactions[7]. In this context, it should be noted that the "theft" of Conti must be somewhat relativized, since several copies of the note already circulated at the time in England, three still existing today[8]. Casting all blame on the sole Conti is thus somewhat exaggerated. In 1726, four essays by father Souciet, the first one focusing on astronomy, also criticized the text of Newton, taking into account a letter from Mr Keil clarifying the reasoning followed by the English scientist. The following year, Souciet completed his astronomical reasoning in a fifth essay, after he read the reply to the unauthorized publication in the Philosophical Transactions. In turn, La Nauze responded point by point to each of these essays in his "Letters to father Souciet" while Halley also opposed to Souciet in two articles in the Philosophical Transactions[9]. The detailed ideas of Newton were finally published in 1728, after his death, under the title of *Chronology of ancient kingdoms amended* (hereafter 'Chronology'). It contains much more detail than the abridged version (which is reproduced in the beginning of the book), allowing a better understanding of the reasoning of the author. Fréret then developed an equally detailed response[10], including the critics by William Whiston[11], Newton's successor as

---

[4] See e.g. F.E. Manuel p94.
[5] F.E. Manuel, p191
[6] A letter from Fréret to Halley also exists, and presents his initial views (see Buchwald & Feingold p366-8).
[7] Newton, Philosophical Transactions, 1725, vol 33, 315-321
[8] F.E. Manuel, p22
[9] Halley, Philosophical Transactions, 1726, vol 34, 205-210 et 1727, vol 35, 296-300
[10] Fréret, Défense de la Chronologie fondée sur les monuments de l'histoire ancienne, contre le système chronologique de M. Newton, 1758 (ci après « Défense ») – the third part contains the astronomical discussion.
[11] Whiston, appendix IX of "A collection of Authentick Records belonging to the Old and New Testament" (1728), which first part is reproduced un the third chapter, section II, §1, of Fréret's Défense. In the following, references

Lucasian professor. It was published after his death. Although comprehensive, this book was not the endpoint, as debates continued until the early 19th century.

Supporters and opponents to this surprising chronology thus existed from the start [12]. Among those who supported Newton were, besides the already mentioned La Nauze [13]: Andrew Reid (1728), Fatio de Duillier (1732), Zachary Pearce (1732), Voltaire (1733), Arthur Ashley Sykes (1744), James Steuart (1757), Edward Gibbon (1758), William Mitford (1784), William Emerson (1770), Robert Wood (1775), and an anonymous "member of the University" (1827)[14]. In the opposite camp, besides the already mentioned Souciet, Fréret and Whiston, were notably Arthur Bedford (1728), James Logan (1728), Samuel Shuckford (1728-37), Jean Hardouin (1726), Thomas Cooke (1731), Jean Masson (1731-2), Arthur Young (1734), Etienne Fourmont (1735), Zachary Grey (1736), Joseph Atwell & Thomas Robinson (1737), Alphonse des Vignolles (1738), Angelo Maria Quirini (1738), Francesco Algarotti (1739), Fréret's colleague Antoine Banier (1740), Thomas Francklin (1741), Samuel Squire (1741), George Costard (1746), William Warburton (1742), Conyers Middleton (1752), L.R. Desh (1755), un certain Rev. Rutherford (1760), Charles de Brosses (1761), Antoine-Yves Goguet (1761), Jean-Sylvain Bailly (1775), Samuel Musgrave (1782), Juan Andrés (1785-1822), Jean-Baptiste Joseph Delambre (1817), Henry Fynes Clinton (1830) and two anonymous authors (1754, 1855)[15]. The issues raised and discussed in these books

---

will be made to the latter.

[12] For a discussion of the situation in France, for example, see C. Grell, Arch. Int. Hist. Sciences, vol 62, n°168, 85-157 (2012); for a general discussion, see chapter 10 of F.E. Manuel.

[13] Besides the Letters to Souciet, there is also his response to Shuckford: La Nauze, Mémoires de Trévoux, article 125, October 1754.

[14] A. Reid (The Chronology of ancient kingdoms amended by sir Isaac Newton, in the present state of the Republick of Letters, vol II, 1728), Fatio de Duillier (letter to J. Conduitt, 10 August 1732), Z. Pearce (A reply to the letter to Dr Waterland, 1732), Voltaire (Letters concerning the English Nation, 1733 and de la chronologie réformée de Newton, qui fait le monde moins vieux de cinq cents ans, in Mélanges de litérature, d'histoire et de philosophie, 1757), A. A. Sykes (An examination of Mr Walburton's account of the conduct of the antient legislators, 1744), J. Steuart (Apologie du sentiment de Mr le chevalier Newton sur l'ancienne chronologie des Grecs contenant des réponses à toutes les objections qui y ont été faites jusqu'à présent, 1757), E. Gibbon (MS34880, British Museum, 1758, reproduced in Miscellaneous works of Edward Gibbon, III, 61-73, 1814), W. Mitford (History of Greece, appendix I, 1784), W. Emerson (appendix « An account of some of the numerous inconsistencies contained in the objections made by the Rev. Dr Rutherford, against sir I. Newton's account of the Argonautic expedition », in A short comment on sir I. Newton principia containing notes upon some difficult places of that excellent book, 1770), R. Wood (an essay on the original genius and writings of Homer, 1775), and an anonymous « member of Cambridge University » (Essays on chronology ; being a vindication of the system of sir Isaac Newton, 1827).

[15] A. Bedford (Animadversions upon Sir Isaac Newton's Book intitled The chronology of ancient kingdoms amended, 1728), J. Logan (1728, in E. Wolf, 1974, The library of James Logan of Philadelphia), S. Shuckford (The sacred and profane history of the world, volume II, 1728-37 then 1752, summarized in Mémoires de Trévoux, article 17, volume of February 1754), J. Hardouin (Le fondement de la Chronologie de Mr Newton, 1726, summarized in Hardouin, Mémoires de Trévoux, article 87 of the volume of Spetember 1729), T. Cooke (The letters of Atticus, 1731), J. Masson (in John Jortin, Miscelleanous Observations upon authors, ancient and modern, vol 2, 1731-2), A. Young (An historical dissertation on idolatrous corruptions in religion, 1734), E. Fourmont (Réflexions critiques sur les histoires des anciens peuples, 1735), Z. Grey (An examination of the fourteenth chapter of sir I. Newton's observations upon the prophecies of Daniel, 1736), J. Atwell & T. Robinson (Hesiodi ascraei quae supersunt cum notis variorum, 1737), A. des Vignolles (1738, in Nouvelle Bibliothèque Germanique, vol 18, partie I, in Steuart 1805, the works, political, metaphysical, and chronological), A. M. Quirini (Primordia Corcyrae, 1738), F. Algarotti (Sir Isaac Newton's Philosophy explained for the use of ladies, 1739), A. Banier (Mythologie des anciens expliquée par l'Histoire, tome 6e, chap XII, p342-6, 1740), T. Francklin (Of the nature of Gods, traduction de Cicéron, 1741), S. Squire (Two essays, the former, a defense of the ancient Greek chronology ; to which is annexed a new chornological synopsis ; the latter, an enquiry into the origin of the Greek language, 1741), G. Costard (Letter to Martin Folkes concerning the rise and progress of astronomy amongst the Antients, 1746), W. Warburton (4e livre de The divine legation of Moses, 1742), C. Middleton (The miscelleanous works, vol 2, 1752), L.R. Desh (Lettre sur la chronologie de M. Newton, Mercure de France, décembre 1755, reproduit dans le livre de Steuart), C. de Brosses (Second mémoire sur la monarchie de Ninive, Histoire et mémoires de l'Académie Royale des Inscriptions, 27, 1-81, 1761), A.-Y. Goguet (The origin of laws, Arts, and Sciences and their progress among the most ancient nations, 1761), J.S. Bailly (Histoire de l'Astronomie Ancienne, 1775 et 1781), S. Musgrave (Two dissertations I. on the graecian mythology II an examination of sir Isaac Newton's objections to the Chronology of the Olympiads, 1782),

are sometimes astronomical in nature (see details below), but they are also theological - with Newton being accused to support Papists or deists, depending of the author...

It should be noted that the quality of responses to Newton's astronomical arguments differ from one author to the next, not only on the level of detail of the analysis, but also on the accuracy of discussions. Souciet particularly stands out here. Rather than mocking Newton[16], it would have been better for him to proof-read his writings since his text is full of errors: there are simple typographical errors, poorly executed copies of text [17], and even various mathematical problems. For example, Souciet almost always confuses distance between colure and Ram's ear (7°36' in 939 BCE) and between Ram's ear – beginning of the zodiacal sign (which is the complement to 15°, i.e. 15° -7°36' = 7°24'). He also sometimes mixes Hipparchus and Chiron, taking the angles referring to the former and the epoch of the latter (which necessarily leads to impossibilities). In addition, he does not correct the coordinates for the tilt of the equinoctial colure[18]. Finally, while he correctly insists on the use of right ascension by the ancients[19] he still continued to use ecliptic longitudes throughout his reasoning... As Halley put it [20], he should "be a little more careful of his numbers […] and inform himself of the Sphericks". La Nauze, while always prompt to point at Souciet's errors, often make similar ones...

An important, although rarely openly declared, element in the debate is the nature of the author. Newton is obviously not any chronologist. Indeed, support or opposition to the new chronology is also directly related to the person of Newton himself, especially since he adorns his temporal edifice with a scientific varnish (in the form of astronomical "evidence"). The aura of Newton can be an advantage[21] or a disadvantage[22], especially in France. On the one hand, some thought that the great man, being right on optical and mechanical issues, cannot be mistaken about the timeline – indeed, the celestial models correctly predicting the future sky configuration are just as good to compute past events. On the other, some hope to destroy Newton's castle by

---

J. Andrés (dell origine, progressi e stato attuale d'ogni litteratura, 1785-1822), J.-B. J. Delambre (Histoire de l'Astronomie Ancienne, 1817), H. F. Clinton (Fasti Hellenici, the civil and literary chronology of Greece and Rome, 1827) ; Rutherford is only quoted by Emerson; an anonymous author reviewing Shuckford's book in mémoires de Trévoux (1754), and an anonymous author reviewing "History of Greece" by George Grote in Dublin University Magazine (vol 45, 1855).

[16] For example Souciet, p127 « l'une est le témoignage des Anciens, l'autre la raison. M. Newton n'a ni l'un ni l'autre ».

[17] For example, the constellations' ends which change position between p161 and p165.

[18] This inclination is 66.5° with respect to the ecliptic, see appendix and Fig. 2 below. See also Halley's remarks in Philosophical Transactions, vol 34, p209: « he ought to have deduced 3 deg 7.5' out of the 15 degrees he assumes for the distance of his colure from the first star of Aries »

[19] He use it elsewhere (Souciet, 5$^e$ dissertation, p 133-135) to try to understand where the value of 7°36' quoted by Newton in his response but without detail (Philosophical Transactions, vol 33, 315-321) could come from. He supposes that it is in fact the ecliptic longitude difference between Fishes' node, which he assumes to be the beginning of Aries sign for Newton, and the Ram's ear (note his usual error: he should have used 7°24' before the ear since the 7°36' corresponds to the distance between the ear and the colure in Newton's work). He then shows that, in right ascension, the angle is very different, the Fishes' node appearing before the ear because of their very different latitudes... La Nauze (5$^e$ lettre, p 404-407) sees in these pages a chimera, a ghost that Souciet tracks. He asserts that this is not the origin of Newton's value of 7°36' but he strangely avoids explaining where the value comes from. Indeed, to find this value, one needs to correct for the colure's inclination, which La Nauze never does in his Letters! These pages linked to right ascension are a good example of the quality level of Souciet's writings: in addition to bad copies and texts (e.g., the Ram's ear used twice to refer to two different stars), Souciet considers a constant precession rate on the equator, not the ecliptic, which is a basic error. He should have precessed his ecliptic longitude and then calculate the right ascension – Buchwald & Feingold also give detail on Souciet's errors with right ascension (p353-362).

[20] Halley, Philosophical Transactions, vol 34, p209 : « be a little more careful of his numbers […] and inform himself of the Sphericks, so as to give us the right ascensions of the stars truly from their given longitudes and latitudes ».

[21] Fréret, Défense, p 419 - « le nom seul de Newton formera toujours un préjugé difficile à détruire. »

[22] Steuart, Apologie, p164 - « les belles découvertes de Mr. Newton dans l'Astronomie et dans l'Optique n'ont pas été moins combattuës dans le commencement, que l'est à présent son sentiment sur la Chronologie. »

undermining a secondary wall, easily attackable because it is questionable in many ways as we will see. The fierce battle between the two camps is thus not at all ethereal.

We will now focus only on the texts related to astronomy.

## 3. *Dating thanks to imperfect calendars*

Elaborating correct calendars is far from easy. The first attempts to do so were simple (e.g. lunar calendar with 12 months) but quickly deviated from the actual solar year and thus from seasons. Regular adjustments were needed, with the insertion of additional months from time to time. Even a 365-day calendar is not perfect, because the actual year is longer: the average value is close to 365.2422 days.

Both *Abrégé* and *Chronology*[23] use this type of error for dating the Egyptian calendar. According to Newton, this calendar was initially using only 360 days, and then five days were added to have a better agreement with the solar year. This calendar was later adopted by the Chaldeans, and this is where a shift can be detected. Newton argues that the initial beginning of the Egyptian year fell on the Spring Equinox, but it deviated slowly from it year after year because of the wrong duration of the calendar year. In the first year of the Chaldean emperor Nabonassar (supposed to be 747 BCE), the beginning of the calendar occurred on 26 February, or 33 days and 5 hours before Equinox, thereby showing that the 365d calendar had been adopted in Egypt 137 years before. Newton was using the Julian calendar and its calculation can be easily checked. In 747 BCE, the spring equinox occurred on March 28 at 22:30 UT[24]. Between February 26 at 0h and March 28 at 22:30, there are almost 31 full days (not 33d5h but close enough). Assuming that the Chaldean calendar was adopted without change, Newton thus places the creation of the Egyptian 365d calendar a little more than a century (on 884BCE) before the first year of Nabonassar. This fixes the succession of Pharaohs since the end of the reign of pharaoh Amenophis coincides with the creation of this calendar in Newton's mind.

We now know that the Chaldean or Mesopotamian calendar influenced the Egyptian one, and not the contrary, and also that the 365d was used already in the 2nd millennium BCE. At the time of Newton, however, only Fréret pointed this problem, both in his *Observations* (p84-89) and in Section I of the third part of his *Défense*[25]. In addition, he also mentions the problems of identifying pharaoh Amenophis with a set of other names – a set well chosen by Newton to condense as much as possible the duration of the Egyptian empire. He also emphasizes a problem in Newton's reasoning: there was a religious festival held at the equinox, that's true, but it was not the first day of the Egyptian calendar[26], known to be linked to the heliacal rising of Sirius (called "Sothis") and the beginning of the Nile flood[27].

---

[23] 884BCE entry of *Abrégé* and with more details *Chronology*, p82-83 : « ils ajoutèrent à la vieille année du Calendrier cinq jours […] sous le règne d'Aménophis, ils ont pu placer le commencement de cette nouvelle année ou l'équinoxe de printemps […] Enfin cette même année luni-solaire fut introduite dans la Chaldée, d'où vint l'ère de Nabonassar ; car les années de Nabonassar et celles d'Egypte commençoient le même jour qu'ils nommoient Thoth, il n'y avoit aucune différence entre elles. La première année de Nabonassar commença le 26 février de l'ancienne année romaine, 747 ans avant l'ère vulgaire de Jésus-Christ, et trente-trois jours et 5h avant l'équinoxe de printemps suivant le mouvement moyen du Soleil […] Or si l'on compte que l'année de 365 jours a cinq heures & 49 minutes de moins que l'année équinoxiale, le commencement de cette année rétrogradera de trente-trois jours et cinq heures en 137 années & par conséquent cette année commença d'abord en Egypte à l'Equinoxe de printemps, suivant le moïen mouvement du Soleil 137 années avant le commencement de l'ère de Nabonassar »

[24] To compute easily the date of solstices and equinoxes between 4000 BCE and 2500, IMCCE website http://www.imcce.fr/fr/grandpublic/temps/saisons.php can be useful. A quick estimate can also be done : Newton says he used a tropical year length of 365j5h49m whereas the Julian year is 365d6h, which makes a difference of 18,6d in 2435 years (interval between 747BCE and 1689) to add to the date of equinox in Newton's time (9 March 22h24 UT in 1689) to derive that date. 1689) – one then finds March 28.

[25] Fréret, Défense, p386

Fréret also noted another problem with Egyptian traditions[28]. The people of the Nile used two calendars, one civil, kept in phase with the seasons, and the other religious, with always 365 days. The latter therefore progressively shifted from the former by a fourth of a day each year. The two calendars come back in phase after about 1460 years[29], a time also called "Sothic cycle". As the two calendars are known to have been in phase again in 138, it follows that a cycle began in 1323 BCE. Fréret even affirms, using various sources including texts of Manetho, that this cycle was not the first one, so that the Egyptian calendar of 365 days is much older than 884 BCE by at least four centuries and a half, if not two millennia. Note that Newton did not understand the scope of Fréret's remark[30]: in his response to *Observations*, he swears that he never placed the beginning of a Sothic cycle in 884 BCE. That is true indeed, but does not solve the problem since Fréret questions the antiquity of the 365d calendar, not whether Newton mentions the Sothic cycle or not: it is indeed impossible, by definition, to begin such a cycle before having fixed the length of the year to 365d.

Fréret made two last remarks on the subject. Newton's only quoted source is Syncellus, but his writings are not reliable because they contain several inconsistencies with both external sources (other sources from the same period), or within the text itself[31]. In addition, the additional 5 days were often not officially counted, even if used in practice – those were often called "stolen days"[32]. Therefore, having a religious text speaking of 360 days is compatible with the practical use of a 365-day calendar at the same time. Newton goes thus too fast by claiming the 365-day calendar a recent invention.[33].

## 4. *Dating using precession*

Dating was an acute problem for historians of the time – archaeometry still lied in a distant future. The analysis of monuments was not clear, and all was left to historians was the analysis of texts, with Scripture in the first place. Biblical generations were counted, with the hope that all written sources will agree with the derived timeline. In this context, astronomy seemed a considerable asset. Celestial events occur at specific times, which can be precisely calculated, even in the 17$^{th}$ century thanks to accurate celestial mechanics calibrated on detailed observations. The precision and compelling aspect of scientific astronomical computing, not available for other methods, exerted a considerable appeal. Indeed, Joseph Scaliger and Denys Pétau used the known cycles of the Sun and the Moon phases to phase ancient calendars with the modern one. Eclipses, rare but predictable events, were also used, especially in the work of Riccioli. The study of the orbits of comets, a new science at the time, pioneered by Halley, provided Whiston with a date for the Flood. Phases of the moon and associated tides allowed Halley to date the invasion of Britain by Caesar.[34].

---

[26] « il n'y a rien dans toute l'antiquité qui puisse nous faire penser que l'année égyptienne ait jamais commencé au printemps. » Fréret, Observations, p85
[27] Fréret, Observations p84-89, and Défense, p391-394
[28] Fréret, Observations p84-89 and Défense p394-400 et p407
[29] because 365.25/0.25=1460, or 1507 yrs if one considers tropical years of 365j 5h 49m
[30] Newton, Philosophical Transactions, vol 33, p320
[31] Fréret, Défense, p405-406
[32] Fréret, Défense, p411-412
[33] It is interesting to note that a supporter of Newton, James Steuart, try to explain Newton's work by underlining the antiquity of the epagomenal days (Apologie, p88). Steuart wants to prove to Shuckford that Ancients could be quite precise, notably on the duration of the year, but without knowing, he goes here against his master's thesis!
[34] J. Scaliger, De emendatione temporum (1583) ; D. Pétau , Opus de doctrina temporum (1627) ; Riccioli, Almagestum Novum (1651); W. Whiston, A new theory of the Earth (1696); E. Halley, A discourse tending to proue at what time and place Julius Cesar made his first descent upon Britain, Philosophical Transactions, 16, 495-501 (1691)

Using astronomy for establishing a timeline was thus not new when Newton became interested in history. However, his choice of the astronomical phenomenon to be used was totally new and potentially revolutionary. Indeed, Newton uses the precession, a well known phenomenon that Newton himself has "demonstrated" in the context of his new physics. It has the advantage of being independent of the latitude of the observer, eliminating one unknown in the calculations. This is clearly the heart of the Newtonian time system[35]. This argument also appears as the most scientific of all in the *Chronology*, and therefore it is the main one to destroy or support when discussing the *Chronology*. This is well understood by contemporaries of Newton (e.g. Fréret[36], Steuart[37], or Bailly[38]), and explains the large number of discussions on the topic.

In the remainder of this article, we split the astronomical reasoning in 4 parts, to ease the understanding of the arguments because the debate often appears confused, despite what Souciet thinks[39]. The last part of the appendix provides celestial charts of the constellations mentioned in the text, as well as colures for the discussed epochs.

One thing should be noted, however: Newton did not really need astronomy. Indeed, his *Original of Monarchies* only uses the average length of reigns to fix the (shortened) timeline[40]. Astronomy is therefore only the cherry on the cake, although it was considered the "best" evidence because of its scientific and demonstrable aspect.

a) Constellation list

To support his chronological argument, Newton first attempts to "prove" the link between constellations and the time of the Argonauts: listing constellations[41], he notes that nearly all of them

---

[35] « c'est la base de la Chronologie » (Fréret, Observations, p56) – It should be noted that the word « precession » does not appear in the Abrégé nor in Newton's answer quoted by Keil (p56 of Souciet first essay). However, the phenomenon under consideration is obvious (saying that solstice shifts by one degree is 72 yrs is sufficiently explicit) and it is thus not Keil's letter that gives the solution, as F.E. Manuel (p23) thougt. It can be added that other authors had already thought about using precession for the chronological problems (e.g., Scaliger see Joseph Scaliger by Anthony Grafton 1993 and Buchwald & Feingold p250), but without using the colures;

[36] « la détermination de l'âge des Argonautes par le lieu des colures dans la sphère réglée sur leur temps, est une de ces idées neuves et brillantes, dont le privilège est de surprendre et de subjuguer les esprits. Je ne m'étonne pas que M. Newton l'ait saisie comme une soudaine inspiration de ce Génie dont il avoit autant de droit que Socrate de se croire assisté dans ses méditations, & que ses partisans se soient jettés avec confiance dans la route nouvelle que leur traçoit ce rayon de lumière. Un fait aussi certain que la précession des équinoxes, employé comme principe, garantissoit à leurs yeux la certitude des conséquences qui devoient en résulter. C'étoit en même temps étendre le ressort de l'Astronomie & faire marcher la Chronologie d'un pas plus sûr, que d'appeler l'une au secours de l'autre. […] Il étoit difficile qu'avec de tels avantages au moins apparens, cette preuve astronomique ne donnât presque l'air d'une démonstration au raisonnement par lequel M. Newton en déduisoit le calcul abrégé. » Fréret, Défense, pxliv-xlv of the preface

[37] « le 3e avantage qu'à eu Mr. Newton consistoit en l'étenduë de ses connaissances de l'Astronomie & dans ce génie créateur dont il étoit doué. Rien de ce qui avoit du rapport à cette science ne pouvoit se dérober à sa pénétration. Il en saisissoit la moindre petite circonstance, pour la tourner à profit dans les autres sciences. Combien la physique n'a-t-elle pas servi entre ses mains pour expliquer les phénomènes de l'Astronomie et de l'Optique combien l'Astronomie à son tour n'a-t-elle pas servi pour expliquer la nature & la voici encore employée pour déterminer la Chronologie » Steuart, Apologie, p116

[38] « l'idée de régler la chronologie par la détermination ancienne des points équinoxiaux et solsticiaux étoit belle, grande et digne d'un homme de génie… mais Newton s'est trompé dans l'application qu'il en a faite & le système qui en résulte est tombé, parce qu'il est contraire aux faits. » Bailly, Histoire de l'Astronomie Ancienne, p509

[39] « la preuve qu'en apporte M. Newton est des plus spécieuses et des plus fortes en apparence. Elle est fondée sur le cours des astres, et sur un calcul astronomique des plus aisez et des plus clairs. » Souciet, 1$^e$ dissertation 1, p51

[40] F.E. Manuel, p122

[41] « Chiron dessina les figures du ciel […] on peut voir par cette Sphère même, qu'elle fut ébauchée au tems de l'expédition des. Argonautes ; car cette expédition s'y trouve marquée parmi les Constellations, aussi bien que differens traits encore plus anciens de l'Histoire Grecque ; il n'y a rien de plus moderne que cette expédition. On voioit sur cette Sphère le *Bélier* d'Or, qui étoit le Pavillon du Navire, dans lequel Phryxus se sauva dans la

refer to the Argonautic expedition, but none to Troy's war – such an event would have been marked in the sky if there was room left, so that Newton advocates for constellations having been defined before Troy's war, and in fact to be used by the Argonauts for their travel.

The main problem with this list is that it contains many non-Greek constellations. At least the twelve constellations of the classic zodiac (Aries, Taurus, Gemini, Cancer, Leo, Virgo, Libra, Scorpio, Sagittarius, Capricorn, Aquarius, Pisces) as well as Hydra, Aquila, and Piscis Austrinus are of Mesopotamian origin[42]. They were transmitted to the Greeks who adopted them - as Fréret mentions[43] these are "constellations dressed in the Greek fashion." Their relationship with the famous expedition is, at best, tenuous. That was perfectly known at the time[44] although La Nauze insists on Greek additions[45], saying that the Greek names for the constellations are different from those given by the Chaldeans (which is not true in many cases) and by the Egyptians (which is correct).

In addition, Fréret noted that "Chiron was not the only one to whom the Greeks thought they owe their Astronomy." [46] They also attributed this title to Prometheus, Atreus, and Palamedes. In this case, why choosing Chiron rather than another? Is it not just because it helps reaching the goal, i.e., the shortening of the conventional chronology? Indeed, Souciet stresses[47] the importance of the Argonautic expedition, to which various events (Trojan War, founding of Rome) are linked by known time intervals. This is an advantage for Chiron, not shared by other characters.

Worse, even assuming that Chiron is the original source, we cannot assume at the same time that he defined the constellations for the Argonauts' use and named these constellations using future events of that expedition. In fact, such events are inherently unpredictable, and are known only once

---

Colchide ; le *Taureau* aux piés d'airain dompté par Jason ; les *Gémeaux* Castor & Pollux, tous deux Argonautes, auprès du *Cigne* de Leda leur mère. La étoient répresentés le *Navire* Argo, & l'*Hydre* ce Dragon si vigilant ; ensuite la *Coupe* de Medée, & un *Corbeau* attaché à des cadavres, qui est le symbole de la mort. D'un autre côté, on remarquoit *Chiron* le maître de Jason, avec son *Autel* & son *sacrifice*. *Hercule* l'Argonaute avec son *Dard* & avec le *vautour* tombant ; le *Dragon*, le *Cancer* & le *Lion* qu'il tua ; la *Lyre* d'Orphée l'Argonaute. C'est aux Argonautes que toutes ces choses ont du rapport. On y avoit encore répresenté *Orion*, fils de Neptune, ou selon d'autres livres, petit-fils de Minos, avec ses *Chiens*, son *Lièvre*, sa *Rivière* & son *Scorpion*. L'Histoire. de Persée est designée par les Constellations de *Persée*, d'*Andromède*, de *Cephée*, de *Cassiopée*, & de la *Baleine* : celle de Callisto, & de son fils Arcas, par la *Grande Ourse*, & le *Gardien de l'Ourse* : celle d'Icare, & de sa fille Erigone, est marquée par le *Bouvier*, le *Chariot*, & la *Vierge*. La *petite Ourse* fait allusion à une des Nourrices de Jupiter, le *Chartier* à Erechthonius, le *Serpentaire* à Phorbas, le *Sagittaire* à Crolus, fils de la Nourrice des Muses, le *Capricorne* à Pan, & le *Verseau* à Ganimède. On y voïoit la *Couronne* d'Ariadne, le *Cheval aîlé* de Bellerophon, le *Dauphin* de Neptune, l'*Aigle* de Ganimede, la *Chèvre* de Jupiter & ses *Chevreaux*, les *Asnons* de Bacchus, les *Poissons* de Venus & de Cupidon, & le *Poisson Austral* leur parent. Ces Constellations & le *Triangle*, sont les anciennes dont parle Aratus: elles font toutes allusion aux Argonautes, à leurs contemporains, & à des gens plus anciens d'une ou de deux Générations. De tout ce qui étoit originairement marqué sur cette Sphère, il n'y avoit rien de plus moderne que cette expédition. *Antinoüs* & la *Chevelure de Bérénice* sont de nouvelle date. Il semble donc que Chiron & Musaeus firent cette sphère, pour l'usage des Argonautes »Newton, Chronologie, p87-89, italics are in the original text ; Some clarifications should be made for this list: Goat and her Kids are in fact part of the constellation Auriga, while Asses are, themselves, placed in Cancer; the River is Eridanus, Chiron the Centaur, Serpentaire is Ophiuchus, the winged horse is Pegasus, the Vulture probably carried the Lyra, and Dart is probably Sagitta. Note that the Cancer is also sometimes called Ecrevisse at that time (a feature often used in French texts). Newton's list also lacks the constellations Equuleus, Libra ("Serres du Scorpion", as elsewhere quoted by Newton) and Lupus (initially linked to the constellation Centaurus), though all of these are duly noted by Ptolemy.

[42] For example: Rogers, JBAA, 1998, 108, 9
[43] « constellations habillées à la grecque » Fréret, Défense, p501 ; see also Défense, p466: « les Grecs aimoient à faire honneur à leur nation de bien des choses qu'ils devoient aux Barbares »
[44] For example Bailly, Histoire de l'Astronomie Ancienne, p512, point XL
[45] La Nauze, 5e lettre au Père Souciet, p433
[46] « Chiron n'étoit pas le seul à qui les Grecs se crussent redevables de leur Astronomie » Fréret, Observations, p73-4
[47] Souciet, 5$^e$ dissertation, p117-118

the expedition is accomplished, as noted by Desh[48]. The (weak) reply to this topic by Steuart[49] is that Chiron has probably given provisional names prior to the expedition, using Egyptian constellations (!), and then he changed them just after the Argonauts' return – just after because a more recent creation would certainly (!) have led to add other events, such as the Trojan War.

Hardouin goes further[50]. He ensures that Chiron as astronomer is a chimera, the centaur being best known for his pharmaceutical-medical art[51]. And even if we accept this "fact", he raises the question of the usefulness of the constellations[52] and of the reference (equinoctial or solsticial) points for navigation, especially since the travel took place along the coasts during a few months: the Pole Star was useful for long travels, but the interest of the rest seems rather doubtful before the invention of the astrolabe, which is much more recent than the famous expedition. There is also a problem of accuracy in stellar positions (see below and notes 70 & 71).

These numerous discussions on Chiron miss one key point: Chiron himself was not important for Newton. Indeed, as his manuscripts show[53], he searched above all a link with the Argonauts because he had an invention without inventor. His initial choice was Palamedes, but after 1700, he changed his mind, preferring Chiron and avoiding any mention to the then fallen hero in his writings.

In summary, we can clearly call into question the definition of the celestial sphere by Chiron or any of his contemporaries. Yet this link is the basic building block on which Newton's thesis relies (see next section).

b) Position of the colures

The central point of the astronomical arguments in Newton's *Chronology* is precession. It is a slow shift (one degree per 72 years) of the equinoctial and solstice points along the ecliptic and throughout the constellations. It can be easily used for dating if you know the exact positions of these points or of a star at two different periods, one of which is of known date (1690 in the case of Newton). The difference of positions, in degrees, then directly gives the interval in years between epochs, after a simple multiplication by 72[54]. What was only needed is a source mentioning the old positions, for example of these remarkable points, with the current positions being known.

Newton thought that these points were located in Chiron's time at the center of the Cancer, Capricorn, Aries and Libra, as he wrote in his *Abrégé* for the year 939 and in the *Chronology*[55]. To clarify what "center" exactly means, he first considers the two extreme stars of Aries (γ Ari, the Ram's ear and « first of Aries », and τ Ari, end of the Ram's tail and « last of Aries »). From their positions in the Flamsteed's catalog[56], he derives the coordinates of the midpoint and look where the equinoctial colure passing through this point intersects the ecliptic. For this calculation, contrary to Souciet, he does not forget that the equinoctial colure is tilted by 66.5° with respect to the ecliptic, leading to a longitude correction for the intersection (this correction increases with distance from the ecliptic, see Fig. 1). He finds a value of 6°44' ♉ for the 1690 ecliptic longitude of the central point, corresponding to an angular offset of 36°44' with the position of the 1690 equinox, leading to an interval of 2645 years between Chiron and 1690. To support his thesis, he then considers various stars from constellations crossed by the equinoctial colure, according to Hipparchus quoting Eudoxus[57]. Repeating the same calculation, he finds an average angular lag of 36°29', which corresponds to 2627 years. Subtracted from 1690, this places the definition of the sphere (occuring shortly before the Argonautic expedition) 43 years after the death of Solomon in Newton's system. He then performed a similar calculation for the solsticial colure, defined by δ Cnc for southern Asses in Cancer, δ Hya for Hydra's neck, ι Argo for the place between stern and mast of the Ship, θ Sge at the tip of the arrow, and η Cap in the middle of Capricorn. Note that this second calculation is simpler because the solsticial colure is perpendicular to the ecliptic and no longitude correction should be applied. The average angle is then 36°29', confirming previous result. Newton mentions that you can also add to the solsticial colure the stars of the neck and the right wing of the Swan (η and κ Cyg) and the left hand of Cepheus (o Cep) and the tail of the Southern Fish - but this time he does not quote any value. Indeed, a comparison with Flamsteed's catalog shows that the stars of Cygnus are further away from the average position than others (yielding 38.5° and 40.5°, respectively instead of 36.5°) and adding o Cep is strange because its coordinates place it at best close to the equinoctial colure[58] though Eudoxus, according to Hipparchus, clearly put Cepheus' left hand in the solsticial colure...

---

[48] Desh, Mercure de France, December 1755, p168-9
[49] Steuart, Apologie, p150
[50] Hardouin, Mémoires de Trévoux, p1569
[51] The same remark is made by Costard p79, Banier et Squire.
[52] Newton, *Abrégé*, entry for 939 : « Chiron définit constellations pour faciliter la navigation » ; Hardouin, Mémoires de Trévoux, p1578 and next ones.
[53] F.E. Manuel, p78-85 and appendix C of Buchwald & Feingold.
[54] This quick calculation is correct, but any integer times 26000 yrs can be added. Indeed, if a shift of 2° is detected, it could correspond to a 2*72=144 yrs interval, or (360+2)*72=26064 yrs if Earth's axis has made one full rotation, or (720+2)*72=51984 yrs for two rotations. No author brought forward this possibility, however.
[55] « Il [Chiron] plaça les points des solstices et équinoxes au 15$^e$ degré de ces constellations, càd vers le milieu des signes du Cancer, Capricorne, Aries et Scorpius. Ces signes n'étoient pas différents des constellations même. » (Abrégé); « Cette année donna lieu aux premiers Astronomes, qui formèrent les Conflellations, de placer les Equinoxes & les Solstices au milieu des Constellations d'Ariès, du Cancer, des Serres du Scorpion [la Balance] & du Capricorne. […] Eudoxe qui fleurissoit environ 60 ans après Meton, & 100 ans avant Aratus, en décrivant la Sphère des Anciens, mit les Solstices & les Equinoxes au milieu des Constellations d'Ariès, du Cancer, des Serres du Scorpion & du Capricorne, comme l'assure Hipparque […] Ainsi du tems de l'expédition des Argonautes, les points cardinaux des Equinoxes & des Solstices, étoient dans le milieu des Constellations d'Ariès, du Cancer, de la Balance & du Capricorne. » Newton, Chronologie, p85-89

---

[56] Flamsteed, Historia Coelestis Britannica, 1725. This catalog is never explicitly quoted by Newton, but he did use it as the coordinates show. At first, Newton used Hevelius catalog but he changed to Flamsteed's after 1700. An evidence of this change is hidden in rhe *Chronology*: η Per, quoted with Hevelius coordinates, is not listed by Flamsteed. Following Buchwald & Feingold (p286), Newton have changed catalog to avoid Flamsteed's critiques, to use the most recent reference, and… to have more stars, since Flamsteed lists more objects than Hevelius. This reflects well the "methods" of Newton, but the same authors, while defending his scepticism and dislike of hypotheses, also provide examples of text manipulations (p217-221, p224-8, p236-8, p279, p284-286, p294, p298, p305-6) and even of text modifications (p134, p162, p199 et p205) by Newton – see also below section 4bII for the omission of some text. All stellar names are from Bayer Uranometria (1603), and are common to all authors discussing the issue.

[57] ν Ari in the middle of the Ram's back, ν and ξ Cet at the Whale's head, ρ Cet for the last turn of Eridanus, τ and η Per for the head and hand of Perseus. It may be noted that Bradley Schaefer (2004, The astronomical lore of Eudoxus, JHA vol 35, p171-3, see also p275-6 and Table 8.3 in Buchwald & Feingold) performed the same exercise as Newton, but independently. He arrives at a similar value (985BC) but with a much larger dispersion (sigma of 468 yrs vs 84 yrs). Newton himself changed opinion on the exact date throughout his work (compare *Abrégé* and *Chronology*, and see also Table 8.6 p287 of Buchwald & Feingold).

[58] His Flamsteed coordinates are 5°41'55" ♉ with a latitude of more than 61°: it thus cannot indicate a solsticial colure, and the equinoctial colure passing by this star crosses the ecliptic at 12.8° ♓, not 6.5° ♉… Whiston notes the problem (see quote in Fréret, Défense, p 433) and Fréret also mentions it (Défense, p434). The latter author proposes that it is a typographical error, o having replaced δ which would better suit the Newtonian system, which is not true (see Fig. A.4). Following Buchwald & Feingold (p285), it is a simple error: if one neglects the sign and the correction due to latitude, the longitude value (5°42') is close to the 6°29' of Newton, so he may have mixed equinox and solstice. However, the error strangely helps Newton's demonstration, and one can find it quite surprising, in view of its simple nature and the harsh reactions to similar errors by Souciet: the error cannot be forgiven so easily, and could have been made on purpose.

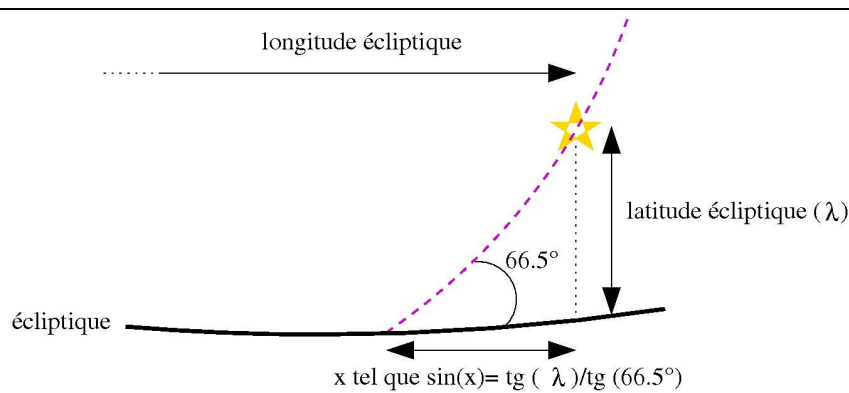

**Figure 1 :** *Correction (x) to apply to ecliptic longitude to find the location of the ancient equinoctial point (which is where the ecliptic and equinoctial colure intersects). The equinoctial colure, the great circle passing by the two equinoctial points and the two ecliptic poles, is tilted by 66.5° over the ecliptic.*

Several remarks can be made, about the choice of stars and the link with Chiron. Let us first consider the choice of stars made by Newton. To begin with, one can notice some obvious missing objects: while Newton repeat several times that the colure passes through the middle of Libra, he quotes no star of this constellation. In fact, there is a reason for this: a colure passing through $\alpha$ and $\beta$ Lib, heart of Libra, intersects the ecliptic at a distance of about 41-48° of the autumn equinox of 1690, not 36.5°! It is therefore impossible, in the Newtonian system to comply with all Eudoxus' requirements[59]. There is another notable absent: Arcturus ($\alpha$ Boo). If we adopt the ancient equinoctial colure of Newton, this easily identifiable bright star is about a degree from the colure[60], but Newton makes no mention of it... because Eudoxus does not say anything about it! In addition, the stars mentioned by Newton also pose various problems[61]. First, they are faint – fourth magnitude at best[62]. They are thus only visible with good eyes and on a night without moon or thin clouds. Would they be truly useful for navigation? The issue is particularly acute because brighter stars exist in most constellations: would the ancients have chosen so faint objects for pointing to celestial positions? Moreover, Newton uses non-traditional constellations. It is known through the descriptions of Eudoxus, Hipparchus or Ptolemy, that the ancient drawings of the constellations are not identical to the modern ones. Whiston quotes several cases: Perseus and Cepheus have changed in configuration, and the Arrow was not only elsewhere (near Aquila, then next to the current Dolphin), but its 17th century drawing did not include the star quoted by Newton, because it was added only a few years before the publication of his work[63]. Finally, even if the least problematic cases, some selected stars do not belong to the body part mentioned by Eudoxus: $\nu$ and $\xi$ Cet are in the neck and mane of the Whale, not in the head[64], $\rho$ Cet never belonged to the Eridan[65], and $\delta$ Hya is in the head and not the neck of Hydra... Emerson[66] and Whiston[67] clearly conclude that Newton chose his stars to fit his scheme.

---

[59] Whiston quoted by Fréret, Défense, p425-427
[60] Whiston quoted by Fréret, Défense, p438-439
[61] Whiston quoted by Fréret, Défense, p428-437
[62] Larger magnitudes correspond indeed to fainter stars.
[63] In fact, $\iota$ Argo and $\theta$ Sge do not exist in Ptolemy's catalog, but $\theta$ Sge is the sole Sagitta star that fits the Newtonian system (see Buchwald & Feingold p279), which is indeed not an excuse for its choice.
[64] However, Ptolemy's description agrees with Newton's choice, it is the definition of the Whale's head that changed with time (p276-7 in Buchwald & Feingold).
[65] It is probably due to Hevelius' maps (cf. Fig 8.13 p280 in Buchwald & Feingold).
[66] Emerson, p144
[67] « cette singulière tentative de M. Newton, de changer la figure ancienne des astérismes pour les ajuster à son hypothèse, jointe aux dix-huit différentes copies du premier chapitre de sa Chronologie, que l'on a trouvées chez lui, sont des preuves du plus fort et du plus long attachement que l'on ait jamais vu parmi les hommes pour une hypothèse. […] [je me] contente de dire que cet argument tant célébré non seulement porte à faux mais qu'il est expressément & directement contraire à la nouvelle Chronologie. » Whiston quoted by Fréret, Défense, p439

Using the same method as Newton, Whiston and Fréret then try to find a solution in better agreement with the descriptions of Eudoxus[68]. They conclude that the best offset would be 42°15', which would put the observations of the celestial sphere reported by Eudoxus in 1353 BCE and not 939. In this case, the equinoctial colure would pass between the star and θ and ε Ari, α and β Lib, near ι, φ, and κ Vir, and not far from Menkar (α Cet); the solsticial colure passes between κ and λ Leo, ς and θ Cap; the Northern tropic passes through the Beehive cluster, then between δ and γ Cnc, and through α Ser, κ and ι Oph, α Her, πCyg, θ And, ι Aur, and κ and ι Gem; the Southern tropic passes between γ and δ Cap, between η and ς Eri, and through ς and α Lep, β CMa, κ Argo, near τ and ψ Cen, λ and ν Sco, and δ Sgr; the celestial equator passes through π, μ, or ν Ari, α Ori, α Hya, δ Crv, α Lib, ς Oph, ν Aql, between Markab (α Peg) and Algenib (γ Peg), and near φ and χ Psc. This is the best solution[69], but it does not mean that Fréret and Whiston necessarily consider that Chiron lived in the 14th century BCE (see below and Appendix for figures).

In addition, an accurate calculation requires accurate data and this rarely exists in Antiquity. Shuckford therefore questions the accuracy of the ancient observations, supposed by Newton[70]: they did not have a star catalog before Hipparchus, and they knew neither the obliquity of the ecliptic or the actual length of the year. Even the stars chosen by Newton appear at two degrees on each side of the correct value[71], and it should be noted that a difference of four degrees corresponds to an interval of three centuries! Even if one "believes" that Chiron is the author of the sphere (see discussion below), how to get an accurate result in this case? La Nauze has a simple answer: the accuracy is based on the average from several stars[72]. However, the choice of the stars is made through modern reconstruction, since no specific star is clearly mentioned in the ancient sources – only vague terms such as "middle", "hand" or "head" are used. They correspond to vast areas (e.g. the back of Aries covers almost 6°!) and can therefore accommodate different solutions separated by several degrees equally well... Whiston also highlights a possible misunderstanding[73]: if their

---

[68] Fréret, Défense, p438-439, et p451-458. Whiston proposes a shift of at least two centuries: he first gives a date of 1353BC (p1010 of his appendix, copied on p439 of Fréret's Defense), then he mentions a shift by two centuries (and not the previous four, p1017 of his appendix, not copied by Fréret), but he never mentions which star he uses or provides details on his calculations (Fréret will provide them for the 4 centuries results, while the calculation for the two centuries is detailed in Buchwald & Feingold p365-6).

[69] It should be noted that there are several stars of the 4th and 5th magnitudes amongst them: their solution is thus only partly better than Newton's on that point. In addition, the declinations quoted by Fréret are imprecise with a few large errors – π Cyg is at more than 10° from the tropic, τ Cen at more than 5°, and φ Psc at more than 5° from equator. Buchwald & Feingold (p248) mention that Newton did not use tropics and equator because it is less simple to use, which is true, but if Fréret can perform the calculation, Newton certainly could do it too. Most probably, it provided no confirmation of his scheme, so that he omit them.

[70] The sacred and profane history of the world, vol II, see also Mémoires de Trévoux, p347-348

[71] It should be noted that, while supporting Newton on many points, Wood also questions the precision of observations in Chiron's time. Voltaire also criticize Newton (chap X, part III of Eléments de la Philosophie de Newton, 1773): since Hipparchus is the first to identify the precession, it « prouve que les Grecs n'avaient pas fait de grands progrès en astronomie » before ; if Meton and Euctemon had found such a big angular difference, they « n'auraient pu s'empêcher de découvrir cette précession des équinoxes… ce silence me fait croire que Chiron n'en avait point tant su que l'on dit, et que ce n'est qu'après coup que l'on crut qu'il avait fixé l'équinoxe du printemps au quinzième degré du Bélier. On s'imagina qu'il l'avait fait parce qu'il l'avait dû faire. Ptolémée n'en dit rien dans son Almageste, et cette considération pourrait, à mon avis, ébranler un peu la chronologie de Newton. »

[72] La Nauze, Mémoires de Trévoux, p2535 : « c'est en faisant servir par le résultat moyen, la diversité et la grossièreté de leurs observations à se corriger les unes par les autres qu'il nous montre aujourd'hui comment ils approchèrent plus ou moins de leur but, le milieu des constellations: ce qu'ils n'étoient nullement capables de vérifier eux-mêmes » For Buchwald & Feingold (chap II), the use of the average constitutes an original point in Newton's work. However, they provide several examples of others using it at the same epoch in demography, meteorology, and even astronomy. Indeed, many points of the *Chronology* have been debated, but the use of the average had never caused any problem and was never discussed, as any novel and revolutionary idea should have been. This text by La Nauze shows that averaging seems quite "natural".

[73] Whiston, quoted by Fréret, Défense, p425-427 – Buchwald & Feingold explain this subject in depth (p289-291) and insist on the fact that Fréret did not understand it. It could have been true at the time of the *Observations*, but it is

names are the same, signs and constellations are not the same thing! Indeed, colures are imaginary circles and asterisms have not been distributed in the sky to mark them... So, passing through the middle of one does not necessarily mean passing through the middle of the other. It is thus impossible to cross at the same time the middle of the *constellations* of Aries and Libra (see previous remark above) though this is easy for their respective *signs*. A little interpretation is thus needed to find when Eudoxus refers to constellation and when he uses the sign - Whiston proposes a quite reasonable solution to try to disentangle them.

Finally, there are assumptions. Although there are some small numerical errors in Newton's writings[74], the calculations may be regarded as essentially correct (unlike the case of Souciet!). The correctness of the result thus only depends on the correctness of the assumptions, as Fréret[75] and La Nauze[76] mention. Fréret identifies two assumptions[77]: Chiron had made the constellations for the use of the Argonauts, and Eudoxus used the same sphere. Even though they are not explicit in Newton's writings, it is indeed what he supposes: the text surreptitiously passes from equinoctial/solsticial points described by Eudoxus (p86) to those of the Argonauts (p89, see excerpt above) without any explanation or justification for the link between the two.

However, assuming that the sphere of Chiron, and the position of colures on it, are identical to those Hipparchus attributed to Eudoxus[78] is a very strong assumption, for which evidence is lacking, although Emerson[79] strangely sees it as "a point of history." Fréret[80] only finds a poetic text mentioning the definition of constellations by Chiron, while many others, by the same author and / or of the same epoch, never talk about constellations in connection with Chiron! The link between Chiron and Eudoxus is not obvious, though it is essential to Newton's system as underlined by Fréret[81]. Halley[82], far from supporting Newton on this point, sees it as "the most questionable part of the entire system." It is indeed possible that the position of the equinoctial and solsticial points were measured in the middle of Aries, Libra, Capricorn and Cancer before the time of Eudoxus, who would have only copied an older description[83], but binding these positions to the Argonauts is an important step, taken blindly by Newton (and Souciet and La Nauze[84] with him!) but Halley is

---

certainly not the case at the time of the *Défense*, as the associated text shows.

[74] A few errors can be found in Newton's writings: calculation errors, erroprs when copying Flamsteed's catalog (for ν Ari, τ Per, ξ Cet, and δ Can), error for the obliquity (the one used is not that of 939 BCE), and even ghost star (not in Flamsteed's catalog, see note 56). They were corrected at the time of publication of complete works by Samuel Horsley in 1795 (see also Table 8.2 in Buchwald & Feingold). However, these errors have no great consequence, because they change the results by a few years only.

[75] « Je ne prétendois point attaquer la justesse de ces calculs […] mais comme ces calculs supposoient des faits, je demandois la preuve de ces faits allégués. » Fréret, Défense, beginning of 3$^{rd}$ part, p384 (talking of *Observations*)

[76] Explaining to Souciet : « vous supposez que le sentiment d'Eudoxe, dont il s'agit dans Hipparque est le sentiment personnel d'Eudoxe, le sentiment qui lui étoit commun avec Méton dans le système de M. Newton. Ce n'est pas cela. L'opinion attribuée à Eudoxe par Hipparque est l'opinion de Chiron. » La Nauze, 1$^e$ lettre, p395

[77] « il dépend entièrement de deux suppositions, sçavoir 1° que Chiron avoit dessiné une sphère céleste pour l'usage des argonautes, 2° que cette sphère étoit celle qu'avoit suivie Eudoxe ; deux choses qui sont avancées gratuitement. » Fréret, Défense, p415

[78] cf. La Nauze, 1$^e$ lettre, p355

[79] Emerson, p139 et 143

[80] Fréret, Défense, p418

[81] « M. Newton n'ayant aucune preuve que la sphère d'Eudoxe fut la même que celle de Chiron, il ne peut rien conclure pour le temps de Chiron » Fréret, Défense p 442

[82] Halley, Philosophical Transactions, p206. Buchwald & Feingold (p328 and p376-9) argue that Halley support Newton's chronology, but this sentence shows the contrary: Halley's support is limited to showing Souciet's errors – in fact, Halley could even have avoided any mention to that hypothesis since Souciet accepted it!

[83] Fréret, Défense, p417

[84] La Nauze, 1$^e$ lettre, p 360 : the only solution to destroy Newton's system is to show that Chiron did not make the sphere, but he judged this possibility unrealistic; La Nauze, Mémoires de Trévoux, p2532 : « comment imaginer que ce n'ait pas été la sphère primitive ? »

careful enough to avoid supporting such a fantasy. Even Souciet[85] acknowledges the fact in his last essay: "no Ancient formally says where Chiron placed the cardinal points."

c) Additional arguments used by Newton
    I. Precession in Meton and Hipparchus times

To support its precessional calculations, Newton also use what was known of two ancient astronomers, Hipparchus and Meton. Being more recent than the famous expedition, the position of the equinoxes and solstices in their time should no longer be in the middle of signs / constellations. Even better, the epochs of these scholars is in fact well known: thanks to Ptolemy, it was known that Meton observed the solstice of 432 BCE and that Hipparchus worked between 158 and 128 BCE. The method of precession-dating can thus be easily checked with them, bringing further support to the chronological system of Newton.

In the *Abrégé*, for year 939, Newton mentions that Meton observed a 7° difference in solstice position compared to Chiron's times. In Keil's letter, Newton insists, stating that there is a difference of 8° between the positions of colures of his time and those in Meton's time and of 4° when considering Hipparchus[86]. The same idea is repeated, with more detail, in the *Chronology*[87].

While the texts referred to by the English scientist leave no room for doubt (though the absence of excerpts from original texts can be regretted), Fréret offers a broader view[88], because all sources should be considered, not only those supporting a given thesis. He takes the case of Eudoxus who "spoke otherwise in his works." In Geminus' calendar, the spring equinox of Eudoxus appears in the 6$^{th}$ degree of the sign of Aries and the winter solstice in the 4$^{th}$ degree of the sign of Capricorn (i.e. a 2° deviation from the expected right angle!) while in his *Enoptron* quoted by Hipparchus, Eudoxus mentions these points in the 15$^{th}$ degree of the sign and in Columella's calendar, Meton and Eudoxus agree to place these points at the 8$^{th}$ degree[89]. Some ancient texts also mentions that Hipparchus differs from Eudoxus by 15° in longitude, but also that Euctemon, Meton's colleague, places the equinoxes in the first degree of the signs in Geminus and Callipus calendars. In short, there is no agreement between ancient texts! According to Fréret, this is not surprising: traditions die hard, and remain in use for a long time before they change. Indeed, a one-degree precision is far from essential to farmers to which these calendars were intended. A good example in this context is the case of Columella[90], who even gives different dates for the (heliacal or achronic) rising of the same stars! Mentioned dates and angles probably come from observations made at different times and then compiled in the same book. Different angles can also be explained by changes in the coordinate system – e.g. definition of sign with respect to the associated constellation or definition of equinoxes or solstices within the sign (at the beginning of the sign, like today, or in the middle), an hypothesis favored by Newton if Fréret is to believe, but not obvious in the *Chronology* (see also La Nauze below).

---

[85] Souciet, 5$^e$ dissertation, p127
[86] Souciet, 1$^e$ dissertation, lettre de Keil reproduite p 56
[87] « Dans l'année de Nabonassar 316 […] Meton & Euctemon observerent le Solstice d'Eté, […] & Columelle nous dit qu'ils le trouvèrent dans le huitième degré du Cancer, qui est au moins de sept degrés plus reculé que la première fois. Or l'Equinoxe rétrograde de sept degrés en 504. ans […] on trouvera l'expédition des Argonautes comme ci-devant, 44. ans ou environ, après la mort de Salomon. […] Hipparque célèbre Astronome, en comparant ses observations avec celles qui avoient été faites avant lui, s'apperçut le premier que les Equinoxes avoient un mouvement par rapport aux étoiles fixes, en s'éloignant d'elles contre la suite des signes. Il crut d'abord que ce mouvement étoit d'environ un degré en cent ans. Il fit les observations des Equinoxes entre les années de Nabonassar 566. & 618. L'année moïenne est 602 qui est 286 ans après l'observation de Meton & d'Euctemon ; or l'Equinoxe rétrograda durant ce nombre d'années, de 4°. Ainsi il fut dans le quatrième degré d'Ariès du tems d'Hipparque,& par consequent il avoit rétrogradé de 11°, depuis l'expédition des Argonautes » Newton, Chronologie, p96-97
[88] Fréret, Observations p62-67, see also Défense, p461-483
[89] By following Newton's arguments, several problems arise (Fréret, Observations, p75-6)... IF the sings begins at γ Ari, the Ram's ear. Indeed, this star has a zero ecliptic longitude in 388BCE, since it has an ecliptic longitude of 29° in 1690. A colure located 15° away implies a time 15*72=1080 yrs before, implying a difference of five centuries with Newton's value. A colure located 8° away implies for Meton a time in the 10$^{th}$ century, while it is known that he lived 5 centuries later. This anachronism is due to the choice of the beginning, and Fréret at first (in his Observations) makes the same error as Souciet by beginning the signs at the Ram's ear. This error will be spotted by Newton (p318 des Philosophical Transactions).

Souciet quotes the same sources as Fréret and recalculated precession shits for the epochs of Meton and Hipparchus. He shows that it is impossible for them to have observed a solstice at 8° or 4° of the signs[91]. Indeed, there is a 29.6° offset between 1700 and 432 BCE, when Meton worked, and the offset amounts to 25.4° for the time of Hipparchus – both values are close to an entire sign, and therefore not compatible with shifts of 8 or 4°. In the same vein, he insists that, when Ptolemy speaks of equinoxes or solstices observations made by Hipparchus or Meton, he never specifies their positions on the sky [92]. He even says that Hipparchus considers almost all mathematicians to put the "cardinal" points at the beginning of the constellations, except Eudoxus – i.e. he does not explicitly exclude Meton, who was well known[93], so Meton should have used the same convention. Souciet proposes a plausible explanation for the 8° problem: since only Columella attributes this value to Meton, why would there not be a copying error[94]? La Nauze sees no problem and reconciles the different sources: for him, it is simply a matter of convention! If we imagine that Meton uses the same convention as Chiron, with the beginning of Aries sign 7° 24 ' before the Ram's ear, then the colure passed in his time in the 8$^{th}$ degree of signs; Euctemon would instead have chosen a different convention, still used today, where equinox begins the Aries sign. This reconciles the two colleagues[95]... Similarly, for Hipparchus' time, an equinox at 4° in Chiron's convention is compatible with an equinox in the first degree if Euctemon's convention is used[96].

The two adversaries also discussed another point brought forward by Hipparchus: the star in the middle of the back of Aries is at least one-third of a sign, or 11°, from the equinoctial colure. Souciet[97] states that this puts the colure near the ear because there are about 11° between ν Ari (mid of the back) and γ Ari (ear), which is correct. This suits Souciet as the ear then mark the beginning of signs in Hipparchus' time, which he advocated (see Fig. 4). However, putting the beginning of the sign some 7.5° west of the Ram's ear, as suggested by Newton, implies the colure to pass close to the head and not the back (see Fig. 2 – it should be noted that Souciet mixes Hipparchus' time with that of Chiron, as he uses the convention for beginning of signs attributed to Chiron by Newton, not the one used by Hipparchus!). La Nauze rather sees in this text an additional indication of the correctness of Newton, i.e. further evidence that the equinox was at 4° in Hipparchus' time, since there is a 15° shift between middle and beginning of sign and a 11° shift between Hipparchus and Chiron hence a 4° difference[98]. Indeed, Newton states in its response to the Fréret in the Philoshical Transactions that the Ram's ear is at 7°36' in ecliptic longitude to the east of the Chiron's colure, so La Nauze concludes that Hipparchus' colure is at 3°24' (= 11° -7°36 ') west of the Ram's ear, which corresponds to the expected position of the vernal equinox in the time of Hipparchus... However, the authors based their discussion on a misunderstanding, as they do not calculate things in the same reference frame: Souciet measure the distance of the stars by simply subtracting ecliptic longitudes while La Nauze here refers to Newton's method, which take into account the inclination of the equinoctial colure on the ecliptic when determining the position of Chiron's equinox (Fig. 2, see also Fig. 3). There is no hope of reconciling views despite the fact that their arguments (colure passing through the ear and colure at 3° 24 ' from ear in ecliptic longitude) are not mutually exclusive, on the contrary! Moreover, both seem to forget in their calculations that Hipparchus worked in right ascension and not ecliptic longitude, an argument on which they both agree (see note 19) ...

---

[90] Fréret, Défense, p481
[91] Souciet, 1$^e$ dissertation, p62-68, see also Fig. 4
[92] Souciet, 1$^e$ dissertation, p64-65
[93] La Nauze (1$^e$ lettre, p392) writes that Hipparchus states in his introduction that he will only quote Aratus and Eudoxus, which would explain the absence of quote to Meton.
[94] Souciet, 1$^e$ dissertation, p 69-70 : a A transformed into H, i.e. a 1 changed into 8 in the Greek number notation.
[95] La Nauze, 1$^e$ lettre, p383-392, see also Fig. 4
[96] La Nauze, 1$^e$ lettre, p 375-376, see also Fig. 4
[97] Souciet, 5$^e$ dissertation, p145
[98] La Nauze, 1$^e$ lettre, p 369-373, and 5$^e$ lettre, p410-414, see also Figs. 4 and 5

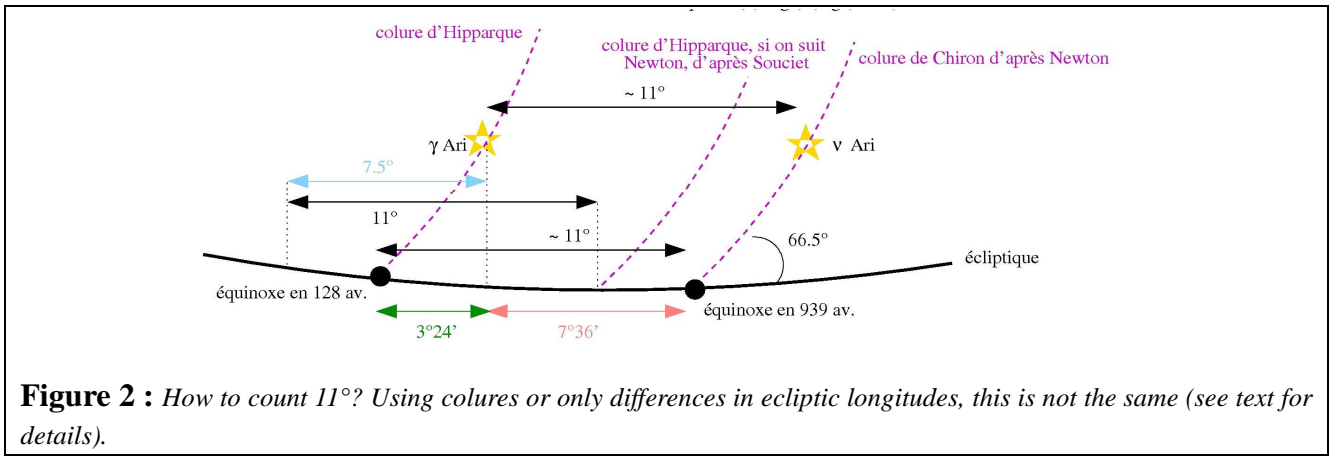

**Figure 2 :** *How to count 11°? Using colures or only differences in ecliptic longitudes, this is not the same (see text for details).*

To settle the issue, Fréret prefers to look at the old text within its global context, analyzing the long descriptions of constellations made by Hipparchus, where he compares his descriptions with those given by Eudoxus[99]. To quote a few examples: the colure passes through the neck of the Hydra (whose first star is ω Hya) for Eudoxus and through the head (δ or σ Hyas) for Hipparchus, which makes a difference of at least 16°; the colure crosses the left hand of Bootes, which covers 6° in longitude (κ, g, and ι Boo) for Eudoxus, while these stars are placed in the 13$^{th}$ degree of Libra at the time of Hipparchus, implying a difference larger than 13° on average. A similar reasoning applies to Southern Fish, the Swan, Cepheus, and the Centaur. However, better than vague regions, Hipparchus quotes the case of four stars, all well identified: α UMa is at the 18$^{th}$ degree of Cancer for Eudoxus and 3$^{rd}$ degree of Lion for Hipparchus, a difference of 15°; η UMa moves, according to Hipparchus, from the 4$^{th}$ to the 18$^{th}$ degree of Libra when passing from a reference frame linked to the beginning of the signs (his case) and a colure in the middle of the signs (Eudoxus' case); γ UMa goes from 10$^{th}$ degree of Virgo to 25$^{th}$ degree of Leo; and the Pole Star from the 18$^{th}$ degree of Pisces to the 3$^{rd}$ degree of Aries. The difference of a third of a sign appears indeed in Hipparchus' text for Aries, but the star under concern cannot be identified so easily: Newton therefore made a rather selective choice, suiting his system ...

Finally, Newton also mentions the determination of the precession by Hipparchus. In fact, he thinks that Hipparchus derived his value of one degree by century for the precession because the Greek thought there were about 11 centuries between his time and that of Argonauts and had measured a precessional shift of 11°[100]. He added that this bad estimate was the cause of the faulty

---

[99] Fréret, Défense, 3$^{rd}$ part, section II, p444-450

[100] « fut dans le quatrième degré d'Ariès du tems d'Hipparque, & par conséquent il avoit rétrogradé de 11°, depuis l'expédition des Argonautes ; c'est-à-dire en 1090. ans, si l'on suit la Chronologie des anciens Grecs, qui étoit alors en usage : ce qui donne environ 99 années, ou en prenant un nombre rond, 100. ans pour un degré, comme Hipparque l'avoit alors déterminé. Mais il est certain que l'Equinoxe rétrograde d'un degré en 72. ans, & de 11° en 792. années. Ainsi comptant ces 792 années en rétrogradant depuis l'année 602 de Nabonassar, qui est l'année d'où nous avons compté les 286 années, on placera par ce calcul l'expédition des Argonautes environ 43. années après la mort de Salomon. Les Grecs avoient donc fait l'expédition des Argonautes de 300. ans plus ancienne qu'elle ne l'étoit effectivement, & cette erreur donna occasion à Hipparque de déterminer la rétrogradation des Equinoxes d'un

Greek chronology[101]. Therefore, the "false" Greek chronology, far from undermining his system, supports it! Obviously, Newton avoids to follow his reasoning to its strange end[102] that Hipparchus would have discovered precession through observations, but then determined the rate by using an uncertain timeline...

Souciet wonders what may well be the origin of this bold assertion by the English scientist. There exists no trace of it in history, and it also seems clear that Hipparchus did not derive the date of Meton's epoch with this method, nor did Ptolemy use it for deriving Hipparchus' time: they would have found much older epochs than what they report, because of the error in the precession estimate[103]. La Nauze answers that Newton was not the source of this idea, but Newton's opponents[104]. Indeed, it can be found in Fréret's *Observations*, for the estimated time of Seneca[105], without any source to support that idea. He then explained that the Greeks did not use this type of reasoning before Hipparchus' discovery and even then, they applied it only for very old times, not Meton's, Hipparchus' or Ptolemy's epochs. Probably seeing the weakness of his argument, La Nauze warns that this idea is not a necessary condition for the Newtonian system, it is only a consequence, a probable conjecture... and he swiftly accuses Seneca of chimerical chronology[106].

In the *Défense*, Fréret decides to play Newton's game[107]. Having established an angular shift of 15° between his observations and those of Eudoxus (see above) and assuming that Chiron lived 1100 years before him, Hipparchus would have had found a value for the precession rate of 15° / 1100 years or 1° every 73 years, something close to the modern value! Since Hipparchus gives a value of one degree per century, Fréret concludes that he had not used the Eudoxus' sphere to determine the value of the precession. However, he argues that the observed difference could have been an inspiration for his subsequent discovery of a discrepancy between his observations and those of Aristille and Timocharis...

<u>II. Thales & Hésiod</u>
To support his shortened timeline, Newton attempts to provide additional elements, a complementary approach to the stellar calculations mentioned above. In the *Chronology*, he explains[108] that Thales determined the cosmic setting of Pleiades to the twenty-fifth day after the

---

degré seulement en cent ans. » Newton, Chronologie, p97-98

[101] « Hipparque comptait un degré par siècle pour la précession, et ils [les Grecs] ont généralement fondé leur chronologie sur cette estimation » letter from Keil, reproduced on p 56 of Souciet, 1$^e$ dissertation : « Hipparcus counted the precession to be a degree in a hundred years, and they are generally founded their Chronology upon that computation »

[102] F.E. Manuel, p74-5

[103] One degree by century rather than one degree by 72 yrs (Souciet, 1$^e$ dissertation, p58)

[104] La Nauze, 1$^e$ lettre, p 363-369

[105] Fréret, Observations, p81-82

[106] La Nauze, 5$^e$ lettre, p398-399

[107] Fréret, Défense, 3$^e$ partie, section I, p415-417

[108] « Après l'expédition des Argonautes on n'entend plus parler de l'Astronomie, jusqu'au tems de Thalès [...] Pline assûre qu'il détermina le coucher cosmique des Pléiades au vingt-cinquiéme jour après l'Equinoxe de l'Automne : d'où le P. Petau calcule la longitude des Pléiades en 23°53'♈ : par consequent la luisante des Pléiades s'étoit éloignée de l'Equinoxe de 4°26'52'' depuis l'expédition des Argonautes : ce mouvement répond à 320 années [...] si l'on compte ces années en rétrogradant depuis le tems que Thalès étoit jeune [...] environ la XLI Olympiade, on trouvera par ce calcul que l'expédition des Argonautes arriva environ 44. ans après la mort de Salomon, comme on a trouvé ci-dessus. » Newton, Chronologie, p95-96 [Note that "cosmic setting is an incorrect translation, it is occasus matutinus in the original English version; see also appendix E in Buchwald & Feingold for detail of the calculations of rising and setting in Newton's time.]. Just before this paragraph, Newton calculates the expected longitude of Pleiades in Chiron's time by subtracting 36°29'. He finds 19°26'8" ♈, and the difference with Pétau's value of 23°53' ♈ gives 4°26'52". Following Buchwald & Feingold (note 26 p256 and p482-4), this star must be η Tau, the brightest star in Pleiades, whose ecliptic longitude is given by Flamsteed to be 25°37'13"♉ in 1686 and 25°40'8"♉ in 1690. However, the reconstructed longitude (Newton never provide the original value) is 19°26'8" ♈ + 36°29' =

Autumn Equinox, which led Pétau determine an ecliptic longitude of about 24° ♈. This reasoning is correct, as a simple estimate shows: a setting while the sun rises corresponds to a longitude at 180° (or six signs) from the Sun, for a star close to the ecliptic; knowing that the Sun travels 360° on the ecliptic in one year (so it moves by about 1° per day), an interval of 25 days corresponds to about 25° of longitude; the position of the Sun at the Autumn equinox is 0°♎ and about 25°♎ 25 days later, and the opposite location then has a longitude of 25°♈. For a star further away from the ecliptic, a correction must be applied, taking into account the latitude of the place of observation, which led to the 24° ♈ of Pétau. Knowing the ecliptic longitude in 1690 and in Thales' time, one can deduce the time interval between the two, and find the date of Thales' observation. Since this happened during the 41$^{st}$ Olympiad, the precession method can be checked. Knowing the coordinates of Pleiades at the time of the sphere definition, Newton also confirms his dating of the original sphere (but there is no link whatsoever with Argonauts in this text).

Newton also writes[109] that Hesiod saw Arcturus rising at sunset 60 days after Winter solstice, placing him a century after Solomon's death. Fréret confirms this estimate[110], stating that Longomontanus had found a date of 970 BCE, Kepler 930 and Riccioli 950 for that observation. Horsley however mentions that, when re-doing the calculation, he arrived at a date close to the Argonautic expedition in Newton's scheme, i.e., before Trojan war, which is confirmed by calculations found in Newton's manuscripts – he probably has somewhat "embellished" the truth here[111]. Celestial simulations with up-to-date softwares confirm that this observation is possible at the time of Hesiod from his workplace. However, deriving an exact date for the observation is difficult because it depends on the horizon along the line-of-sight (flat or mountainous), but also of the height above the horizon to actually "see" Arcturus (the stars on the horizon are strongly extinguished and therefore unobservable). Of course, this method enables us to date only Hesiod's time, the date of the Argonauts is only indirectly derived, knowing that Hesiod lived one generation after Trojan war which occurred a known amount of time after the Argonautic expedition – the use of "generation" with the meaning of a few decades has been debated. It should be noted that Hesiod shows the selective choice of texts by Newton. Indeed, the same text by Pliny relating Thales' observation also mentions that the event occurred *at* the equinox in Hesiod's time (i.e., the Pleiades had a zero longitude at that epoch). Since Newton finds that the Pleiades were in the 20$^{th}$ degree of Aries when the sphere was defined, this 20° shift implies that Hesiod lived more than 1400 yrs before the sphere definition, which leads to a problem for the dating of the expedition in Newton's system[112].

d) Additional points not discussed by Newton

If Newton discussed only the points adressed above in the *Abrégé* and *Chronology*, his detractors or his supporters sought to find every consequence of the Newtonian system, and discuss a few additional points. This section focuses on them.

The main objection to the *Abrégé* by Fréret in his *Observations* and (especially) Souciet concerns the beginning of the signs. For them, the signs begin with the first star of Aries, which they identify with γ Ari, the Ram's ear. They add 15° to the coordinates of that star to find the position of Chiron's colure, and derived that this position is at 44° from the position of the Spring

---

25°55'8" ♉, which do not correspond to n Tau, nor any star in Flamsteed's catalog (star s Tau is the closest).
[109] « Hesiode nous dit que de son tems soixante jours après le Solstice d'hiver, l'étoile Arcturus se levoit précisément quand le Soleil se couchoit : on voit par là qu'Hesiode fleurissoit environ 100. ans après la mort de Salomon, ou dans la Génération ou l'Age qui suivit immédiatement la guerre de Troye, comme Hesiode le déclare aussi lui-même. » Newton, Chronologie, p98
[110] Fréret, Défense, p 460
[111] Opera Omnia, Newton (edited by S. Horsley, 1795), p75 and Buchwald & Feingold, note 121 p296
[112] Opera Omnia, Newton (edited by S. Horsley, 1795), p73

equinox of 1700, which gives a date of 1470 BCE for the expedition[113]. In the Philosophical Transactions, Newton answered Fréret that the colure is at 7° 36 ' east of the ear[114]. Souciet then deduces that the sign of Aries begins in the middle of nowhere in Newton's system, because there is no star located at 15° of colure, or at 7°24' west of the ear. He accuses Newton to "adjust Astronomy to his system"[115] and he believes his solution is best because it considers more useful to start a reference frame with a bright star. In addition, Sextus Empiricus and Macrobius mention that signs begin at remarkable stars – even if they refer to Egyptians and Chaldeans; this is not a problem for Souciet since the Greeks have inherited many things from them[116].

The answer to this objection is twofold. First, beginning the sign of Aries at the Ram's ear is not compatible with having a colure passing in the middle of the back of Aries - a colure located 15° away from the ear rather crosses the tail[117]. Halley and then Newton in the *Chronology* (p91) indeed showed that the star ν Ari, located in the middle of the back, is compatible with Newtonian calculations. On the other hand, the beginning of the signs do not seem very important because, unlike the Egyptians and Chaldeans mentioned by Sextus Empiricus and Macrobius, the Greek texts only mention the middle of signs[118]. Of course, the constellations being of variable widths while each sign covers 30°, it seems easier to spot the center of a constellation that virtual points located at 15° on each side. La Nauze (for once) mentions that Souciet does not take into account the inclination of the colure nor the fact that Ancients worked in right ascension, as noted by Souciet himself. However, La Nauze himself corrects none of the calculations, and thus simply places his arguments in Souciet's erroneous steps.

La Nauze and Souciet also discusses the case of Spica (α Vir). According to Souciet[119], Hipparchus places this star at 24° ♍, i.e. near the colure following Riccioli. To check, Souciet places the beginning of the signs at the Ram's ear and found that the 1700 equinoxes would then appear at 59' ♓ (for Spring) and 59' ♎ (for Autumn, see Fig. 3). Going back to 128 BCE, Hipparchus' time, he found the equinoxes at 26° 22' of the same signs, which he considers as not too far from the theoretical value of 30°. It then returns to Spica's longitude: 24° ♍ as stated by Pétau (one can find the same result by precessing Flamsteed's catalog). Again, he considers this value as not far from 26° 22'. Spica was therefore almost on the colure, and the colure was near the beginning of the signs. Souciet believes that an error of a few degrees, as observed, is not impossible, given the lack of specific and precise astronomical instruments at the time. Unfortunately, he makes some mistakes in his calculation (see Fig. 3): Pétau's longitude of Spica is valid in the usual system, starting at the Spring equinox, not in Souciet's system beginning at the ear; if that problem is corrected, then Spica is located at 20.5° ♍ or more than 6° from the colure, which is a rather large error. La Nauze rather places the beginning of the signs at about 3° from the Ram's ear: in this case, the error on the position of the equinox is lower (a few degrees), but it remains high for Spica (about 6°), but La Nauze considers that mistake as less severe than a mistake on the equinoctial position. La Nauze does not notice Souciet's coordinate error[120], however, and he is not very fair[121]: first, if we accept the Souciet's calculations and a positioning error of a few degree, there is no incompatibility in Souciet's writings; second, he is mistaken when stating that Spica is misplaced at 24° ♍.

---

[113] Souciet, 1ᵉ dissertation, p 52 ; Fréret, Observations, p75-6
[114] Newton, Philosophical Transastions, p318
[115] Souciet, 5ᵉ dissertation, p120
[116] Souciet, 5ᵉ dissertation, p129 et 150-155
[117] Halley, Philosophical Transactions, vol 34, p207-209 ; La Nauze, 1ᵉ lettre, p354-360
[118] La Nauze, 1ᵉ lettre, p354-360 ; 5ᵉ lettre, p400-401 et 431-440
[119] Souciet, 1ᵉ lettre, p60-63
[120] He however makes another error, quoting a surprising 17° ♍ for Spica, in his note p 377.
[121] La Nauze, 1ᵉ lettre, p377 et 383

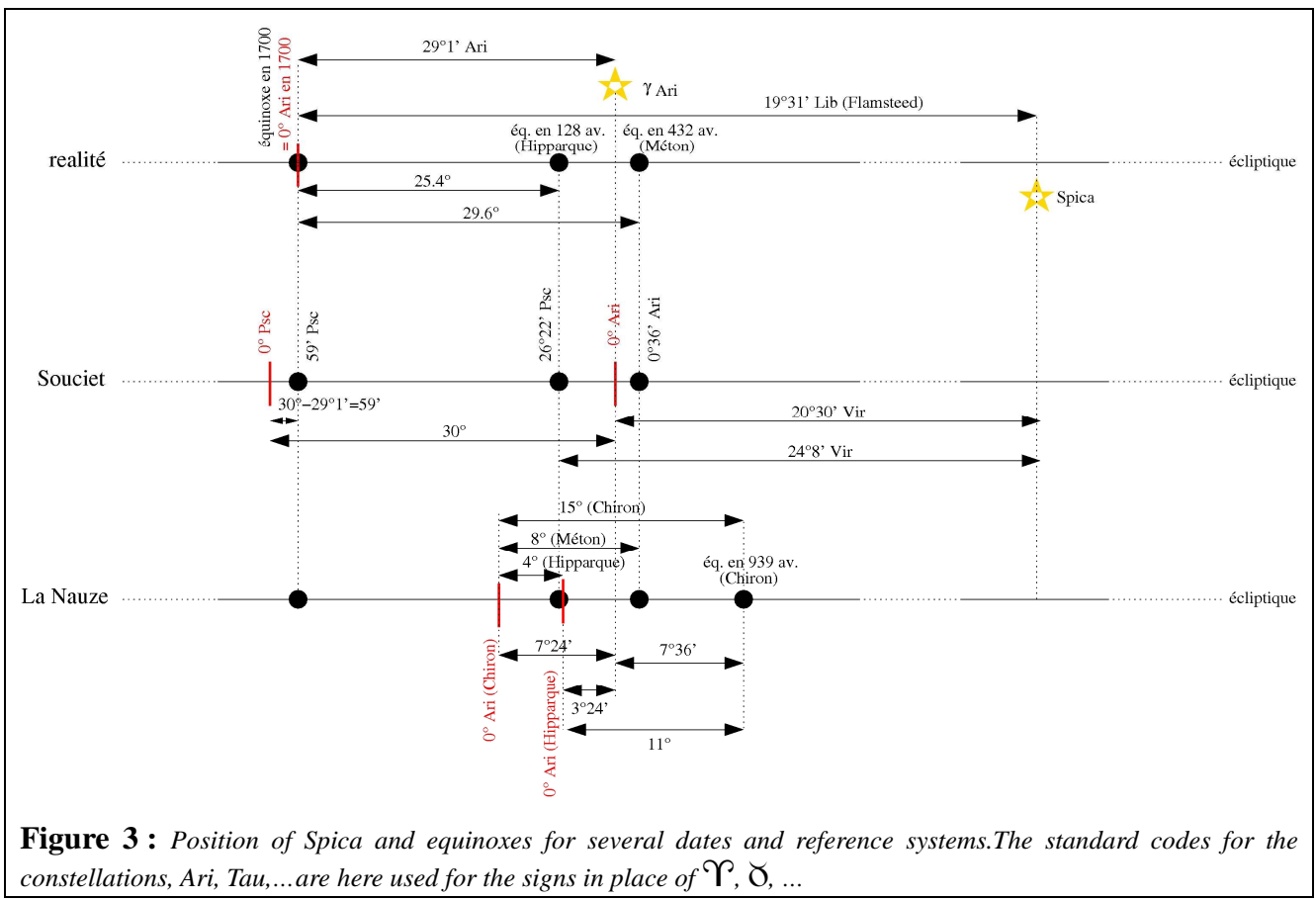

**Figure 3 :** *Position of Spica and equinoxes for several dates and reference systems. The standard codes for the constellations, Ari, Tau,…are here used for the signs in place of ♈, ♉, …*

The identification of the first star of Aries is another problem, dear to Souciet. If the ear is chosen by both Newton and Souciet (in his first essay), another star seems to replace it in Antiquity: according to Hyginus, there was a first magnitude star in the front foot of Aries[122]. Souciet guesses that it was very close to the ear[123], i.e. with a negligible coordinate difference. La Nauze corrects him on that point[124]. For this star to rise before that of the ear, as noted by the Ancients, there needs to be a few degrees difference in coordinates (not only 45 '), because the Ram's foot is further south than Ram's head and would rise second if the coordinate difference was smaller. This appears consistent with beginning of signs located at a few degrees west of the ear in the time of Hipparchus, as in the Newtonian system. Souciet's 45' can only refer to a difference of right ascension[125]. In addition, he considers that Souciet misinterpreted "in pede of posteriobus primo unam" from Hyginus' text reporting the existence of this mysterious star. He could not have mentioned first magnitude because it is Hipparchus who invented that magnitude system much later, so he just means the star of the first front foot. La Nauze does not identify this star, but Halley does, thanks to an accurate calculation where all hypotheses are demonstrated[126]: this star is η Psc, to the north of Pisces' node.

There is also the question of the head of the Whale or sea monster: does the Newtonian colure cross it, as specified in ancient texts? As Whiston does (see above), Souciet remarks that this is not the case[127]. La Nauze is not of the same opinion[128] but he contradicts himself[129] by ensuring that the head is narrow, contrary to what Souciet states, just after stating five pages earlier that the head covered a wide angular range between 29.5° ♈ and 11° ♉. Since the Ram's ear being at 29° ♈ and the beginning of the tail at 14° ♉, La Nauze deduces that what passes in the middle of the Ram passes in the middle of the head... which is far from true! La Nauze does not take into account the effect of the equinoctial colure inclination (Fig. 1, he will finally realize his error, as shown by his later text in the Mémoires de Trévoux, p2537). In addition, the identification of the stars is problematic: that at 11° ♉ is probably x or λ Cet, that at 29° ♈ γ Ari and that at 14° ♉ ε Ari, but where did he find a star of the Whale's head at 29° ♈? There is none. Stars of the Whale's head in Bayer's Uranometria are α, κ and λ Cet, μ Cet being already in the neck. Using their Flamsteed's coordinates, we find that the equinoctial colure passing near these stars intersects the ecliptic in 15.6°, 20.9°, 14.2°, 10.0° ♉, respectively. Newton's solution (6° 29' ♉) does not agree with these values, while that proposed by Whiston and Fréret (12° 15' ♉) appears much better.

---

[122] Souciet, 5$^e$ dissertation, p119
[123] He determines an angular distance of 45' by using another remark by Hipparchus (Souciet, 5$^e$ dissertation, p120), but he is mistaken because Hipparchus mentions a 45' distance between colure and first star, without stating that that star is the Ram's ear, which is only Souciet's hypothesis! He will again quote that text (p150) asserting this time that Hipparchus mentions that the first of the three stars of the Ram's head, which is the ear in Souciet's mind, is distant by the 20$^{th}$ of an hour (or 45') from the colure. La Nauze (5$^e$ lettre, p429-431) considers that distance to be in right ascension, not in ecliptic longitude.
[124] La Nauze, 1$^e$ lettre, p 343-345 ; 5$^e$ lettre p 381
[125] La Nauze, 5$^e$ lettre, p429-431
[126] Halley, Philosophical Transactions, vol 35

Souciet also discusses some additional stars[130]: the head of Aries, again, the foot of Bootes, and various parts of the Centaur. He uses Hipparchus' texts while considering things attributed by Newton to Chiron's time! The result is an absurd mishmash, which indeed doesn't prove anything. La Nauze realizes the problem and reports it[131]. He also corrects Souciet on his bald assertion that a star in the first degree of the sign in Hipparchus' time would appear in 1700 at 29° of the same sign. It is obviously impossible, because it would put Hipparchus in the 4$^{th}$ century BCE instead of the 2$^{nd}$ century BCE (since $29 \times 72 = 2088$ years, and $1700 - 2088 = 388$ BCE).

Souciet, still tormented by the problem of the beginning of the signs, finally compares the definition of the zodiac in the Newtonian system (with a start more than 7° away from of the Ram's ear) with his own (with a start at the Ram's ear). Indeed, constellations and signs differ: the constellations cover 20° to 50° of ecliptic longitude, while each sign, by definition, occupies 30°. An overlap between signs and other constellations than the one with the same name is therefore unavoidable in some cases. However, the overlap can be minimized by choosing the best positions. Souciet takes the example of Aries[132]: as it covers only 20°, it is necessary to insert 10° somewhere. One can either put 5° on either side of the constellation, or put 10° before the constellation's start, or put 10° after or split the 10° into unequal intervals such as Newton's 7.5°. Reviewing the signs and constellations, Souciet shows that his system leads to a lower overlap than Newton. La Nauze disputes these findings. Not only does he consider the extension of Aries constellation to be 25°[133], but he also proposes to look at overlaps on both sides of the constellations, not only the beginnings as Souciet did[134]. He then shows that Newton's arrangement seems more balanced than Souciet's. Incidentally, La Nauze makes several errors, notably in the choice of "last" stars associated of the constellations. Table 1 and Fig. 4 compares the two systems: the Newtonian solution is indeed more balanced than Souciet's.

---

[127] Souciet, 5$^e$ dissertation, p144
[128] La Nauze, 1$^e$ lettre, p359
[129] La Nauze, 5$^e$ lettre – p416 and p421 are here compared
[130] Souciet, 5$^e$ dissertation, p148-150
[131] La Nauze, 5$^e$ lettre, p 422 and p426-429
[132] Souciet, 5$^e$ dissertation, p124-125 and p158-168
[133] La Nauze, 5$^e$ lettre, p394-396
[134] La Nauze, 1$^e$ lettre, p346-353 ; 5$^e$ lettre, p440-464

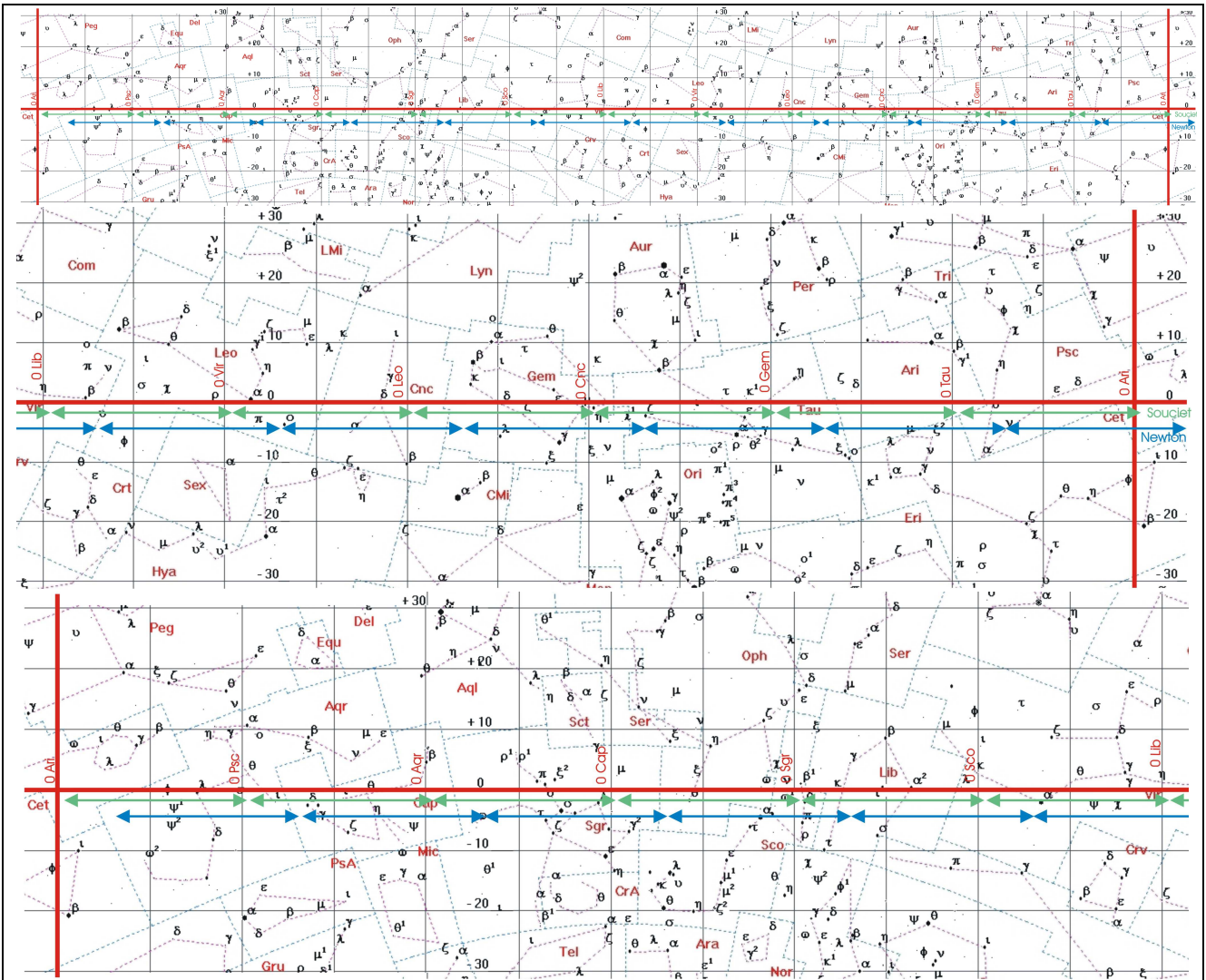

**Figure 4 :** *Splitting the zodiac following Souciet (green upper arrows) and Newton blue lower arrows) – the two lower panels are close-up of the top panel. The beginning of each sign for year 1700 is noted using the standard constellation codes, e.g. Ari (for ♈), Tau (for ♉). Ecliptic and meridian of zero longitude appear as thick orange lines.*

**Table 1 :** *Ecliptic longitudes for limits of signs and of constellations, using the Newtonian system (starting at 7° 24 'west of the Ram's ear) and Souciet's system (starting at the Ram's ear). The values provided by La Nauze appear in italics, the inconsistencies between the two tables of Souciet are shown in red and underlined. As a point of comparison, the last column gives the extreme longitudes of stars of these constellations in the Flamsteed's catalog. Note that Souciet and La Nauze consider a star for the signs in the Newtonnian system at 21° 25 ' in 1700, whereas 29° 1'-7° 24' = 21° 37 '. This comes from the usual confusion by Souciet between 7° 24 ' (star of sign-ear) and 7° 36' (ear-colure). Longitudes reproduced here correspond to the equinox 1700 for the first two columns and 1690 for the last.*

| Sign and constellation | Newton | Souciet | Flamsteed |
|---|---|---|---|
| **ARIES** ♈ | | | |
| First star of the constellation | *29°1' ♈* | *29°1' ♈* | 26°36' ♈ |
| Last star of the constellation | 19°13' ♉ | 19°<u>1'</u> ♉ | 21°6' ♉ |
| Beginning of sign | 21°25' ♈ | 29°1' ♈ | |
| End of sign | 21°25' ♉ | 29°1' ♉ | |
| **TAURUS** ♉ | | | |
| First star of the constellation | *14°7' ♉* | *14°7' ♉* | 16°49' ♉ |
| Last star of the constellation | 20°16' ♊ | 20°16' ♊ | 26°3' ♊ |
| Beginning of sign | 21°25' ♉ | 29°1' ♉ | |
| End of sign | 21°25' ♊ | 29°1' ♊ | |
| **GEMINI ♊** | | | |
| First star of the constellation | *29°14' ♊* | *29°14' ♊* | 26°37' ♊ |
| Last star of the constellation | 24°27' ♋ | <span style="color:red">27°27'</span> ♋ | 22°43' ♋ |
| Beginning of sign | 21°25' ♊ | 29°1' ♊ | |
| End of sign | 21°25' ♋ | 29°1' ♋ | |
| **CANCER ♋** | | | |
| First star of the constellation | *18°0' ♋* | *18°0' ♋* | 22°49' ♋ |
| Last star of the constellation | <span style="color:red">25°39'</span> ♌ | 12°41' ♌ | 12°20' leo |
| Beginning of sign | 21°25' ♋ | 29°1' ♋ | |
| End of sign | 21°25' ♌ | 29°1' ♌ | |
| **LEO ♌** | | | |
| First star of the constellation | *11°1' ♌* | *11°1' ♌* | 10°57' ♌ |
| Last star of the constellation | 20°50' ♍ | 20°50' ♍ | 20°42' ♍ |
| Beginning of sign | 21°25' ♌ | 29°1' ♌ | |
| End of sign | 21°25' ♍ | 29°1' ♍ | |
| **VIRGO ♍** | | | |
| First star of the constellation | *14°45' ♍* | *14°45' ♍* | 17°30' ♍ |
| Last star of the constellation | 29°33' ♎ | 29°33' ♎ | 8°17' ♏ |
| | *5°54' sco* | *5°54' sco* | |
| Beginning of sign | 21°25' ♍ | 29°1' ♍ | |
| End of sign | 21°25' ♎ | 29°1' ♎ | |
| **LIBRA ♎** | | | |
| First star of the constellation | *26°40' ♎* | *26°40' ♎* | 3°5' ♏ |
| Last star of the constellation | 29°23' ♏ | <span style="color:red">16°41'</span> ♏ | 27°4' ♏ |
| | 26°29' ♏ | 26°29' ♏ | |
| Beginning of sign | 21°25' ♎ | 29°1' ♎ | |
| End of sign | 21°25' ♏ | 29°1' ♏ | |
| **SCORPIUS ♏** | | | |
| First star of the constellation | *16°28' ♏* | *16°28' ♏* | 26°48' ♏ |
| Last star of the constellation | <span style="color:red">29°7'</span> ♐ | 23°31' ♐ | 20°15' ♐ |
| Beginning of sign | 21°25' ♏ | 29°1' ♏ | |
| End of sign | 21°25' ♐ | 29°1' ♐ | |
| **SAGITTARIUS ♐** | | | |
| First star of the constellation | *26°56' ♐* | *26°56' ♐* | 19°23' ♐ |
| Last star of the constellation | 24°18' ♑ | <span style="color:red">22°6'</span> ♑ | 26°38' ♑ |
| | *0°32' ♒* | *0°32' ♒* | |
| Beginning of sign | 21°25' ♐ | 29°1' ♐ | |
| End of sign | 21°25' ♑ | 29°1' ♑ | |
| **CAPRICORNUS ♑** | | | |
| First star of the constellation | *28°34 ♑* | *28°34 ♑* | 28°6' ♑ |
| Last star of the constellation | 21°53' ♒ | 21°53' ♒ | 21°29' ♒ |
| Beginning of sign | 21°25' ♑ | 29°1' ♑ | |
| End of sign | 21°25' ♒ | 29°1' ♒ | |
| **AQUARIUS ♒** | | | |
| First star of the constellation | *7°35' ♒* | *7°35' ♒* | 7°24' ♒ |
| Last star of the constellation | 24°52' ♓ | 24°52' ♓ | 15°58' ♓ |
| Beginning of sign | 21°25' ♒ | 29°1' ♒ | |
| End of sign | 21°25' ♓ | 29°1' ♓ | |
| **PISCES ♓** | | | |
| First star of the constellation | *14°25' ♓* | *14°25' ♓* | 11°6' ♓ |
| Last star of the constellation | 25°33' ♈ | 25°33' ♈ | 28°22' ♈ |
| Beginning of sign | 21°25' ♓ | 29°1' ♓ | |
| End of sign | 21°25' ♈ | 29°1' ♈ | |

## 5. *Summary and Conclusion*

In the 17th century and early 18th century, the chronology is a booming "science". Attempts to date events mostly depend on text analyses, but astronomy begins to be used. However, an accurate chronometer is lacking - dating by isotope still being science fiction at the time. Newton identifies such a chronometer: the use of the precession of the equinoxes.

Attributing the description of the celestial sphere by Eudoxus to the time of the Argonauts (in particular to Chiron), he carefully selects stars potentially described by these texts, and finds that the expedition occurs in the 10th century BCE. In addition, he uses the difference between the actual solar year and the calendar of 365 days to connect the Egyptian calendar to the Chaldean one. Coupled to the shortening of the average length of reigns, this leads to a timeline shortened by five centuries. Such a timeline "proves" the precedence of the Hebrews and their religion, and postulates the relative youth of the Egyptian and Chinese civilizations.

However, there are several problems. For work on calendars, Newton misidentifies Pharaohs, reverses calendar inheritance, and misplaced the beginning of the year at Equinox. Regarding the precession, the idea is original and innovative, it also has the appearance of a demonstration... which would be valid if the assumptions were correct (i.e. the sphere described by Eudoxus is that of Chiron) and if the stars used for the calculation were well chosen (i.e. sufficiently bright and corresponding to the description of Eudoxus). In both cases, Newton often gives the impression to choose only what supports his thesis, which explains the ensuing debates. While the original concept is great, the conditions for its application did not exist until recently: an organized civilization, which has already observed a lot (knowing ecliptic, zodiac, colures…) and observes with a high precision (better than a degree) and in a known system (no ambiguity about the identification of the stars, the reference system, ...). Newton presumptuously supposes that he has clear facts while he only has unclear observations ("middle") in a poorly known system (definition of signs and constellations): his work is logically doomed to fail.

While Newton's approach is inscribed in the tradition of his time, his surprising chronology will provoke debates until the early 19th century. Some discussions focus on the duration of the reigns and interpretation of texts, but other sources closely consider the astronomical arguments: Whiston, Souciet, Fréret and La Nauze. The clearest, most comprehensive and most accurate analysis is that by Fréret, although he is not astronomer, but he quotes part of the Whiston's work. The Jesuit Souciet multiplies errors, remarked by Halley, he is also draft and goes into parallel directions, sometimes interesting but little or badly explored. La Nauze is virulent about Souciet, but does not appear to work much better, making the same errors than the ones he criticizes. The debate here is not about the accuracy of Newton's work: it is the person of Newton which is at sake. *Chronology* is indeed its Achilles' heel, his "scientific" work which is the most questionable. By attacking it, it is the entire work of the English scientist that opponents hoped to put into question. This however will not be the case: *Chronology* will indeed fall, but the mechanical and optical works will keep their place amongst scientists.

In the end, the *Chronology* simply sunk into oblivion, as the alchemical work, not to dent the image of the great scientist. However, neglecting these pseudo-scientific aspects results in having a false image of Newton's personality. He worked for years at these things today considered as minor: they were nevertheless very important to him. Like him (in the *Chronology*), we must not forget facts simply because they do not suit our desires...

***Acknnowledgments :*** The author thanks C. Grell, R. Halleux, E.R. Parkin, and L. Désamoré.

## *6.    Appendix*

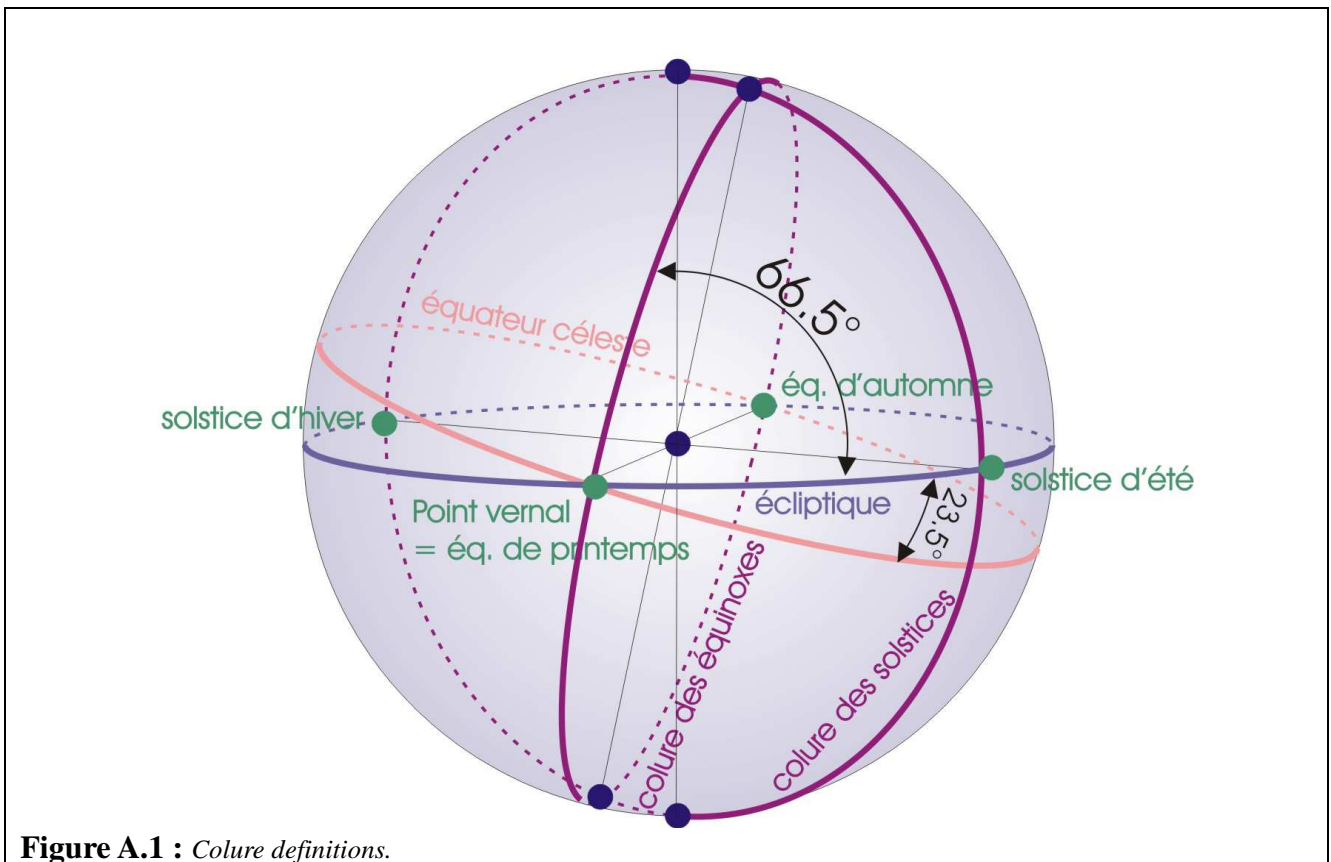

**Figure A.1 :** *Colure definitions.*

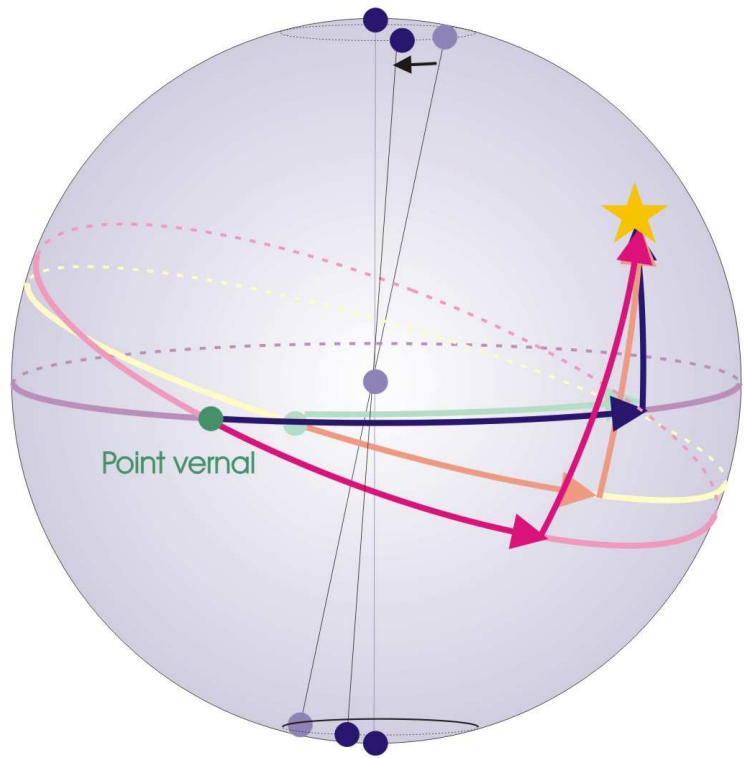

**Figure A2 :** *Change in coordinates because of precession: both right ascension and declination changes with time, but only ecliptic longitude is affected.*

bayer*.jpg

**Figure A3 :** *Excerpts from Bayer's Uranometria with the stars discussed by Newton highlighted in yellow, and those discussed by Fréret, Whiston and Souciet in red. Scans come from http://www.lindahall.org/services/digital/ebooks/bayer/*

sky_939.jpg + sky_1353.jpg

**Figure A4 :** *Position of stars associated with colures (dark orange lines) in 939BCE, with problematic objects in blue, and in 1353BCE, with Whiston couples of stars in blue. Coordinates are right ascension and declination. The stars chosen by Fréret to indicate the tropics and equator are shown in green for 1353BCE. Argo stars have not been shown.*

# Astronomie et chronologie chez Newton
# Arguments astronomiques à l'appui
# de la Chronologie de Newton[1]

Yaël Nazé, Chargé de recherches FNRS
Département AGO, Université de Liège, Allée du 6 Août 17, Bât B5C, B4000-Liège, Belgique

*Dans sa Chronologie, Newton utilise des « preuves » astronomiques pour appuyer son rajeunissement extrême des époques anciennes. Ces éléments, au vernis scientifique, donnent une crédibilité certaine à l'ensemble. Ils ont donc été âprement discutés, les débats sapant petit à petit les hypothèses du savant anglais pour finalement porter un coup mortel à l'ensemble. Cela n'a toutefois pas entamé le prestige du savant anglais.*

« la partie qui me reste à examiner est ce que l'on a fait passer pour la plus importante et la plus ingénieuse des découvertes dont on a prétendu que son Livre étoit rempli » (Fréret, *Défense*, p415)

## *1.    Introduction*

L'on retient de Newton les *Principia* et l'*Opticks*, œuvres majeures considérées comme fondatrices, avec d'autres, de la science moderne. Mais l'esprit du savant anglais ne se limita pas à ces deux sujets. Ce que d'aucuns considèrent comme sa « face sombre » comporte notamment des études historico-chronologiques poussées. Certains, à travers les âges, y ont vu un hobby, une petite manie de vieil homme, ou encore une conséquence d'un problème nerveux ayant eu lieu en 1693 (comme Laplace ou Biot). Pourtant, la correspondance, les manuscrits, et la bibliothèque de Newton montrent qu'il ne s'agit nullement de l'intérêt passager d'un esprit malade[2]. Certes, la Chronologie fut sa préoccupation majeure au cours de ses dernières années. Cependant, Newton y consacra du temps tout au long de sa vie. Il essaya même diverses voies d'interprétation, changeant parfois d'opinion sur un point ou l'autre, pour finir par élaborer une « Chronologie » publiée de manière posthume.

Par cette œuvre, Newton s'inscrit dans la tradition de son époque. En effet, les études chronologiques constituent alors un sujet historique crucial, âprement débattu. Newton assure ainsi que « l'histoire sans chronologie est confuse »[3], et aucun historien actuel ne peut lui donner tort sur ce point. À son époque, deux thèses s'affrontaient : l'une plaçant la Création 4000 ans avant notre ère, l'autre plusieurs siècles avant. L'Antiquité des civilisations égyptiennes et chinoises venait faire trembler ce saint édifice. Dans ce contexte, Newton croyait fermement en la justesse des Écritures et l'ancienneté (donc la place primordiale) de la civilisation juive. La « Chronologie » sera un moyen de « démontrer » cette suprématie, tout en éloignant l'Apocalypse. Cela ne se fera pas sans mal.

Tout d'abord, Newton prendra un peu de liberté avec son célèbre « Hypotheses non fingo ». S'il reproche à certains chronologistes leurs preuves uniques, leur non-neutralité vis-à-vis d'une thèse à démontrer, et leur utilisation irréfléchie des textes anciens, il semble ne pas retenir ses beaux principes et se contente lui-même d'un choix sélectif de sources et d'un scepticisme minimal quand

---

[1] An English version of this article is available on arXiv:1212.4943
[2] Emerson mentionne ce travail comme celui de sa vie ('work of his whole lifetime') dans sa "Defence of the Chronology" (annexe de "A short comment on Sir I. Newton Principia" (1760, p150); voir aussi "Newton Historian" de F.E. Manuel (Cambridge Univ. Press, 1693).
[3] "History without chronology is confused", New College MSS II fol 72, cite par F.E. Manuel p37.

sa thèse est appuyée par un texte donné. Un bel exemple est l'attribution de la sphère d'Eudoxe à Chiron, sur la base d'un seul vers poétique, au mieux…

Ensuite Newton soutient une chronologie radicale, extrêmement courte – plus courte que la solution la plus courte d'alors – en retranchant cinq siècles à la ligne du temps classique. Ses arguments principaux reposent sur la durée moyenne d'un règne, et sur l'astronomie. L'un comme l'autre ne sont pas, en soi, des arguments originaux[4], mais Newton se devait d'être original dans ce domaine-là comme dans les autres… et c'est l'utilisation de la précession qui est ici « un acte de pur génie »[5].

Cet article se propose d'analyser les preuves astronomiques apportées par Newton et les répliques qui y ont été faites. Après une brève introduction rappelant le contexte de l'ouvrage et les débats qu'il engendra, cet article continue avec l'examen de chaque point « astronomique » avancé par Newton (durée du calendrier et de l'année tropique, utilisation de la précession pour dater des événements), mais aussi par les débatteurs en présence. Une annexe introduit les notions de repérage céleste nécessaires à la compréhension de l'ensemble.

Cet article a été écrit avant la sortie de « Newton and the origin of civilization » (J.Z. Buchwald & M. Feingold, 2013). Cependant, vu sa publication postérieure, quelques références à cet ouvrage lui ont été ajoutées. Le livre susmentionné se centre sur Newton et, de la *Chronologie*, discute surtout de la précession, de la durée des règnes et de l'empire égyptien. Cet article court ne considère que l'astronomie utilisée par Newton, soit la précession mais aussi les problèmes calendaires, tout en discutant également des autres éléments astronomiques avancés lors du débat autour de la *Chronologie*. Il complémente donc le livre de Buchwald & Feingold.

## 2.    *Publication et réception de la « Chronologie »*

Newton écrivit beaucoup, tout au long de sa vie, sur la Chronologie et l'histoire du monde. Cependant, comme souvent, il fallut un peu le forcer à publier. Dans ce cas-ci, ce n'est pas Halley qui joua un rôle moteur, mais bien une intrigue un peu rocambolesque. Tout commence avec l'*Abrégé de la Chronologie de M. le chevalier Isaac Newton* (ci-après « Abrégé »). Note courte destinée à la princesse de Galles, elle fut ramenée en France par l'abbé Conti qui avait longuement discuté de chronologie avec le savant anglais. Bien qu'elle ne soit pas censée être diffusée, Conti montre la note à plusieurs personnes, de sorte que l'Abrégé finit par être publié en France en 1725, sans l'autorisation de Newton mais avec les observations critiques de Nicolas Fréret (ci-après « Observations »)[6]. Newton y répondit de manière acerbe – quoique plutôt limitée – dans les Philosophical Transactions[7]. Remarquons à ce sujet que le « vol » de Conti doit être un peu relativisé, sachant qu'il existait alors en Angleterre plusieurs copies de cette Chronologie abrégée, dont trois existent toujours[8]. En 1726, quatre dissertations du père Souciet, dont la première est centrée sur l'astronomie, critiquent également le texte de Newton, en tenant compte également d'une lettre de Mr Keil apportant des précisions sur le raisonnement suivi par l'Anglais. L'année suivante, Souciet complète son raisonnement astronomique, suite à la réponse du savant anglais dans les Philosophical Transactions, dans une cinquième dissertation. La Nauze répond point par point à chacune de ces dissertations dans ses « lettres au père Souciet », tandis qu'Halley vient lui aussi s'opposer à Souciet[9]. Le détail des idées chronologiques est finalement publiée en 1728, après

---

[4] Voir par exemple F.E. Manuel, p94.
[5] "An act of sheer genius", F.E. Manuel, p191
[6] Il existe également une letter de Fréret à Halley reprenant ses vues initiales (voir Buchwald & Feingold p366-8).
[7] Newton, Philosophical Transactions, 1725, vol 33, 315-321
[8] F.E. Manuel, p22
[9] Halley, Philosophical Transactions, 1726, vol 34, 205-210 et 1727, vol 35, 296-300

la mort de Newton, sous le titre de la *Chronologie des anciens royaumes corrigée* (ci-après « Chronologie »). Elle comporte bien plus de détails que sa version abrégée (qui est reproduite en début d'ouvrage), ce qui permet de mieux cerner le raisonnement de l'auteur. Fréret élabore alors une réponse tout aussi détaillée[10], comprenant les critiques de Whiston[11], successeur de Newton à la chaire de professeur lucasien ; elle sera publiée après sa mort. Bien que très complet, cet ouvrage ne sera pas le point final de l'affaire, qui continuera à être discutée jusqu'au début du 19e siècle

Les débats autour de cette chronologie surprenante n'ont donc pas manqué, et ce dès le début[12]. Parmi ceux qui soutiennent Newton, on compte, outre La Nauze[13] déjà cité, Andrew Reid (1728), Fatio de Duillier (1732), Zachary Pearce (1732), Voltaire (1733), Arthur Ashley Sykes (1744), James Steuart (1757), Edward Gibbon (1758), William Mitford (1784), William Emerson (1770), Robert Wood (1775), et un anonyme « membre de l'Université » (1827)[14]. Dans le camp opposé, outre Souciet, Fréret et Whiston déjà cités, les adversaires se nomment notamment Arthur Bedford (1728), James Logan (1728), Samuel Shuckford (1728-37), Jean Hardouin (1726), Thomas Cooke (1731), Jean Masson (1731-2), Arthur Young (1734), Etienne Fourmont (1735), Zachary Grey (1736), Joseph Atwell & Thomas Robinson (1737), Alphonse des Vignolles (1738), Angelo Maria Quirini (1738), Francesco Algarotti (1739), le collègue de Fréret Antoine Banier (1740), Thomas Francklin (1741), Samuel Squire (1741), George Costard (1746), William Warburton (1742), Conyers Middleton (1752), L.R. Desh (1755), un certain Rev. Rutherford (1760), Charles de Brosses (1761), Antoine-Yves Goguet (1761), Jean-Sylvain Bailly (1775), Samuel Musgrave (1782), Juan Andrés (1785-1822), Jean-Baptiste Joseph Delambre (1817), Henry Fynes Clinton (1830) et deux anonymes (1754, 1855)[15]. Les problèmes soulevés et discutés sont d'ordre

---

[10] Fréret, Défense de la Chronologie fondée sur les monuments de l'histoire ancienne, contre le système chronologique de M. Newton, 1758 (ci après « Défense ») – la 3e partie est celle comportant les arguments astronomiques.

[11] Whiston, annexe IX de "A collection of Authentick Records belonging to the Old and New Testament" (1728), dont la partie I est reproduite dans la troisième partie, section II, § I de Fréret. Par la suite, les références se feront à cette copie.

[12] Pour une discussion de la situation en France, voir C. Grell, , Arch. Int. Hist. Sciences, vol 62, n°168, 85-157 (2012), et pour une vue plus globale, voir chapitre 10 du livre de F.E. Manuel.

[13] Il faut y ajouter ln réponse à Shuckford : La Nauze, Mémoires de Trévoux, article 125, octobre 1754.

[14] A. Reid (The Chronology of ancient kingdoms amended by sir Isaac Newton, in the present state of the Republick of Letters, vol II, 1728), Fatio de Duillier (lettre à J. Conduitt, 10 août 1732), Z. Pearce (A reply to the letter to Dr Waterland, 1732), Voltaire (Letters concerning the English Nation, 1733 et de la chronologie réformée de Newton, qui fait le monde moins vieux de cinq cents ans, in Mélanges de litérature, d'histoire et de philosophie, 1757), A. A. Sykes (An examination of Mr Walburton's account of the conduct of the antient legislators, 1744), J. Steuart (Apologie du sentiment de Mr le chevalier Newton sur l'ancienne chronologie des Grecs contenant des réponses à toutes les objections qui y ont été faites jusqu'à présent, 1757), E. Gibbon (MS34880, British Museum, 1758, reproduit in Miscellaneous works of Edward Gibbon, III, 61-73, 1814), W. Mitford (History of Greece, annexe I, 1784), W. Emerson (annexe « An account of some of the numerous inconsistencies contained in the objections made by the Rev. Dr Rutherford, against sir I. Newton's account of the Argonautic expedition », in A short comment on sir I. Newton principia containing notes upon some difficult places of that excellent book, 1770), R. Wood (an essay on the original genius and writings of Homer, 1775), et un anonyme « membre de l'Université de Cambridge » (Essays on chronology ; being a vindication of the system of sir Isaac Newton, 1827).

[15] A. Bedford (Animadversions upon Sir Isaac Newton's Book intitled The chronology of ancient kingdoms amended, 1728), J. Logan (1728, in E. Wolf, 1974, The library of James Logan of Philadelphia), S. Shuckford (The sacred and profane history of the world, préface du volume II, 1728-37 puis 1752, résumées dans les Mémoires de Trévoux, article 17 du volume de février 1754), J. Hardouin (Le fondement de la Chronologie de Mr Newton, 1726, résumé dans Hardouin, Mémoires de Trévoux, article 87 du volume de septembre 1729), T. Cooke (The letters of Atticus, 1731), J. Masson (in John Jortin, Miscelleanous Observations upon authors, ancient and modern, vol 2, 1731-2), A. Young (An historical dissertation on idolatrous corruptions in religion, 1734), E. Fourmont (Réflexions critiques sur les histoires des anciens peuples, 1735), Z. Grey (An examination of the fourteenth chapter of sir I. Newton's observations upon the prophecies of Daniel, 1736), J. Atwell & T. Robinson (Hesiodi ascraei quae supersunt cum notis variorum, 1737), A. des Vignolles (1738, in Nouvelle Bibliothèque Germanique, vol 18, partie I, in Steuart 1805, the works, political, metaphysical, and chronological), A. M. Quirini (Primordia Corcyrae, 1738), F. Algarotti (Sir Isaac Newton's Philosophy explained for the use of ladies, 1739), A. Banier (Mythologie des anciens expliquée par l'Histoire, tome 6e, chap XII, p342-6, 1740), T. Francklin (Of the nature of Gods, traduction de Cicéron, 1741), S. Squire (Two essays, the former, a defense of the ancient Greek chronology ; to which is annexed a new

astronomique (voir ci-dessous), mais aussi théologiques – avec l'accusation pour Newton de soutenir tantôt les papistes, tantôt les déistes…

Il faut remarquer que les réponses aux arguments astronomiques ne diffèrent pas seulement sur le niveau de détail de l'analyse, la justesse des discussions varie aussi fortement d'un cas à l'autre. Souciet se distingue ici particulièrement. S'il parsème son texte de piques superfétatoires[16], il aurait mieux valu qu'il relise son texte car on y trouve des erreurs un peu partout. Il s'agit parfois de simples erreurs typographiques, ou de recopiages mal exécutés[17], mais surtout de divers problèmes mathématiques. Ainsi, Souciet confond quasi systématiquement la distance colure-oreille (7°36' en 939 avant notre ère) et oreille-début de signe (son complément à 15°, soit 15°-7°36'=7°24'). Il mélange même parfois Hipparque et Chiron, prenant les angles du premier et l'époque du second (ce qui forcément conduit à des impossibilités). De plus, il ne corrige pas les coordonnées de l'inclinaison du colure équinoxial[18]. Enfin, il insiste (avec justesse) sur l'ascension droite comme coordonnée utilisée par les anciens[19] mais continue pourtant à faire ses raisonnements en longitude écliptique... Comme le dit Halley[20], il « aurait dû être plus prudent avec ses nombres, et s'informer de la trigonométrie sphérique… ». La Nauze, pourtant prompt à reprendre Souciet sur ses erreurs, en commet souvent des similaires...

Un élément important, quoique sous-jacent, du débat tient à la nature même de l'auteur. Newton n'est évidemment pas le premier quidam chronologiste venu. Le soutien, ou au contraire l'opposition, qui accueille la nouvelle chronologie vise aussi spécifiquement la personne de Newton et ce d'autant plus qu'il orne son édifice temporel d'un vernis scientifique sous la forme de

---

chornological synopsis ; the latter, an enquiry into the origin of the Greek language, 1741), G. Costard (Letter to Martin Folkes concerning the rise and progress of astronomy amongst the Antients, 1746), W. Warburton (4e livre de The divine legation of Moses, 1742), C. Middleton (The miscelleanous works, vol 2, 1752), L.R. Desh (Lettre sur la chronologie de M. Newton, Mercure de France, décembre 1755, reproduit dans le livre de Steuart), C. de Brosses (Second mémoire sur la monarchie de Ninive, Histoire et mémoires de l'Académie Royale des Inscriptions, 27, 1-81, 1761), A.-Y. Goguet (The origin of laws, Arts, and Sciences and their progress among the most ancient nations, 1761), J.S. Bailly (Histoire de l'Astronomie Ancienne, 1775 et 1781), S. Musgrave (Two dissertations I. on the graecian mythology II an examination of sir Isaac Newton's objections to the Chronology of the Olympiads, 1782), J. Andrés (dell origine, progressi e stato attuale d'ogni litteratura, 1785-1822), J.-B. J. Delambre (Histoire de l'Astronomie Ancienne, 1817), H. F. Clinton (Fasti Hellenici, the civil and literary chronology of Greece and Rome, 1827) ; Rutherford est cité par Emerson; . l'anonyme écrivant la revue du livre de Shuckford dans les mémoires de Trévoux (1754), et l'anonyme responsable de la revue de "History of Greece" de George Grote (in Dublin University Magazine, vol 45, 1855).

16 Exemple Souciet, p127 « l'une est le témoignage des Anciens, l'autre la raison. M. Newton n'a ni l'un ni l'autre ».
17 Par exemple les fins de constellations qui changent de position entre la p161 et la p165.
18 Cette inclinaison est de 66.5° sur l'écliptique, voir annexe et Fig. 3 ci-dessous. Voir aussi les remarques de Halley (Philosophical Transactions, vol 34, p209) : « he ought to have deduced 3 deg 7.5' out of the 15 degrees he assumes for the distance of his colure from the first star of Aries »
19 Il l'utilise d'ailleurs (Souciet, 5e dissertation, p 133-135) pour tenter de comprendre la valeur de 7°36' donnée par Newton dans sa réponse (Philosophical Transations, vol 33, 315-321) sans explication. Il suppose en fait qu'il s'agit de la distance sur l'écliptique entre le nœud des poissons, qui servirait donc de début de signe chez Newton, et l'oreille du Bélier (à noter son erreur habituelle : il faudrait ici utiliser 7°24' avant l'oreille car les 7°36 se réfèrent chez Newton à la distance oreille-colure...). Il montre ensuite qu'en ascension droite, l'angle est très différent, le nœud des poissons se trouvant même devant l'oreille plutôt que derrière vu leurs latitudes très différentes... La Nauze (5e lettre, p 404-407) voit dans ces pages une chimère, un fantôme que poursuit Souciet. Il assure que les 7°36' ne viennent pas de là, mais curieusement, il se garde bien d'expliquer leur origine... car pour trouver cette valeur, il faut tenir compte de l'inclinaison du colure équinoxial, et La Nauze ne le fait pas dans ses lettres ! Ces pages liées à l'ascension droite sont un bel exemple du niveau du livre de Souciet : outre les recopiages erronés et le texte inadéquat (par ex. « oreille du bélier » citée deux fois en référence à deux astres différents), Souciet considère un taux de précession constant sur l'équateur, et non l'écliptique, ce qui constitue une erreur fondamentale (voir Annexe b). Il aurait dû modifier la longitude écliptique pour l'année en question, et recalculer alors les coordonnées équatoriales (voir Annexe a). Buchwald & Feingold détaillent aussi les erreurs de calcul de Souciet quand il travaille en ascension droite (p353-362).
20 Halley, Philosophical Transactions, vol 34, p209 : « be a little more careful of his numbers […] and inform himself of the Sphericks, so as to give us the right ascensions of the stars truly from their given longitudes and latitudes ».

« preuves » astronomiques. Fréret[21] précise ainsi que « le nom seul de Newton formera toujours un préjugé difficile à détruire. » Pour être juste, l'aura de Newton joue aussi à son désavantage, car le combat contre ses idées est rude, surtout en France. Steuart le souligne[22] : « les belles découvertes de Mr. Newton dans l'Astronomie et dans l'Optique n'ont pas été moins combattuës dans le commencement, que l'est à présent son sentiment sur la Chronologie. » D'un côté, on pense que le grand homme, ayant eu raison en optique et mécanique, ne peut se tromper sur la chronologie – les modèles célestes prédisant correctement le futur étant tout aussi aptes à retrouver les événements passés. Son traitement même élève désormais la chose au range d'une « science », car l'astronomie passe pour une preuve irréfutable. De l'autre, on espère détruire la cité en minant une muraille annexe, plus facilement attaquable car sujette à caution en de nombreux points dès qu'on y regarde de plus près. Le combat acharné que se livrent les deux camps n'est donc pas seulement un débat éthéré sans arrière-pensée équivoque.

Puisque le sujet de cet article est l'astronomie utilisée dans cette Chronologie, nous nous focaliserons par la suite uniquement sur les textes qui y en font mention.

## 3. *Datation grâce aux imperfections calendaires*

Elaborer un calendrier exact est loin d'être une mince affaire. Les premières tentatives anciennes, simples (par ex. Calendrier de 12 mois lunaires), déviaient rapidement de l'année solaire réelle et donc des saisons. Il fallait réajuster régulièrement le calendrier, notamment en intercalant des mois supplémentaires de temps à autre. Même un calendrier de 365 jours n'est pas parfait, car l'année réelle est plus longue : sa valeur moyenne est proche de 365,2422 jours. C'est d'ailleurs pour cela que nous avons un calendrier grégorien, comportant régulièrement des années bissextiles de 366 jours.

L'Abrégé et la Chronologie exploitent ce type d'erreur pour dater le calendrier égyptien. D'après Newton, ce dernier était initialement de 360 jours, auxquels on finit par ajouter 5 jours, dits épagomènes, pour avoir un meilleur accord avec l'année solaire. Ce calendrier fut ensuite adopté par les Chaldéens, sans aucun discernement. Les deux écrits de Newton suivant exactement le même raisonnement, en des termes très similaires, on ne recopie ici que le texte de la Chronologie, légèrement plus détaillé[23] : « ils ajoutèrent à la vieille année du Calendrier cinq jours […] sous le règne d'Aménophis, ils ont pu placer le commencement de cette nouvelle année ou l'équinoxe de printemps […] Enfin cette même année luni-solaire fut introduite dans la Chaldée, d'où vint l'ère de Nabonassar ; car les années de Nabonassar et celles d'Egypte commençoient le même jour qu'ils nommoient Thoth, il n'y avoit aucune différence entre elles. La première année de Nabonassar commença le 26 février de l'ancienne année romaine, 747 ans avant l'ère vulgaire de Jésus-Christ, et trente-trois jours et 5h avant l'équinoxe de printemps suivant le mouvement moyen du Soleil […] Or si l'on compte que l'année de 365 jours a cinq heures & 49 minutes de moins que l'année équinoxiale, le commencement de cette année rétrogradera de trente-trois jours et cinq heures en 137 années & par conséquent cette année commença d'abord en Egypte à l'Equinoxe de printemps, suivant le moïen mouvement du Soleil 137 années avant le commencement de l'ère de Nabonassar », soit en 884 avant notre ère.

---

[21] Fréret, Défense, p 419
[22] Steuart, Apologie, p164
[23] Newton, Chronologie, p82-83 – voir aussi entrée 884 de l'Abrégé.

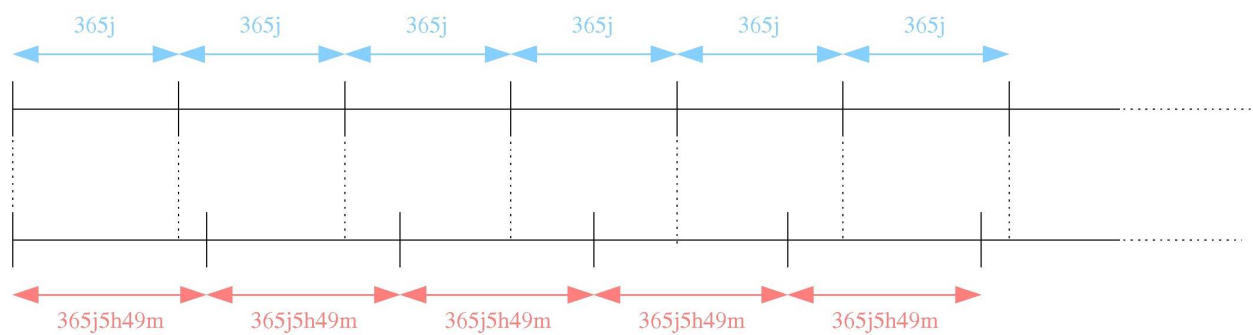
**Figure 1 :** *Décalage progressif des années de 365 jours par rapport à l'année solaire vraie.*

Newton suit ici le calendrier julien et son calcul peut aisément être vérifié. En 747 avant notre ère, première année de Nabonassar, l'équinoxe de printemps se produisit le 28 mars à 22h30 UT[24]. Entre le 26 février à 0h, début de la première année de Nabonnassar, et le 28 mars à 22h30, il y a quasiment 31 jours complets (et non 33j5h – il y a ici une petite erreur de calcul). Newton attribue ce décalage à l'erreur du calendrier de 365 jours. En effet, dans un tel système, on perd chaque année presque un quart de jour (5h49m selon Newton, voir aussi Fig. 1) : pour prendre 31 jours d'avance, il faut donc attendre 128 ans (~4×31). En supposant que le calendrier chaldéen ait été adopté sans modification aucune, Newton place donc l'établissement du calendrier égyptien de 365 jours – et la fin de règne d'Aménophis – un peu plus d'un siècle avant la première année de Nabonassar. Comme cette dernière possède une date bien connue dans les calendriers modernes, on peut alors dater facilement la succession des pharaons, via Aménophis.

On sait aujourd'hui que le calendrier chaldéen ou mésopotamien a influencé l'égyptien, et non l'inverse : le peuple du Nil doit en effet à ces voisins l'année de 365 jours, établie au 2[e] millénaire avant notre ère. Le postulat de base est donc pour le moins caduc, et il existe d'autres problèmes dans le raisonnement. Cependant, à l'époque de Newton, Fréret sera le seul à mettre en doute ce calcul, dans ses Observations (p84-89) ainsi que dans la section I de la troisième partie de sa Défense. Outre les problèmes d'identification du pharaon Aménophis – aux identités bien choisies chez Newton pour condenser la chronologie, Fréret attaque également l'argument calendaire. Il souligne tout d'abord la question de l'héritage, loin d'être certain, bien au contraire[25]. Il insiste ensuite sur le problème du commencement chez Newton : s'il y avait bien une fête religieuse organisée à l'équinoxe, cela n'en fait pas pour autant le premier jour du calendrier égyptien, bien connu pour être fixé au lever héliaque de Sirius (baptisé « Sothis »), soit en phase avec le début de la crue du Nil[26]. En fait, il affirme avec raison[27] qu'« il n'y a rien dans toute l'antiquité qui puisse nous faire penser que l'année égyptienne ait jamais commencé au printemps. »

Fréret observe aussi un problème lié aux usages égyptiens[28]. Le peuple du Nil utilisait deux calendriers, l'un civil, maintenu en phase avec les saisons, et l'autre religieux. Ce dernier comportait toujours 365 jours, sans aucun ajout ou correction. Il s'ensuit un décalage progressif d'avec l'année civile réelle d'un quart de jour par an. On retrouve donc les deux calendriers en phase tous les 1460 ans environ[29], un intervalle de temps aussi appelé « cycle sothiaque » : comme les deux années coïncidaient parfaitement en 138, il en découle qu'un cycle commença en 1323 avant notre ère. Fréret affirme même, en interprétant diverses sources dont des textes de Manéthon, que ce cycle n'était pas le premier. Le calendrier égyptien de 365 jours serait donc bien plus ancien que 884 avant notre ère – au moins de 4 siècles et demi, si pas de deux millénaires. A noter que Newton n'avait ici pas compris la portée de la remarque de Fréret : en réponse aux Observations, il jure ses grands dieux qu'il n'a jamais placé le commencement d'un cycle sothiaque en 884 avant notre ère[30]. S'il est vrai qu'il ne parle pas de ce cycle, il oublie que l'argument de Fréret porte sur l'antiquité de l'utilisation de 365 jours – impossible en effet de commencer un cycle sothiaque avant d'avoir fixé la durée de l'année à 365 jours...

---

24 Pour connaître la date des solstices et équinoxes entre 4000 avant notre ère et 2500, le site de l'IMCCE http://www.imcce.fr/fr/grandpublic/temps/saisons.php peut être utile. Sinon, il suffit d'une date (Newton connaissait probablement la date de l'équinoxe de printemps à son époque, qui se produisit le 9 mars à 22h24 UT en 1689) et de la longueur des années. En effet, il précise utiliser 365j5h49m pour l'année tropique, alors que l'année julienne vaut 365j6h, soit une différence de onze minutes : en 2435 ans (intervalle entre 747 avant notre ère et 1689), cela fait une différence de 18,6 jours, à ajouter à la date de l'équinoxe de 1689 pour trouver le 28 mars.
25 Fréret, Défense, p386
26 Fréret, Observations p84-89, et Défense, p391-394
27 Fréret, Observations, p85
28 Fréret, Observations p84-89 et Défense p394-400 et p407

Les deux dernières remarques de Fréret dans le domaine calendaire portent sur les sources de Newton. Sa seule source mentionnée est Syncelle, mais il semble évident que ses écrits ne sont pas fiables car ils comportent diverses contradictions tant avec l'extérieur (d'autres sources de la même époque), qu'avec lui-même[31]. D'autre part, un coup d'œil plus aiguisé aurait facilement permis à Newton de se rendre compte que les 5 jours épagomènes n'étaient généralement pas comptés officiellement, même si utilisés en pratique – on parle de jours « dérobés »[32]. Avoir un texte religieux parlant de 360 jours est donc tout à fait compatible avec l'utilisation pratique d'un calendrier de 365 jours à la même époque... Newton va un peu vite en besogne en faisant du calendrier de 365 jours une invention récente[33].

## *4. Datation grâce à la précession*

La datation constituait un sérieux problème aux historiens de l'époque, car la géologie et l'archéométrie faisaient encore partie d'un lointain futur… L'analyse des monuments n'étant pas claire, il ne restait aux historiens que l'analyse des textes, l'écriture sainte en premier lieu. Ainsi, on comptait les générations décrites, en essayant de faire correspondre les mesures tirées des diverses sources écrites. Dans ce contexte, l'astronomie paraît à première vue un atout considérable. Les événements célestes se produisent en effet des moments précis, reconstituables au 17$^e$ siècle grâce à l'établissement de théories enfin précises, basées sur des observations détaillées. Cette précision et l'aspect irréfutable du calcul scientifique, introuvables ailleurs, exercent alors un attrait non négligeable. Joseph Scaliger et Denys Pétau utilisent les cycles connus du Soleil et de la Lune pour mettre en phase les différents calendriers anciens. Les éclipses, événements rares mais prédictibles, servent elles aussi, notamment dans les travaux de Riccioli. L'étude des orbites de comètes, science nouvelle dont Halley est pionnier, fournit même à Whiston de quoi dater le déluge. Les phases de la Lune, et les marées associées à cet astre, permettent à Halley de dater l'invasion de l'Angleterre par César[34].

L'utilisation de l'astronomie en chronologie n'est donc pas nouvelle quand Newton s'intéresse à l'histoire. Toutefois, le choix du phénomène astronomique est totalement nouveau et potentiellement révolutionnaire. En effet, Newton, lui, utilise la précession, un phénomène précis et bien connu que Newton lui-même a « démontré » dans le cadre de sa nouvelle physique (voir annexe). Elle possède en outre l'avantage de ne pas dépendre de la latitude du lieu d'observation, supprimant une inconnue gênante. Il s'agit clairement du cœur du système newtonien[35]. C'est aussi l'argument en apparence le plus scientifique de toute la Chronologie, et donc celui qu'il faut détruire pour faire crouler l'ensemble, ou qu'il faut soutenir pour consolider l'édifice temporel. Cela est bien compris des contemporains de Newton, ce qui explique le nombre de discussions sur le sujet.

---

[29] car 365.25/0.25=1460, ou 1507 ans si l'on considère des années solaires de 365j 5h 49m
[30] Newton, Philosophical Transactions, vol 33, p320
[31] Fréret, Défense, p405-406
[32] Fréret, Défense, p411-412
[33] Il est ici piquant de constater qu'un défenseur de Newton, James Steuart, tente de justifier Newton en soulignant l'antiquité des cinq jours épagomènes (Apologie, p88). Steuart veut ainsi prouver à Shuckford que les Anciens n'avaient pas une si mauvaise précision quant à la durée de l'année... mais, sans s'en rendre compte, il va ici contre les enseignements de son « maître » !
[34] J. Scaliger, De emendatione temporum (1583) ; D. Pétau , Opus de doctrina temporum (1627) ; Riccioli, Almagestum Novum (1651); W. Whiston, A new theory of the Earth (1696); E. Halley, A discourse tending to proue at what time and place Julius Cesar made his first descent upon Britain, Philosophical Transactions, 16, 495-501 (1691)

Fréret explique ainsi[36] : « la détermination de l'âge des Argonautes par le lieu des colures dans la sphère reglée sur leur temps, est une de ces idées neuves et brillantes, dont le privilège est de surprendre et de subjuguer les esprits. Je ne m'etonne pas que M. Newton l'ait saisie comme une soudaine inspiration de ce Génie dont il avoit autant de droit que Socrate de se croire assisté dans ses méditations, & que ses partisans se soient jettés avec confiance dans la route nouvelle que leur traçoit ce rayon de lumière. Un fait aussi certain que la précession des équinoxes, employé comme principe, garantissoit à leurs yeux la certitude des conséquences qui devoient en résulter. C'étoit en même temps étendre le ressort de l'Astronomie & faire marcher la Chronologie d'un pas plus sûr, que d'appeler l'une au secours de l'autre. […] Il étoit difficile qu'avec de tels avantages au moins apparens, cette preuve astronomique ne donnât presque l'air d'une démonstration au raisonnement par lequel M. Newton en déduisoit le calcul abrégé. »

Steuart abonde dans son sens[37] : « le 3e avantage qu'à eu Mr. Newton consistoit en l'étenduë de ses connaissances de l'Astronomie & dans ce génie créateur dont il étoit doué. Rien de ce qui avoit du rapport à cette science ne pouvoit se dérober à sa pénétration. Il en saisissoit la moindre petite circonstance, pour la tourner à profit dans les autres sciences. Combien la physique n'a-t-elle pas servi entre ses mains pour expliquer les phénomènes de l'Astronomie et de l'Optique combien l'Astronomie à son tour n'a-t-elle pas servi pour expliquer la nature & la voici encore employée pour déterminer la Chronologie ».

Bailly lui aussi souligne[38] : « l'idée de régler la chronologie par la détermination ancienne des points équinoxiaux et solsticiaux étoit belle, grande et digne d'un homme de génie ». Il ajoute cependant « mais Newton s'est trompé dans l'application qu'il en a faite & le système qui en résulte est tombé, parce qu'il est contraire aux faits. » Tout le problème est là : le raisonnement est erroné. Toutefois, il ne s'agit pas d'un obscur savant de province, c'est du grand Newton dont on parle ici.

Dans la suite de cet article, l'examen du raisonnement chronologico-astronomique sera coupé en 4 parties, pour faciliter la compréhension des arguments avancés (dans un sens ou dans l'autre) car le débat n'est pas simple, même si Souciet affirme[39] : « la preuve qu'en apporte M. Newton est des plus spécieuses et des plus fortes en apparence. Elle est fondée sur le cours des astres, et sur un calcul astronomique des plus aisez et des plus clairs. ». La dernière partie de l'annexe fournit les cartes célestes des constellations évoquées dans le texte, ainsi que celles des colores aux époques discutées.

---

[35] « c'est la base de la Chronologie » (Fréret, Observations, p56) – à noter que le terme de précession n'apparaît pas dans l'Abrégé ni même dans la réponse de Newton reprise dans la lettre de Keil (p56 de la première dissertation de Souciet). Toutefois, on ne peut se tromper sur l'identité du phénomène choisi (les termes de recul du solstice d'un degré en 72 ans sont suffisamment explicites), et ce n'est donc pas la lettre de Keil qui donne la solution de l'énigme comme le pense F.E. Manuel (p23). Ajoutons que d'autres avaient déjà réfléchi à l'utilisation de la précession pour les questions chronologiques avant Newton (notamment Scaliger, voir Joseph Scaliger, Anthony Grafton, 1993, et Buchwald & Feingold p250), mais sans utiliser les colures.
[36] Fréret, Défense, pxliv-xlv de la préface
[37] Steuart, Apologie, p116
[38] Bailly, Histoire de l'Astronomie Ancienne, p509
[39] Souciet, 1ᵉ dissertation 1, p51

Une chose doit cependant être notée : Newton n'a pas vraiment besoin de l'astronomie. En effet, dans son « Original of Monarchies », il utilise seulement la durée moyenne des règnes pour fixer la ligne du temps[40]. L'astronomie ne constitue donc qu'une cerise sur le gâteau, quoiqu'elle soit finalement considérée comme sa « meilleure » preuve vu son aspect scientifique et démontrable.

a) Liste des constellations

Pour appuyer son raisonnement chronologique, Newton tente d'abord de « prouver » le lien entre constellations et l'époque des Argonautes[41] : « Chiron dessina les figures du ciel [...] on peut voir par cette Sphère même, qu'elle fut ébauchée au tems de l'expédition des. Argonautes ; car cette expédition s'y trouve marquée parmi les Constellations, aussi bien que differens traits encore plus anciens de l'Histoire Grecque ; il n'y a rien de plus moderne que cette expédition. On voioit sur cette Sphère le *Bélier* d'Or, qui étoit le Pavillon du Navire, dans lequel Phryxus se sauva dans la Colchide ; le *Taureau* aux piés d'airain dompté par Jason ; les *Gémeaux* Castor & Pollux, tous deux Argonautes, auprès du *Cigne* de Leda leur mère. La étoient réprésentés le *Navire* Argo, & l'*Hydre* ce Dragon si vigilant ; ensuite la *Coupe* de Medée, & un *Corbeau* attaché à des cadavres, qui est le symbole de la mort. D'un autre côté, on remarquoit *Chiron* le maître de Jason, avec son *Autel* & son *sacrifice*. *Hercule* l'Argonaute avec son *Dard* & avec le *vautour* tombant ; le *Dragon*, le *Cancer* & le *Lion* qu'il tua ; la *Lyre* d'Orphée l'Argonaute. C'est aux Argonautes que toutes ces choses ont du rapport. On y avoit encore réprésenté *Orion*, fils de Neptune, ou selon d'autres livres, petit-fils de Minos, avec ses *Chiens*, son *Lièvre*, sa *Rivière* & son *Scorpion*. L'Histoire. de Persée est designée par les Constellations de *Persée*, d'*Andromède*, de *Cephée*, de *Cassiopée*, & de la *Baleine* : celle de Callisto, & de son fils Arcas, par la *Grande Ourse*, & le *Gardien de l'Ourse* : celle d'Icare, & de sa fille Erigone, est marquée par le *Bouvier*, le *Chariot*, & la *Vierge*. La *petite Ourse* fait allusion à une des Nourrices de Jupiter, le *Chartier* à Erechthonius, le *Serpentaire* à Phorbas, le *Sagittaire* à Crolus, fils de la Nourrice des Muses, le *Capricorne* à Pan, & le *Verseau* à Ganimède. On y voïoit la *Couronne* d'Ariadne, le *Cheval aîlé* de Bellerophon, le *Dauphin* de Neptune, l'*Aigle* de Ganimede, la *Chèvre* de Jupiter & ses *Chevreaux*, les *Asnons* de Bacchus, les *Poissons* de Venus & de Cupidon, & le *Poisson Austral* leur parent. Ces Constellations & le *Triangle*, sont les anciennes dont parle Aratus: elles font toutes allusion aux Argonautes, à leurs contemporains, & à des gens plus anciens d'une ou de deux Générations. De tout ce qui étoit originairement marqué sur cette Sphère, il n'y avoit rien de plus moderne que cette expédition. *Antinoüs* & la *Chevelure de Bérénice* sont de nouvelle date. Il semble donc que Chiron & Musaeus firent cette sphère, pour l'usage des Argonautes ».

Quelques précisions sont à apporter au niveau de cette liste : la Chèvre et les Chevreaux dont il est question font en fait partie de la constellation du Cocher (ou Chartier ci-dessus), tandis que les Ânons se trouvent, eux, dans le Cancer ; la Rivière correspond à l'Eridan actuel, le Serpentaire à Ophiuchus (qui est accompagné d'un Serpent), Chiron au Centaure, le gardien de l'Ourse à Arcturus, le Cheval Aîlé à Pégase, le Vautour tenait probablement la Lyre, et le Dard est probablement la Flèche. A noter que le Cancer est aussi parfois appelé Ecrevisse à cette époque. Il manque par contre les constellations du Petit Cheval, de la Balance (ou Serres du Scorpion, citées ailleurs par Newton) et du Loup (liée au départ à la constellation du Centaure), pourtant dûment notées par Ptolémée.

Le problème principal de cette liste est qu'elle contient de nombreuses constellations d'origine non grecques. Au minimum, les douze constellations classiques du zodiaque (Bélier, Taureau, Gémeaux, Cancer, Lion, Vierge, Balance, Scorpion, Sagittaire, Capricorne, Verseau, Poissons) ainsi qu'Hydre, Aigle, et Poisson Austral sont d'origine mésopotamienne[42]. Elles ont été transmises aux Grecs qui les adoptèrent – comme le dit Fréret[43], il y a des « constellations habillées à la grecque ». Leur lien avec la célèbre expédition est donc, au mieux, plus que ténu. La chose était d'ailleurs connue dans l'Occident moderne[44], même si La Nauze insiste sur l'apport grec en assurant[45] que les noms grecs des constellations sont différents de ceux donnés par les Chaldéens (ce qui est faux dans de nombreux cas, voir ci-dessus) et les Egyptiens (ce qui est correct).

---
[40] F.E. Manuel, p122
[41] Newton, Chronologie, p87-89, italiques reproduites ci-dessous

En outre, Fréret remarque[46] que « Chiron n'étoit pas le seul à qui les Grecs se crussent redevables de leur Astronomie ». Ils attribuaient aussi ce titre à Prométhée, Atrée, et Palamedes. Dans ce cas, pourquoi favoriser Chiron plutôt qu'un autre ? N'est-ce pas simplement parce que cela concourt au but recherché, soit une réduction de la chronologie classique ? En effet, Souciet souligne l'importance de l'expédition des Argonautes[47], à laquelle divers événements majeurs (guerre de Troie, fondation de Rome) sont liés par des intervalles temporels connus. C'est un avantage de Chiron que ne possèdent pas d'autres personnages « fondateurs ».

Pire, en supposant même que Chiron soit la source originelle, on ne peut supposer en même temps qu'il ait défini les constellations à l'usage des Argonautes et que ces constellations soient liées à des événements de leur expédition. En effet, ces derniers sont par essence imprévisibles, et n'ont donc été connus qu'une fois l'expédition accomplie, comme le remarque L.R. Desh[48]. La (faible) réplique à ce sujet par J. Steuart[49] est que Chiron a probablement donné des noms provisoires avant l'expédition, tirés des constellations égyptiennes (!), pour en changer juste après le retour des aventuriers, car une création plus récente aurait certainement (!) conduit à retrouver des traces célestes d'événements marquants, comme la guerre de Troie.

Hardouin va plus loin encore dans la critique[50]. Il assure que Chiron astronome est une chimère, le centaure étant plutôt connu pour son art pharmaceutico-médical[51]. Et même si l'on acceptait ce « fait », il pose la question de l'utilité de ces constellations[52] et de la position des points de référence (solsticiaux ou équinoxiaux) pour la navigation des Argonautes, un simple « cabotage le long des côtes » de quelques mois : que la Polaire fut utile aux voyageurs, soit, mais l'intérêt du reste semble assez douteux, avant l'invention de l'astrolabe, lui-même bien postérieur à la célèbre expédition. Il existe en outre un problème de précision dans le repérage stellaire (voir ci-dessous et notes 70 et 71).

Ces nombreuses discussions sur Chiron oublient un point essentiel : Chiron lui-même n'est pas important pour Newton. En effet, comme le démontrent ses manuscrits[53], il cherchait surtout un lien avec les Argonautes car il avait une invention sans inventeur… Son choix s'était au départ porté plutôt sur Palamedes mais, après 1700, il change d'avis en préférant Chiron, évitant alors jusqu'à la moindre mention du héros déchu dans ses écrits.

---

[42] voir par exemple : Rogers, JBAA, 1998, 108, 9
[43] Fréret, Défense, p501 ; on peut aussi relever cette piquante remarque (Défense, p466) : « les Grecs aimoient à faire honneur à leur nation de bien des choses qu'ils devoient aux Barbares »
[44] voir par exemple Bailly, Histoire de l'Astronomie Ancienne, p512, point XL
[45] La Nauze, 5e lettre au Père Souciet, p433
[46] Fréret, Observations, p73-4
[47] Souciet, 5$^e$ dissertation, p117-118
[48] Desh, Lettre sur la chronologie de M. Newton, Mercure de France, décembre 1755, p168-9
[49] Steuart, Apologie, p150
[50] Hardouin, Mémoires de Trévoux, p1569
[51] La meme Remarque est faite par Costard p79, Banier et Squire.
[52] Newton, Abrégé, entrée 939 : « Chiron définit constellations pour faciliter la navigation » ; Hardouin, Mémoires de Trévoux, p1578 et suivantes
[53] F.E. Manuel, p78-85 et annexe C de Buchwald & Feingold.

En résumé, on peut clairement mettre en doute la définition de la sphère céleste par Chiron ou l'un de ses contemporains. Ce lien constitue pourtant la brique de base sur lequel repose l'édifice précessionnel (voir section suivante).

b) Position des colures

Le point central de la justification astronomique de la chronologie newtonienne repose sur la précession. Il s'agit d'un lent décalage (un degré par 72 ans) des points solsticiaux et équinoxiaux par rapport aux constellations (voir annexe et Fig. 2). Ce phénomène peut servir à la datation à condition de connaître les positions précises de ces points à deux époques, dont l'une est de date connue (1690 dans le cas de Newton). La différence de positions, en degrés, donne alors directement l'intervalle de temps en années, après une simple multiplication par 72[54]. Il suffisait donc de trouver une source parlant des positions anciennes de ces points remarquables, les positions actuelles étant connues. Ces sources, Newton les trouve dans des textes anciens, et il en tire le substrat qui l'intéresse.

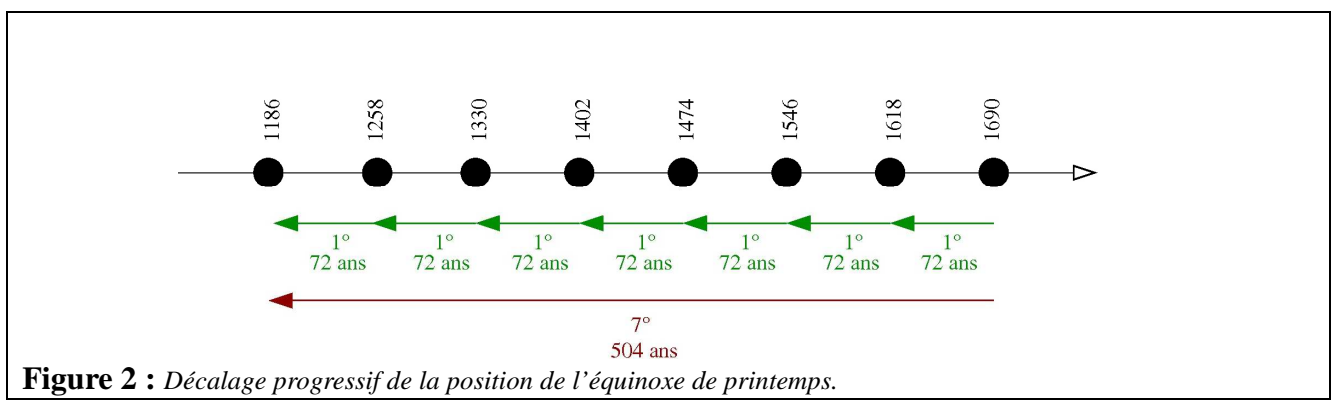

**Figure 2 :** *Décalage progressif de la position de l'équinoxe de printemps.*

Dans l'Abrégé, il écrit ainsi pour l'entrée de l'année 939 : « Il [Chiron] plaça les points des solstices et équinoxes au 15$^e$ degré de ces constellations, càd vers le milieu des signes du Cancer, Capricorne, Aries et Scorpius. Ces signes n'étoient pas différents des constellations même. »

Dans la Chronologie, Newton détaille son raisonnement[55] : « Cette année donna lieu aux premiers Astronomes, qui formèrent les Conflellations, de placer les Equinoxes & les Solstices au milieu des Constellations d'Ariès, du Cancer, des Serres du Scorpion [la Balance] & du Capricorne. […] Eudoxe qui fleurissoit environ 60 ans après Meton, & 100 ans avant Aratus, en décrivant la Sphère des Anciens, mit les Solstices & les Equinoxes au milieu des Constellations d'Ariès, du Cancer, des Serres du Scorpion & du Capricorne, comme l'assure Hipparque […] Ainsi du tems de l'expédition des Argonautes, les points cardinaux des Equinoxes & des Solstices, étoient dans le milieu des Constellations d'Ariès, du Cancer, de la Balance & du Capricorne. » Pour préciser ces points-milieu, Newton considère d'abord les deux étoiles extrêmes du Bélier (γ Ari, oreille du Bélier et « première d'Ariès », et τ Ari, fin de la queue du Bélier et « dernière d'Ariès »). Trouvant leurs positions dans le catalogue de Flamsteed[56], il trouve les coordonnées du point milieu et

---

54 A noter que ce rapide calcul est correct... à un nombre de tours complets près. En effet, si l'on observe un décalage angulaire de 2°, il peut indiquer un intervalle temporel de 2*72=144 ans, mais aussi (360+2)*72=26064 ans si l'axe de la Terre a effectué une rotation d'un tour complet plus deux degrés, ou (720+2)*72=51984 ans pour deux tours complets, etc. Aucun des débatteurs n'a cependant considéré cette possibilité.
55 Newton, Chronologie, p85-89
56 Flamsteed, Historia Coelestis Britannica, 1725. Ce catalogue n'est jamais cité explicitement par Newton, mais ce dernier l'a pourtant bel et bien utilisé, comme le prouvent les coordonnées stellaires citées ! Au début de ses travaux, regarde où le colure équinoxial passant par cet astre coupe l'écliptique. Pour ce calcul, il n'oublie pas que ce colure est incliné de 66.5° sur l'écliptique, d'où une correction à appliquer à la longitude des étoiles pour trouver la longitude réelle de l'intersection, correction d'autant plus grande que l'astre est éloigné de l'écliptique (voir Fig. 3). Il trouve une valeur de 6°44' ♉, soit un décalage angulaire total de 36°44', ou 2645 ans, entre Chiron et lui (en 1690 pour être précis). Pour appuyer cette thèse, il va ensuite prendre diverses étoiles appartenant aux endroits des constellations où Eudoxe fait passer le colure équinoxial, d'après un commentaire d'Hipparque[57]. A chaque fois, il refait le même calcul, et trouve un décalage moyen de 36°29', qui correspond à 2627 ans à retrancher à 1690, soit la définition de la sphère (précédant de peu l'expédition) ayant eu lieu 43 ans après la mort de Salomon dans son système. Il recommence ensuite avec le colure solsticial, défini par δ Cnc pour l'Ânon méridional du Cancer, δ Hya pour le cou de l'Hydre, ι Argo situé entre poupe et mât du Navire, θ Sge à la pointe de la Flèche, et η Cap au milieu du Capricorne. Le calcul est ici plus simple, car le colure est perpendiculaire à l'écliptique et aucune correction ne doit être appliquée. Il trouve ici aussi un décalage moyen de 36°29', et voit ainsi son raisonnement confirmé. Il précise que l'on peut aussi ajouter pour le colure des solstices les étoiles du cou et de l'aile droite du Cygne (η et κ Cyg) ainsi que celle de la main gauche de Cephée (ο Cep) et la queue du Poisson austral – mais il ne donne pas les valeurs ici, et pour cause : une comparaison avec Flamsteed montre que les astres du Cygne sont plus éloignés de la position moyenne que les autres (8.5° et 10.5° au lieu de 6.5°) ; et l'ajout de ο Cep est étrange, car ses coordonnées le font appartenir, au mieux, au colure équinoxial[58] alors qu'Eudoxe, d'après Hipparque, met pourtant bien la main gauche de Cephée dans le colure solsticial...

---

Newton utilisait le catalogue d'Hévélius, comme le démontrent ses manuscrits (voir F.E. Manuel p71), mais il adopta celui de Flamsteed après 1700. Une preuve de ce changement est caché dans la Chronologie : η Per, cité dans le texte avec des coordonnées provenant d'Hévélius – l'étoile n'est pas reprise par Flamsteed. Selon Buchwald & Feingold (p286), Newton aurait procédé à ce changement pour désamorcer les critiques de Flamsteed, prendre la référence la plus récente et surtout… avoir plus d'astres à disposition pour faire son choix, car Flamsteed liste plus d'objets qu'Hévélius, ce qui en dit long sur les « méthodes » du savant anglais, À ce sujet, les mêmes auteurs, tout en soulignant le « scepticisme » de Newton et son absence d'hypothèses, fournissent aussi de nombreux exemples de manipulations de textes (voir p217-221, p224-8, p236-8, p279, p284-286, p294, p298, p305-6) voire de modifications (p134, p162, p199 et p205) – voir également un cas d' « oubli » flagrant pour Hésiode à la section 4.b.II ci-dessous. Les noms des étoiles comportent une lettre grecque suivie d'un code de trois lettres spécifique à une constellation (voir http://www.iau.org/public/constellations/ ) : ils proviennent de l'Uranometria de Bayer (1603), et sont utilisés par tous les débatteurs.

[57] ν Ari au milieu du dos du Bélier, ν et ξ Cet à la tête de la Baleine, ρ Cet pour le dernier pli de l'Eridan, τ et η Per pour la tête et la main de Persée. À noter que Bradley Schaefer (2004, The astronomical lore of Eudoxus, JHA vol 35, p171-3, voir aussi p275-6 et Table 8.3 de Buchwald & Feingold) a fait le même exercice que Newton, indépendamment, à partir du même passage d'Hipparque. Il arrive à une conclusion similaire (985BC pour la définition de la sphère originelle) mais avec une dispersion bien plus élevée (déviation standard de 468 ans à la place de 84 ans). Newton lui-même changea d'opinion au cours du temps à propos de la date exacte (il suffit de comparer l'Abrégé avec la Chronologie, voir aussi Table 8.6, p287 de Buchwald & Feingold).

[58] Ses coordonnées dans Flamsteed sont 5°41'55'' ♉ avec une latitude de plus de 61° : il ne peut donc passer par un colure solsticial, et le colure équinoxial passant par ce point coupe l'écliptique en 12.8° ♓ et non 6.5° ♉… Whiston cite le problème (voir citation dans Fréret, Défense, p 433) et Fréret lui-même en parle (Défense, p434). Ce dernier propose une erreur typographique, un ο ayant remplacé le δ qui siérait mieux selon lui au système newtonien, ce qui n'est pas exact (voir Fig. A.5 en annexe). Selon Buchwald & Feingold (p285), il s'agit d'une simple erreur de calcul : si l'on néglige le signe et la correction due à la latitude, la valeur de la longitude (5°42') est proche du 6°29' de Newton, ils supposent donc qu'il a confondu solstice et équinoxe. Si l'erreur arrange bien Newton, on peut cependant s'en étonner, vu sa simplicité et les réactions aux erreurs de Souciet du même ordre : elle n'est donc pas si facilement excusable et pourrait être intentionnelle.

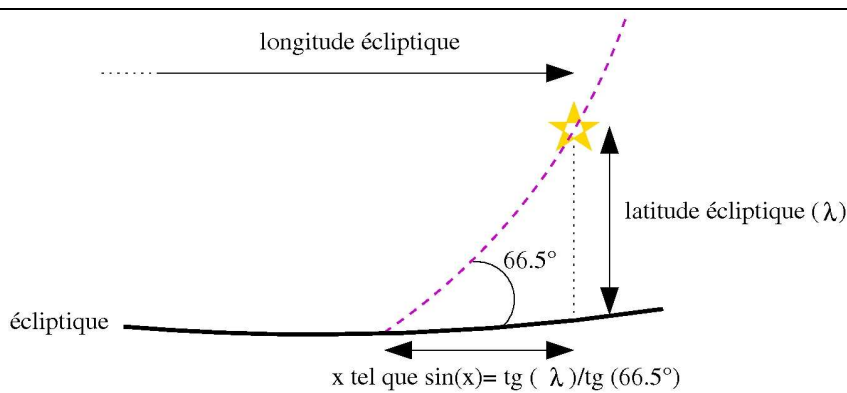

**Figure 3 :** *Correction (x) à apporter aux longitudes écliptiques pour trouver le lieu de l'équinoxe ancienne, qui est l'intersection entre écliptique et colure équinoxial incliné de 66,5°.*

Plusieurs choses sont ici à examiner : le choix des étoiles et l'attribution à Chiron. Etudions tout d'abord le choix des étoiles fait par Newton. Pour commencer, on peut remarquer une absence criante : alors que Newton répète plusieurs fois que le colure équinoxial passe par le milieu des « Serres » (soit la Balance), il ne cite aucune étoile de cette constellation. En fait, cet oubli est loin d'être anodin... Un colure passant par les étoiles $\alpha$ et $\beta$ Lib, cœur de la Balance, coupe l'écliptique à une distance d'environ 41-48° de l'équinoxe d'automne de 1690, et non de 36,5° ! Il est donc impossible, dans le système newtonien, de respecter toutes les prescriptions d'Eudoxe[59]. Autre absente remarquable : Arcturus ($\alpha$ Boo). Si l'on adopte le colure équinoxial antique de Newton, cette étoile très brillante et facilement repérable passait à environ un degré du colure[60], pourtant, Newton n'en fait aucune mention... et Eudoxe non plus (et pour cause !). D'autre part, les étoiles citées par Newton posent elles aussi divers problèmes[61]. Tout d'abord, elles sont peu brillantes car de la quatrième grandeur[62], au mieux. Cela implique qu'elles ne sont visibles qu'avec de bons yeux et dans une nuit sans Lune ni fins nuages... Peut-on dès lors les considérer utiles pour servir au repérage ? La question est d'autant plus aigue qu'il existe dans la plupart des constellations citées des étoiles de première ou troisième grandeur : les anciens voulant désigner des positions célestes claires auraient-ils choisis des étoiles faiblardes et peu souvent détectables ? De plus, Newton dessine les constellations de manière non traditionnelle. On sait en effet, par les descriptions d'Eudoxe, Hipparque ou Ptolémée, que les dessins antiques des constellations ne sont pas identiques aux modernes. Whiston en cite plusieurs : Persée et Céphée ont changé de disposition, et la Flèche non seulement se trouvait ailleurs (près de l'Aigle, puis au niveau du dauphin actuel), mais en plus son tracé moderne n'incluait pas l'étoile citée par Newton, ajoutée quelques années seulement avant ses travaux[63]. Enfin, même dans les cas les moins douteux, les étoiles choisies n'appartiennent pas à la partie du corps citée par Eudoxe : ainsi, $\nu$ et $\xi$ Cet se trouvent dans le col et la crinière de la Baleine, et non sur la tête[64], $\rho$ Cet n'a jamais fait partie de l'Eridan[65] et $\delta$ Hya se trouve dans la tête et non le cou de l'Hydre... Emerson le dit clairement : Newton a choisi ses étoiles[66]. Whiston conclut[67] : « cette singulière tentative de M. Newton, de changer la figure

---

[59] Whiston cité par Fréret, Défense, p425-427
[60] Whiston cité par Fréret, Défense, p438-439
[61] Whiston cité par Fréret, Défense, p428-437
[62] Rappelons ici le système des magnitudes inventé par Hipparque. Les étoiles les plus brillantes sont de première grandeur, celles à peine visible à l'œil nu de sixième grandeur. Cette échelle de luminosité est une échelle logarithmique inversée.
[63] En fait, $\iota$ Argo et $\theta$ Sge n'existent même pas dans le catalogue de Ptolémée, mais $\theta$ Sge est la seule étoile de la Flèche convenant au système newtonien (cf. Buchwald & Feingold p279), ce qui n'est évidemment pas une excuse.
[64] Toutefois, la description de Ptolémée s'accorde avec le choix de Newton, c'est la definition de la tête de la Baleine qui a changé avec le temps (p276-7 de Buchwald & Feingold).
[65] La consultation des cartes d'Hévélius pourrait être à l'origine de cette erreur (cf. Fig 8.13 p280 de Buchwald & Feingold).
[66] Emerson, p144

ancienne des astérismes pour les ajuster à son hypothèse, jointe aux dix-huit différentes copies du premier chapitre de sa Chronologie, que l'on a trouvées chez lui, sont des preuves du plus fort et du plus long attachement que l'on ait jamais vu parmi les hommes pour une hypothèse. […] [je me] contente de dire que cet argument tant célébré non seulement porte à faux mais qu'il est expressément & directement contraire à la nouvelle Chronologie. »

Se plaçant dans le système newtonien et en utilisant les rouages précessionnels à la manière du savant anglais, Whiston et Fréret tentent alors de trouver un système plus en accord avec les descriptions d'Eudoxe[68]. Ils concluent que le meilleur décalage serait de 42°15', ce qui placerait les observations de la sphère céleste rapportées par Eudoxe en 1353 avant notre ère et non 939. Dans ce cas, le colure équinoxial passerait entre les étoiles θ et ε Ari, α et β Lib, du côté de ι, φ, et κ Vir, et non loin de Menkar (α Cet) ; le colure solsticial passerait lui entre κ et λ Leo, θ et ς Cap ; le tropique septentrional passerait par l'amas de la Crèche, puis entre δ et γ Cnc, par α Ser, ι et κ Oph, α Her, π Cyg, θ And, ι Aur, κ et ι Gem ; le tropique méridional entre δ et γ Cap, entre η et ς Eri, par ς et α Lep, β CMa, κ Argo, non loin de τ et ψ Cen, λ et ν Sco, δ Sgr ; et l'équateur céleste par π, μ, ou ν Ari, α Ori, α Hya, δ Crv, α Lib, ς Oph, ν Aql, entre Markab (α Peg) et Algenib (γ Peg), vers φ et χ Psc. Ce serait la meilleure solution[69], mais cela ne veut pas dire que Fréret et Whiston insinuent que Chiron a vécu au 14$^e$ siècle avant notre ère (voir en effet ci-dessous).

De plus, un calcul précis nécessite des données précises, et ce genre de choses n'existe que rarement dans l'Antiquité. Shuckford met ainsi en doute la précision des anciens, tels Chiron, dans sa préface[70] : ils ne possédaient pas de catalogue stellaire avant Hipparque, et ne connaissaient ni l'obliquité de l'écliptique ni la longueur réelle de l'année. Les étoiles choisies par Newton apparaissent d'ailleurs à deux degrés, en plus ou en moins, de la valeur correcte[71], or une différence de quatre degrés implique un décalage temporel de trois siècles ! Dans ce cas, même si on « croit » que Chiron est bien l'auteur de la sphère (voir cependant discussion ci-dessous), comment obtenir un résultat précis ? Pour La Nauze, c'est très simple : la précision repose sur la moyenne que fait Newton à partir de plusieurs étoiles[72]. Cependant, le choix des étoiles se fait par reconstruction

---

[67] Whiston cité par Fréret, Défense, p439

[68] Fréret, Défense, p438-439, et p451-458. Whiston propose de reculer d'au moins deux siècles la chronologie de Newton : il donne d'abord la date de 1353 BC en p1010 de son annexe (recopiée p439 de la *Défense* de Fréret), puis propose un recul de deux siècles à la place des quatre précédemment utilisés en p1017 de son annexe (non recopiée par Fréret), suite aux remarques de Halley sur la « première » du Bélier, mais il ne cite aucune étoile ni ne donne de détail sur ses calculs (celui à la base des quatre siècles est explicité par Fréret dans sa *Défense*, celui à la base des deux siècles est explicité dans Buchwald & Feingold p365-6).

69 Au passage, on peut noter qu'il existe de nombreuses étoiles de 4$^e$, voire 5$^e$, grandeur parmi les étoiles choisies : leur solution n'est donc que partiellement meilleure au niveau de la luminosité des astres. D'autre part, les déclinaisons données par Fréret sont peu précises, avec quelques erreurs importantes – π Cyg se trouve à plus de 10° du tropique, τ Cen à plus de 5°, et φ Psc à plus de 5° de l'équateur. Buchwald & Feingold (p248) précisent que Newton n'utilise pas les tropiques et équateur car c'est moins simple à calculer. C'est effectivement le cas, mais si Fréret le fait sans problème, Newton devait en être capable aussi. Il est plus probable qu'il n'a pas trouvé de confirmation de sa datation par ce moyen, d'où le fait qu'il néglige d'en parler.

[70] The sacred and profane history of the world, vol II, voir aussi Mémoires de Trévoux, p347-348

[71] À noter que, bien que soutenant Newton de manière générale, Wood met lui aussi en doute la précision des observations à l'époque de Chiron. Voltaire a une démarche similaire (voir chap X, partie III, Eléments de la philosophie de Newton, 1773) : puisqu'Hipparque est le premier à s'apercevoir de la précession, cela « prouve que les Grecs n'avaient pas fait de grands progrès en astronomie » auparavant ; si Méton et Euctémon avaient trouvé une différence angulaire aussi grande, ils « n'auraient pu s'empêcher de découvrir cette précession des équinoxes… ce silence me fait croire que Chiron n'en avait point tant su que l'on dit, et que ce n'est qu'après coup que l'on crut qu'il avait fixé l'équinoxe du printemps au quinzième degré du Bélier. On s'imagina qu'il l'avait fait parce qu'il l'avait dû faire. Ptolémée n'en dit rien dans son Almageste, et cette considération pourrait, à mon avis, ébranler un peu la chronologie de Newton. »

[72] La Nauze, Mémoires de Trévoux, p2535 : « c'est en faisant servir par le résultat moyen, la diversité et la grossièreté de leurs observations à se corriger les unes par les autres qu'il nous montre aujourd'hui comment ils approchèrent plus ou moins de leur but, le milieu des constellations: ce qu'ils n'étoient nullement capables de vérifier eux-

moderne, aucun astre en particulier n'ayant été clairement cité dans les sources anciennes, qui parlent juste de « milieu » ou d'éléments tels une main ou une tête. De telles zones sont grandes (le dos du Bélier couvre ainsi plus de 6° !), et peuvent donc accommoder des solutions éloignées de plusieurs degrés... Whiston souligne[73] également un problème possible de compréhension car si les noms sont communs, constellations et signes ne sont pas identiques ! Il ne faut en effet pas oublier que les colures sont des cercles imaginaires, et qu'astres et astérismes n'ont pas été distribués dans le ciel pour les marquer... Du coup, passer par le milieu de l'un n'implique pas nécessairement passer par le milieu d'un autre. Il est ainsi impossible de passer en même temps au milieu des *constellations* du Bélier et de la Balance (voir aussi ci-dessus) alors que cela est naturel, vu leur définition, pour les *signes* associés. Il faut donc interpréter le texte d'Eudoxe, pour voir quand il parle de constellation et quand il désigne plutôt le signe – Whiston propose une solution, certes raisonnable, pour tenter d'identifier l'un et l'autre.

Enfin, tout ce calcul précessionnel repose sur des hypothèses. Même s'il y a quelques petites erreurs numériques de-ci, de-là[74], les calculs peuvent être considérés comme essentiellement corrects dans les écrits de Newton (contrairement au cas de Souciet !). La justesse du résultat dépend donc de la réalité des hypothèses. Fréret le souligne d'ailleurs[75] : « Je ne prétendois point attaquer la justesse de ces calculs [...] mais comme ces calculs supposoient des faits, je demandois la preuve de ces faits allégués. » Fréret entreprend ensuite de préciser ces faits[76] : « il dépend entièrement de deux suppositions, sçavoir 1° que Chiron avoit dessiné une sphère céleste pour l'usage des argonautes, 2° que cette sphère étoit celle qu'avoit suivie Eudoxe ; deux choses qui sont avancées gratuitement. » La Nauze tente d'ailleurs d'expliquer ces fondements à Souciet[77] : « vous supposez que le sentiment d'Eudoxe, dont il s'agit dans Hipparque est le sentiment personnel d'Eudoxe, le sentiment qui lui étoit commun avec Méton dans le système de M. Newton. Ce n'est pas cela. L'opinion attribuée à Eudoxe par Hipparque est l'opinion de Chiron. » Même si ce n'est pas explicite, c'est effectivement ce que fait Newton dans sa Chronologie : on passe subrepticement des points-repères décrits par Eudoxe (p86) à ceux des Argonautes (p89, voir extrait ci-dessus) sans aucune explication ni justification du lien entre les deux.

Cependant, supposer que la sphère de Chiron, et la position des colures sur celle-ci, sont identiques à ce qu'Hipparque attribue à Eudoxe[78] est une hypothèse extrêmement forte, pour laquelle les preuves manquent, même si Emerson y voit « un point d'histoire »[79]. Fréret[80] ne trouve au mieux qu'un seul ver poétique parlant de la définition des constellations par Chiron, un seul alors que bien d'autres vers, du même auteur et/ou de la même époque, parlent des constellations mais

---

mêmes ». Pour Buchwald & Feingold (chap II), l'utilisation de la moyenne constitue un apport extrêmement original et inédit de la part de Newton, point sur lequel ils insistent. Toutefois, ils précisent eux-mêmes que d'autres s'en servaient à l'époque (en démographie, météorologie, artillerie, voire... astronomie). Si bien des points de la *Chronologie* ont été débattus, l'utilisation de la moyenne n'a jamais posé de problème ni n'a été combattue comme l'est toujours une nouvelle idée révolutionnaire. Ce passage de La Nauze montre d'ailleurs clairement que la moyenne semble « naturelle » et ne pose pas de problème à l'époque.

[73] Whiston, cité par Fréret, Défense, p425-427 – Buchwald & Feingold expliquent longuement ce sujet (p289-291) en insistant sur le fait que Fréret ne comprend pas ce sujet. Cela fut peut-être exact au temps des *Observations*, mais ne l'est certainement plus au moment de la *Défense*, comme le montre le passage associé.

[74] Quelques erreurs parsèment les écrits de Newton: erreurs de calcul, de recopiage des coordonnées de Flamsteed (ν Ari, τ Per, ξ Cet, et δ Can), d'obliquité (celle utilisée n'est pas celle de 939 avant notre ère), et même étoile fantôme (non reprise dans le catalogue de Flamsteed, voir note 56). Elles ont été corrigées lors de l'édition des œuvres complètes par Samuel Horsley en 1795 (voir aussi Table 8.2 de Buchwald & Feingold). Cependant, ces erreurs sont sans grande conséquence sur l'âge final, car elles ne changent les résultats que de quelques années...

[75] Fréret, Défense, début de la 3e partie, p384 (à propos des Observations)
[76] Fréret, Défense, p415
[77] La Nauze, 1e lettre, p395
[78] cf. La Nauze, 1e lettre, p355
[79] Emerson, p139 et 143
[80] Fréret, Défense, p418

jamais en rapport avec Chiron ! Quant au lien entre Chiron et Eudoxe, il n'est pas non plus évident, alors qu'il est nécessaire au système, comme le souligne Fréret[81] : « M. Newton n'ayant aucune preuve que la sphère d'Eudoxe fut la même que celle de Chiron, il ne peut rien conclure pour le temps de Chiron ». Halley, loin de soutenir Newton sur ce point, y voit d'ailleurs « la partie la plus blâmable de tout le système »[82]. Que ces points se soient trouvés là avant Eudoxe, qui n'aurait que recopié la chose, soit[83], mais de là à les lier spécifiquement aux Argonautes, il y a un pas que Newton franchit allègrement (et Souciet ou La Nauze[84] avec lui !), mais Halley se garde bien de soutenir un tel fantasme. Souciet reconnaît d'ailleurs dans sa dernière dissertation[85] : il n'y a « aucun Ancien qui dise formellement le lieu où Chiron plaça des points cardinaux ».

c) Arguments supplémentaires avancés par Newton
    I. Précession chez Méton et Hipparque

Pour appuyer son calcul précessionnel, Newton va utiliser ce que l'on sait de deux astronomes anciens, Méton et Hipparque. Postérieurs à la célèbre expédition, ils doivent observer un écart dans la position des équinoxes et solstices : à leur époque, ces points ne doivent plus se trouver au milieu des signes/constellations. Mieux, l'époque de ces savants est en fait bien connue : grâce à Ptolémée, on sait que Méton a observé le solstice de 432 avant notre ère et qu'Hipparque a travaillé entre 158 et 128 avant notre ère. La datation par précession peut donc ici être facilement vérifiée, ce qui appuiera d'autant plus le système chronologique de Newton. Faire référence à ces « autorités » anciennes n'est donc pas anodin.

Dans l'Abrégé, pour l'entrée 939, Newton précise ainsi que « Méton, en l'an 316 de Nabonassar observa que le solstice avoit reculé de 7 degrés depuis le temps auquel Chiron l'avoit fixé. » Dans la lettre de Keil, l'Anglais insiste en précisant qu'il y a une différence de 8° entre la position des colures de son époque et celle de l'époque de Méton et de 4° pour Hipparque[86]. On retrouve la même idée détaillée dans la Chronologie[87] : « Dans l'année de Nabonassar 316 […] Meton & Euctemon observèrent le Solstice d'Eté, […] & Columelle nous dit qu'ils le trouvèrent dans le huitième degré du Cancer, qui est au moins de sept degrés plus reculé que la première fois. Or l'Equinoxe rétrograde de sept degrés en 504. ans […] on trouvera l'expédition des Argonautes comme ci-devant, 44. ans ou environ, après la mort de Salomon. […] Hipparque célèbre Astronome, en comparant ses observations avec celles qui avoient été faites avant lui, s'apperçut le premier que les Equinoxes avoient un mouvement par rapport aux étoiles fixes, en s'éloignant d'elles contre la suite des signes. Il crut d'abord que ce mouvement étoit d'environ un degré en cent ans. Il fit les observations des Equinoxes entre les années de Nabonassar 566. & 618. L'année moïenne est 602 qui est 286 ans après l'observation de Meton & d'Euctemon ; or l'Equinoxe rétrograda durant ce nombre d'années, de 4°. Ainsi il fut dans le quatrième degré d'Ariès du tems d'Hipparque, & par conséquent il avoit rétrogradé de 11°, depuis l'expédition des Argonautes »

Alors que les écrits cités par l'Anglais ne laissent aucune place au doute (on peut cependant regretter l'absence d'extraits des textes originaux), Fréret propose une vision plus large[88] car il faut considérer l'ensemble des sources, et non juste celles qui nous arrangent. Il prend le cas d'Eudoxe, qui « parloit autrement dans ses ouvrages ». Ainsi, dans le calendrier de Geminus, il place l'équinoxe de printemps au 6e degré du signe du Bélier et le solstice d'hiver au 4e degré du signe du Capricorne (soit 2° de différence par rapport à l'angle droit attendu !) tandis que dans son *Enoptron* cité par Hipparque, Eudoxe place ces points au 15e degré des signes et que dans le calendrier de Columelle, Méton et Eudoxe s'accordent à les placer au 8e degré[89]... Dans les textes anciens, on trouve aussi qu'Hipparque diffère d'Eudoxe de 15° en longitude, mais aussi qu'Euctémon, collègue de Méton, place dans les calendriers de Geminus et Callipus l'équinoxe au premier degré des signes. Bref, difficile de dégager un esprit commun des différents textes anciens ! Selon Fréret, cela n'est pas étonnant : les traditions ont la peau dure, et subsistent longtemps avant qu'on ne puisse les changer. Il ne faut pas oublier non plus qu'une précision d'un degré est loin d'être indispensable aux agriculteurs à qui ces calendriers étaient destinés. Un bel exemple est le cas de Columelle[90], qui donne des dates différentes pour le lever (héliaque ou achronique) des mêmes étoiles ! Ces dates et les angles mentionnés proviennent probablement d'observations faites à des dates différentes et compilées ensuite dans le même ouvrage, quoique les angles différents puissent aussi s'expliquer par des changements de système de coordonnées (hypothèse privilégiée par Newton selon Fréret quoique ce ne soit pas apparent dans la Chronologie, voir aussi La Nauze ci-dessous).

---

[81] Fréret, Défense p 442
[82] Halley, Philosophical Transactions, p206 : « the most exceptionable part of the whole system » – généralement recopié en utilisant « questionable » (discutable) alors que le texte original est bien « exceptionable » (blâmable, inadmissible). Buchwald & Feingold (p328 et p376-9) assurent qu'Halley soutient Newton, mais ce passage montre bien le contraire : son soutien se borne à montrer les déficiences de Souciet, contradicteur de Newton – en fait, Halley aurait même pu éviter toute mention de l'hypothèse, puisque Souciet n'y trouve rien à redire !
[83] Fréret, Défense, p417
[84] La Nauze, 1$^e$ lettre, p 360 : la seule solution pour éradiquer le système newtonien est de montrer que Chiron n'est pas l'auteur de la sphère céleste, mais il juge cette solution peu plausible ; La Nauze, Mémoires de Trévoux, p2532 : « comment imaginer que ce n'ait pas été la sphère primitive ? »
[85] Souciet, 5$^e$ dissertation, p127
[86] Souciet, 1$^e$ dissertation, lettre de Keil reproduite p 56
[87] Newton, Chronologie, p96-97

Souciet cite les mêmes sources que Fréret et refait les calculs de précession pour les époques de Méton ou Hipparque, de manière à montrer qu'il est impossible que ces derniers aient pu observer un solstice à 8° ou 4° des signes[91] : il y a en effet 29.6° de décalage pour l'équinoxe entre 1700 et 432 avant notre ère, époque de Méton, et 25.4° pour l'époque d'Hipparque – tout cela est proche d'un signe entier, et donc non compatible avec 8° ou 4°. Dans la même veine, il insiste, en précisant que si Ptolémée parle des observations d'équinoxes ou solstices faites par Hipparque ou Méton, il ne précise jamais le lieu céleste où cela se passe[92]. Il précise même qu'Hipparque affirme que presque tous mathématiciens ont mis les points « cardinaux » au commencement des constellations, sauf Eudoxe – il n'exclut donc pas explicitement Méton, pourtant bien connu[93]. Souciet trouve même une explication plausible au problème des 8° : puisqu'il n'y a que Columelle qui l'attribue à Méton, pourquoi n'y aurait-il pas une erreur de copiste[94] ? La Nauze ne voit, lui, aucun problème. Il réconcilie même les différentes sources entre elles : il s'agit simplement d'une question de convention ! Si l'on imagine que Méton utilise la convention de Chiron, avec un début des signes 7°24' avant l'oreille du Bélier, alors le colure passait bien de son temps au 8e degré – Euctémon aurait, lui, plutôt choisi une convention devenue aujourd'hui classique, qui met l'équinoxe au départ des signes, ce qui permet de réconcilier les deux collègues[95]... De même, une équinoxe à 4° dans la convention de Chiron est compatible avec une équinoxe au premier degré

---

[88] Fréret, Observations p62-67, voir aussi Défense, p461-483
[89] En acceptant même les arguments de Newton, on tombe sur divers problèmes (Fréret, Observations, p75-6)... si l'on commence les signes à γ Ari, l'oreille du Bélier. Ainsi, cette étoile possède une longitude écliptique nulle en 388 avant notre ère puisqu'elle a une longitude écliptique d'environ 29° en 1690. Un colure 15° plus loin implique une date 15*72=1080 ans plus ancienne pour Chiron, soit 5 siècles de différence avec Newton. Un colure 8° plus loin impose pour Méton une date située au 10e siècle avant notre ère, alors qu'on sait qu'il observait cinq siècles plus tard. Cet anachronisme flagrant repose tout entier sur le choix du commencement, et Fréret commet au départ la même erreur que Souciet, pensant que Newton commence son signe à l'oreille du Bélier. Cette erreur lui vaudra une remontrance de Newton dans sa réponse (p318 des Philosophical Transactions).
[90] Fréret, Défense, p481
[91] Souciet, 1e dissertation, p62-68, voir aussi Fig. 5
[92] Souciet, 1e dissertation, p64-65
[93] La Nauze (1e lettre, p392) assure que l'introduction d'Hipparque précise qu'il ne citera les noms que d'Aratus et d'Eudoxe, ce qui expliquerait la disparition de Méton ici...
[94] Souciet, 1e dissertation, p 69-70 : un A transformé en H, soit un 1 en 8 dans la notation grecque des nombres
[95] La Nauze, 1e lettre, p383-392, voir aussi Fig. 5

chez Hipparque, si ce dernier adopte une convention « classique », comme Euctémon[96].

Les deux adversaires palabrent aussi quant à une autre remarque d'Hipparque, pour qui l'étoile au milieu du dos du Bélier est au moins à un tiers de signe, soit 11°, du colure équinoxial. Souciet[97] précise que cela met le colure près de l'oreille car il y a environ 11° entre ν Ari (milieu du dos) et γ Ari (oreille), ce qui est correct et arrange bien Souciet puisque l'oreille marque alors le début des signes (voir Fig. 4). Par contre, en utilisant un début de signe 7.5° à l'ouest de l'oreille comme suggéré par Newton, un colure 11° plus loin passerait près de la tête et non dans le dos (voir Fig. 4, notez que Souciet confond ici l'époque d'Hipparque et celle de Chiron, puisqu'il utilise la convention de début de signe de Chiron/Newton, non utilisée par Hipparque !). La Nauze voit plutôt dans cet extrait un indice supplémentaire de la justesse d'une équinoxe à 4° chez Hipparque, puisque ce dernier voit un décalage précessionaire de 11° entre Chiron et lui (15°-4°=11°)[98]. Comme Newton précise dans sa réponse à Fréret que l'oreille du Bélier se trouve, en longitude, à 7°36' à l'est du colure de Chiron, La Nauze en déduit que le colure d'Hipparque passe à 3°24'(=11°-7°36') à l'ouest de l'oreille, ce qui correspond à la position attendue de l'équinoxe de printemps au temps d'Hipparque... Cependant, les deux auteurs ne calculent pas les choses dans le même repère : Souciet mesure la distance des astres en longitude écliptique simple, La Nauze reprend les valeurs de Newton qui a, lui, tenu compte de l'inclinaison du colure équinoxial pour déterminer la position de l'équinoxe de Chiron, pour les comparer à des longitudes écliptiques simples (Fig. 4, voir aussi Fig. 5). On assiste là à un véritable dialogue de sourds, sans espoir de réconcilier les vues... alors que leurs arguments (colure passant par l'oreille et colure à 3°24' de l'oreille en longitude écliptique) ne sont pas mutuellement exclusifs, au contraire ! Par contre, ils semblent oublier dans tous leurs calculs « précis » qu'Hipparque travaillait en ascension droite et non en longitude écliptique, alors qu'ils en conviennent pourtant tous deux (voir note 19)...

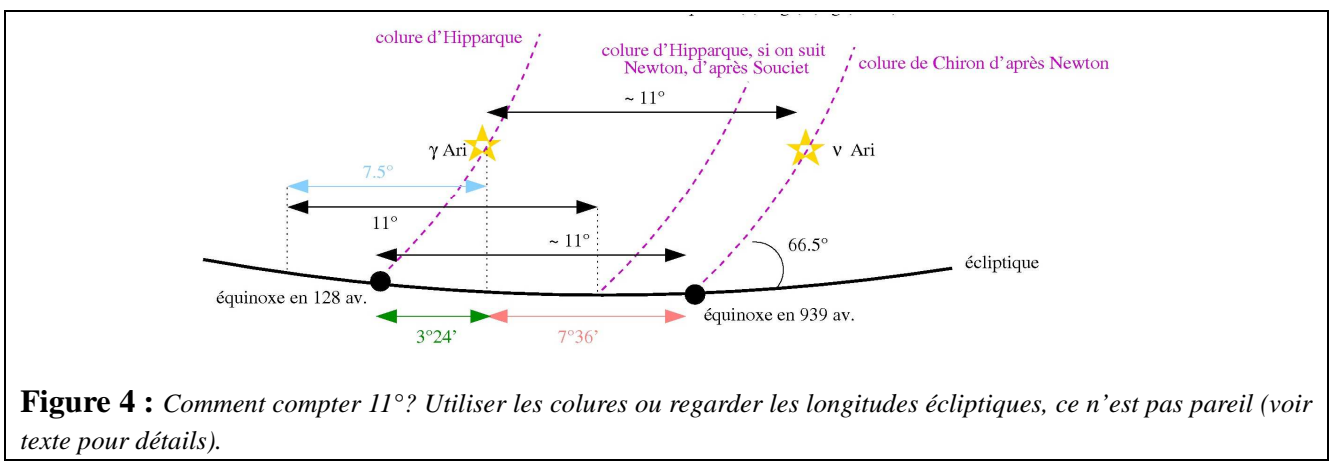

**Figure 4 :** *Comment compter 11°? Utiliser les colures ou regarder les longitudes écliptiques, ce n'est pas pareil (voir texte pour détails).*

Pour régler la question, Fréret préfère placer l'extrait litigieux dans son contexte global. Il préfère donc analyser l'ensemble des longues descriptions de constellations faites par Hipparque, où il compare ses descriptions avec celles données par Eudoxe[99]. Ainsi, pour citer quelques exemples, le colure passe dans le col de l'Hydre (dont la première étoile est ω Hya) chez Eudoxe et sur sa tête (étoiles δ ou σ Hya) chez Hipparque, ce qui fait une différence d'au moins 16° ; le colure rencontre chez Eudoxe la main gauche du Bouvier, qui couvre 6° en longitude (étoiles κ, ι et g Boo), tandis que ces étoiles se placent au 13e degré de la Balance du temps d'Hipparque, ce qui implique une différence plus grande que 13° en moyenne. Un raisonnement similaire s'applique au Poisson Austral, au Cygne, à Céphée, et au Centaure. Toutefois, mieux que des lieux vagues, Hipparque précise le cas de quatre étoiles, bien identifiées : α UMa se trouve ainsi au 18$^e$ degré du Cancer chez Eudoxe et au 3$^e$ degré du Lion chez Hipparque, soit une différence de 15° ; η UMa passe, selon Hipparque, du 4$^e$ au 18$^e$ degré de la Balance lorsqu'on passe d'un repère avec colure au début des signes (son propre cas) à un repère avec colure au milieu des signes (cas d'Eudoxe), γ UMa fait le chemin du 10$^e$ degré de la Vierge au 25$^e$ degré du Lion, et l'étoile polaire (actuelle) du 18$^e$ degré des Poissons au 3$^e$ degré du Bélier. Certes, Hipparque parle bien d'une différence d'un tiers de signe environ (les fameux 11° qui tarabustent Souciet) pour le Bélier, mais l'astre concerné est moins clair à identifier que dans d'autres cas : Newton a donc fait un choix plutôt sélectif, mais qui avait l'avantage de convenir à son système...

---

[96] La Nauze, 1e lettre, p 375-376, voir aussi Fig. 5
[97] Souciet, 5e dissertation, p145
[98] La Nauze, 1e lettre, p 369-373, et 5e lettre, p410-414, voir aussi Figs. 5 et 6
[99] Fréret, Défense, 3e partie, section II, p444-450

Enfin, Newton rebondit aussi sur la détermination et l'usage de la précession par les Grecs en général et Hipparque en particulier. Ainsi, il écrit dans la Chronologie[100] : l'équinoxe « fut dans le quatrième degré d'Ariès du tems d'Hipparque, & par conséquent il avoit rétrogradé de 11°, depuis l'expédition des Argonautes ; c'est-à-dire en 1090. ans, si l'on suit la Chronologie des anciens Grecs, qui étoit alors en usage : ce qui donne environ 99 années, ou en prenant un nombre rond, 100. ans pour un degré, comme Hipparque l'avoit alors déterminé. Mais il est certain que l'Equinoxe rétrograde d'un degré en 72. ans, & de 11° en 792. années. Ainsi comptant ces 792 années en rétrogradant depuis l'année 602 de Nabonassar, qui est l'année d'où nous avons compté les 286 années, on placera par ce calcul l'expédition des Argonautes environ 43. années après la mort de Salomon. Les Grecs avoient donc fait l'expédition des Argonautes de 300. ans plus ancienne qu'elle ne l'étoit effectivement, & cette erreur donna occasion à Hipparque de déterminer la rétrogradation des Equinoxes d'un degré seulement en cent ans. » Il continue, mettant la précession à toutes les sauces[101] : « Hipparque comptait un degré par siècle pour la précession, et ils [les Grecs] ont généralement fondé leur chronologie sur cette estimation ». La « fausse » chronologie grecque, loin de saper le système, vient le soutenir ! Évidemment, Newton évite la conclusion étrange de ce raisonnement[102] : Hipparque aurait découvert par la précession par des observations, mais pour déterminer son taux, il les aurait délaissées au profit d'une chronologie incertaine…

Souciet se demande quelle peut bien être l'origine de cette affirmation péremptoire du savant anglais. Il n'en existe aucune trace claire, et il semble d'ailleurs certain qu'Hipparque n'a pas estimé sa distance temporelle d'avec Méton par cette méthode ni Ptolémée pour sa distance à Hipparque car ils auraient trouvé des intervalles plus grands que les valeurs réelles, vu l'erreur dans l'estimation de la précession à l'époque[103]. La Nauze répond ici que ce sont les adversaires mêmes de Newton qui l'ont affirmé[104] : on trouve en effet cette idée dans les Observations de Fréret, à propos d'une estimation chronologique de Sénèque[105] – mais il ne fournit aucune source ancienne pour appuyer l'idée (et pour cause!). Il explique ensuite que, forcément, les Grecs n'ont utilisé ce type de raisonnement qu'après la détermination d'Hipparque et ils ne l'ont appliqué que pour les faits très anciens, pas pour ceux qu'ils pouvaient dater facilement (comme les époques de Méton, Hipparque ou Ptolémée). Voyant probablement la faiblesse de son raisonnement, La Nauze prévient cependant que la chose n'est pas une condition nécessaire au système newtonien, elle n'en est qu'une conséquence, une conjecture probable... Il s'empresse d'ailleurs d'accuser Sénèque de chronologie chimérique[106].

---

[100] Newton, Chronologie, p97-98
[101] Lettre de Keil, reproduite p 56 de Souciet, 1$^e$ dissertation : « Hipparcus counted the precession to be a degree in a hundred years, and they are generally founded their Chronology upon that computation »
[102] F.E. Manuel, p74-5
[103] un degré par siècle au lieu d'un degré par 72 ans (Souciet, 1$^e$ dissertation, p58)
[104] La Nauze, 1$^e$ lettre, p 363-369
[105] Fréret, Observations, p81-82
[106] La Nauze, 5$^e$ lettre, p398-399

Fréret, lui, est plus direct dans sa Défense, et prend Newton à son propre jeu[107]. Hipparque établit un décalage de 15° entre ses observations et celles d'Eudoxe (voir ci-dessus), or on pensait alors que Chiron vivait 1100 ans avant, Hipparque aurait donc dû trouver une valeur pour la précession de 15°/1100 ans, soit 1° par... 73 ans, proche de la valeur moderne ! Puisqu'il donne une valeur d'un degré par siècle, c'est qu'il n'a pas utilisé la sphère d'Eudoxe pour déterminer la valeur du décalage précessionnel. Ceci dit, il suppose que la différence observée a pu l'inspirer pour sa découverte ultérieure d'un décalage avec les observations d'Aristille et Timocharis...

  II. Thalès & Hésiode

  Pour soutenir sa chronologie réduite, Newton tente d'apporter des éléments supplémentaires, avec une approche complémentaire aux calculs stellaires évoqués ci-dessus. La Chronologie explique ainsi[108] : « Après l'expédition des Argonautes on n'entend plus parler de l'Astronomie, jusqu'au tems de Thalès [...] Pline assûre qu'il détermina le coucher cosmique[109] des Pléiades au vingt-cinquiéme jour après l'Equinoxe de l'Automne : d'où le P. Petau calcule la longitude des Pléiades en 23°53'♈ : par conséquent la luisante des Pléiades s'étoit éloignée de l'Equinoxe de 4°26'52'' depuis l'expédition des Argonautes[110] : ce mouvement répond à 320 années [...] si l'on compte ces années en rétrogradant depuis le tems que Thalès étoit jeune [...] environ la XLI Olympiade, on trouvera par ce calcul que l'expédition des Argonautes arriva environ 44. ans après la mort de Salomon, comme on a trouvé ci-dessus. » Ce raisonnement est relativement correct, et un calcul simple permet de s'en rendre aisément compte : si un astre n'est pas très éloigné de l'écliptique, son coucher alors que le Soleil se lève implique une longitude écliptique à 180° de celle du Soleil (soit 6 signes) ; sachant que le Soleil parcourt 360° sur l'écliptique en un an (il se déplace donc d'environ 1° par jour), un intervalle de 25 jours correspond à 25° environ de longitude ; la position du Soleil à l'équinoxe d'automne vaut 0°♎ donc environ 25°♎ 25 jours plus tard, et la position opposée vaut alors 25° ♈. Pour un astre éloigné de l'écliptique, il faut corriger un peu cette valeur, en tenant compte de la latitude du lieu d'observation, ce qu'a fait Pétau, d'où sa valeur de 24° environ. Connaissant la longitude écliptique en 1690 et au temps de Thalès, on en déduit l'intervalle de temps entre les deux, donc la date de l'observation de Thalès. Sachant que son époque correspond à la 41[e] olympiade, dont la date est connue par ailleurs, cela permet de vérifier la validité de la méthode précessionaire. Utilisant le décalage calculé pour la sphère originelle pour calculer les coordonnées écliptiques des Pléiades, Newton y trouve aussi une confirmation de la date de sa confection, qu'il assimile à celle de l'époque des Argonautes (point sur lequel ce passage n'apporte évidemment aucune preuve).

---

[107] Fréret, Défense, 3[e] partie, section I, p415-417

[108] Newton, Chronologie, p95-96 ; voir aussi annexe E de Buchwald et Feingold pour les détails de tels calculs à l'époque newtonienne.

[109] Le texte français utilise « coucher cosmique », ce qui correspond (aujourd'hui) au coucher d'un astre avec le Soleil, alors que le texte original anglais utilise « occasus matutinus », soit le coucher achronique ou acronyque (l'astre se couche quand le Soleil se lève – voir définitions de ce genre d'événements sur le site de l'IMCCE http://www.imcce.fr/fr/grandpublic/systeme/promenade/pages6/725.html). Le terme de coucher cosmique est donc inapproprié pour l'équinoxe d'automne : en effet, le signe du Bélier correspond à la position du Soleil au printemps, un astre se couchant en même temps que le Soleil et se trouvant donc dans ce signe ne peut le faire qu'en cette saison et non six mois plus tard en automne. De plus, puisqu'il est ici question de « voir » le phénomène, il faudrait tenir compte de la réfraction atmosphérique, du relief de l'horizon, et de la hauteur au-dessus de l'horizon pour observer l'événement (voir paragraphe suivant).

110 Juste avant ce paragraphe, Newton calcule la longitude attendue des Pléiades en retirant 36°29' de la longitude écliptique des Pléiades. Il trouve 19°26'8'' ♈, et la différence avec les 23°53' ♈ de Pétau donne bien 4°26'52''. Selon Buchwald & Feingold (note 26 p256 et p482-4), cet astre doit être η Tau, la « lucide/luisante » des Pleiades (soit la plus brillante), dont la longitude écliptique est estimée par Flamsteed à 25°37'13''♉ en 1686 et 25°40'8''♉ en 1690. Cependant, la longitude reconstruite pour 1690 (Newton ne donne en effet pas la valeur de départ) vaut 19°26'8'' ♈ + 36°29' = 25°55'8'' ♉, ce qui ne correspond pas à η Tau ni à aucun objet du catalogue de Flamsteed (l'étoile s Tau est la plus proche de cette valeur).

Newton affirme aussi[111] : « Hesiode nous dit que de son tems soixante jours après le Solstice d'hiver, l'étoile Arcturus se levoit précisément quand le Soleil se couchoit : on voit par là qu'Hesiode fleurissoit environ 100. ans après la mort de Salomon, ou dans la Génération ou l'Age qui suivit immédiatement la guerre de Troye, comme Hesiode le déclare aussi lui-même. » Fréret confirme la datation, en précisant[112] que Longomontanus en tire une date de 970 avant notre ère, Kepler 930 et Riccioli 950. Horsley précise cependant qu'en refaisant le calcul, il arrive à une date proche de celle de l'expédition des Argonautes chez Newton, soit avant la guerre de Troie, ce que semble confirmer le calcul trouvé dans un manuscrit de Newton – ce dernier aurait donc ici embelli le résultat[113]. Une simulation confirme que cette observation est possible à l'époque d'Hésiode depuis sa région. Cependant, trouver une date précise est difficile car l'observation en elle-même dépend du relief de l'horizon (présence de montagnes ou de collines), mais aussi de la hauteur au-dessus de l'horizon nécessaire pour « voir » réellement Arcturus (les astres à l'horizon étant très éteints et donc inobservables). Bien sûr, cette méthode permet juste de dater Hésiode, la date de l'expédition n'étant confirmée qu'indirectement, en supposant qu'il vit une génération après la guerre de Troie qui, elle, suivit l'expédition des Argonautes d'un intervalle de temps connu – à ce propos, l'emploi d'une « génération » au sens propre (soit quelques décennies) sera d'ailleurs débattu. Il faut également noter que le cas d'Hésiode met en lumière le choix plutôt sélectif des textes par Newton. En effet, le passage de Pline relatif à Thalès mentionne que le même phénomène avait lieu à l'équinoxe (et non 25 jours plus tard) du temps d'Hésiode. Newton calcule que les Pléiades étaient au 20$^e$ degré du signe du Bélier lors de la définition de la sphère par Chiron (voir ci-dessus et note 110), or le passage de Pline implique une longitude nulle, et ces vingt degrés de différence impliquent qu'Hésiode vivait à une époque située plus de 1400 ans avant la définition de la sphère, ce qui conduit à un hiatus quant à la date de l'expédition[114].

d) Points additionnels non abordés par Newton

Si Newton s'est contenté dans l'Abrégé et la Chronologie des points abordés ci-dessus, ce ne fut pas le cas des écrits de ses détracteurs ou de ses soutiens, qui cherchèrent à trouver jusqu'à la dernière conséquence du système newtonien. Cette section s'attarde sur ces points additionnels, centrés sur la position des astres, qui poussent le modèle dans ses derniers retranchements et par là même furent âprement débattus.

L'objection principale de Fréret dans ses Observations mais surtout de Souciet au texte concis de l'Abrégé porte sur le début du signe. Pour eux, le signe commence à la première étoile du Bélier, qu'ils identifient à γ Ari, l'oreille de la bête. S'ils ajoutent 15° aux coordonnées de cette étoile pour trouver la position du colure de Chiron, ils trouvent cette position éloignée de 44° de l'équinoxe de 1700, ce qui lui donne une date de 1470 avant notre ère pour l'expédition[115]. Newton ayant répondu aux observations de Fréret en précisant que le colure se trouve à 7°36' à l'est de l'oreille en longitude écliptique[116], Souciet en déduit que le signe du Bélier commence au milieu de nulle part chez Newton, car il n'y a aucune étoile située à 15° du colure, soit 7°24' à l'ouest de l'oreille. Il assure ainsi que Newton « ajuste l'Astronomie à son système »[117]. Il estime sa solution meilleure car il juge plus utile de démarrer un repère à une étoile brillante. De plus, Sextus Empiricus et Macrobe racontent que les signes commencent à des étoiles remarquables et s'ils ne parlent pas des Grecs, mais des Egyptiens et des Chaldéens, ce n'est pas un problème pour Souciet

---

[111] Newton, Chronologie, p98
[112] Fréret, Défense, p 460
[113] Opera Omnia, Newton (édité par S. Horsley, 1795), p75 et Buchwald & Feingold, note 121 p296
[114] Opera Omnia, Newton (édité par S. Horsley, 1795), p73
[115] Souciet, 1$^e$ dissertation, p 52 ; voir aussi Fréret, Observations, p75-6
[116] Newton, Philosophical Transastions, p318
[117] Souciet, 5$^e$ dissertation, p120

car les Grecs ont hérité de ces peuples[118].

La réponse à cette objection est double. Tout d'abord, on ne peut commencer les signes à l'oreille du Bélier et avoir un colure qui passe au milieu du dos du Bélier – un colure situé 15° plus loin passe plutôt du côté de la queue[119]. Halley, puis Newton dans la Chronologie (p91), montrent d'ailleurs que l'étoile ν Ari, située au milieu du dos, est compatible avec les calculs newtoniens. D'autre part, le début des signes ne semble pas si important car, contrairement aux Egyptiens et Chaldéens évoqués par Sextus Empiricus et Macrobe, les textes grecs ne parlent que du milieu des signes[120]. En soi, les constellations étant de largeur variable alors que les signes ont tous 30° d'extension (voir aussi ci-dessous), il semble en effet plus simple de repérer le milieu d'une constellation que les points virtuels situés à 15° de part et d'autre. La Nauze remarque enfin que les calculs de Souciet ne tiennent pas compte de l'inclinaison du colure équinoxial, ni du fait que les anciens travaillaient plutôt en ascension droite, comme le souligne d'ailleurs Souciet lui-même. Cependant, La Nauze lui-même ne corrige aucun des calculs, et place donc ses raisonnements dans les pas erronés de Souciet.

Souciet et La Nauze vont aussi discuter du cas de Spica, l'Epi de la Vierge (α Vir). Selon Souciet[121], Hipparque place cette étoile en 24°♍, soit près du colure selon Riccioli. Pour vérifier la chose, Souciet place le début des signes à l'oreille du Bélier, à son habitude. Il trouve alors que les équinoxes de 1700 se placent en 59'♓ pour le printemps et 59'♎ pour l'automne (voir Fig. 5). Remontant alors jusqu'en 128 avant notre ère, époque d'Hipparque, il trouve alors les équinoxes en 26°22' des mêmes signes, ce qui n'est pas loin de la valeur théorique de 30°. Il reprend alors la longitude de Spica selon le père Pétau, 24° ♍ (on trouve le même résultat en précessionnant le catalogue de Flamsteed) : cette valeur est peu éloignée des 26°22'. Spica se trouvait donc bien près du colure, et ce colure se trouvait près du commencement des signes. Souciet estime qu'une erreur de quelques degrés, comme observé, n'est pas impossible vu l'absence d'instruments précis à cette époque. Hélas, il commet quelques erreurs dans son calcul (voir Fig. 5) : cette longitude de Spica est valable dans un repère de type moderne, commençant à l'équinoxe, et non dans le système de Souciet commençant à l'étoile de l'oreille : si l'on corrige ce problème et que l'on se place dans le repère de Souciet, Spica se positionne alors en 20.5° ♍, soit à plus de 6° du colure, une erreur plutôt importante. La Nauze place plutôt le début des signes à l'époque d'Hipparque à environ 3° de l'oreille : dans ce cas, l'erreur sur la position de l'équinoxe est moindre (quelques degrés), mais celle de Spica reste élevée (6° environ), ce qui semble à La Nauze moins grave qu'une erreur sur la position équinoxiale. La Nauze ne remarque cependant pas l'erreur de coordonnées de Souciet[122], et il fait une double preuve de mauvaise foi[123] : tout d'abord, si l'on accepte les calculs de Souciet et une erreur dans le positionnement de quelques degrés, il n'y a aucune incompatibilité chez ce dernier ; ensuite, il assure que Spica est « mal marqué » à 24° ♍, ce qui est faux.

---

[118] Souciet, 5ᵉ dissertation, p129 et 150-155
[119] Halley, Philosophical Transactions, vol 34, p207-209 ; La Nauze, 1ᵉ lettre, p354-360
[120] La Nauze, 1ᵉ lettre, p354-360 ; 5ᵉ lettre, p400-401 et 431-440
[121] Souciet, 1ᵉ lettre, p60-63
[122] Il en commet par contre une autre, donnant un inexplicable 17° ♍ pour Spica, dans sa note p 377.
[123] La Nauze, 1ᵉ lettre, p377 et 383

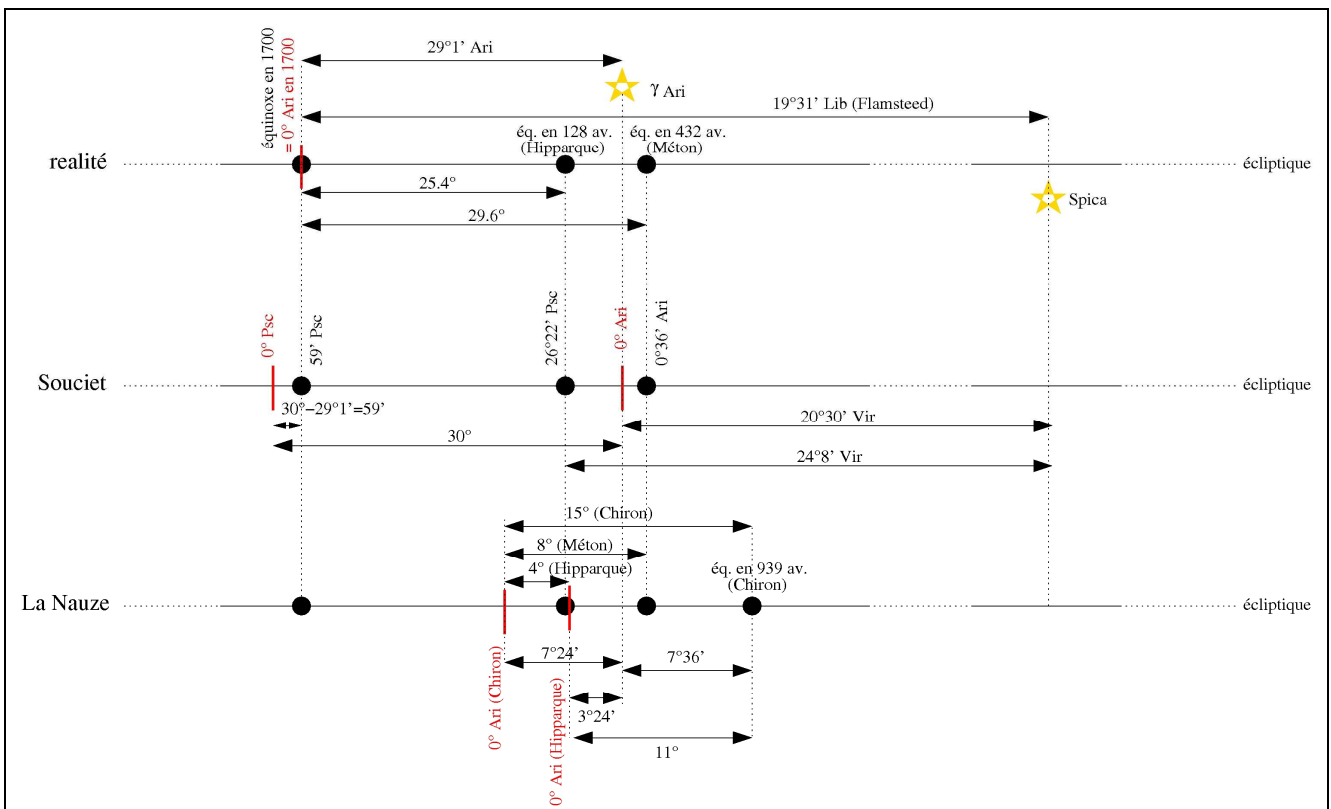

**Figure 5 :** *Position de Spica, des équinoxes à diverses dates, et systèmes de coordonnées adoptés par La Nauze et Souciet (les codes standards des constellations, Ari, Tau, etc., sont ici utilisés pour les signes, à la place de ♈, ♉, etc.).*

Une autre pierre d'achoppement est l'identification de la première étoile du Bélier, si chère à Souciet. Si l'oreille est choisie tant par Newton que Souciet (dans sa première dissertation), un autre astre semble lui damer le pion dans l'Antiquité : selon Hygin, « il y avoit au pied de devant du Bélier une étoile de première grandeur »[124]. Souciet la suppose très proche de l'oreille[125], donc avec une différence de coordonnées négligeable pour ses calculs. La Nauze le reprend sur ce point[126]. Pour que cet astre puisse se lever avant l'oreille, comme le notent les anciens, il faut quelques degrés de différence avec l'oreille (et non 45' seulement), car le pied est plus au sud que la tête : toute différence moindre entraîne un lever de l'oreille en premier depuis les latitudes moyennes... Tout cela concorde avec un début des signes quelques degrés à l'ouest de l'oreille du temps d'Hipparque, comme dans le système newtonien – à la limite, le 45' peut se rapporter à une différence d'ascension droite[127]... De plus, Souciet a mal interprété le « in pede de posteriobus primo unam », texte d'Hygin rapportant l'existence de cette étoile mystérieuse : il ne parle pas de première grandeur (il ne pouvait le faire car c'est Hipparque qui inventa le système de magnitudes, bien plus tard !), il précise juste qu'il s'agit de l'étoile du premier pied de devant... La Nauze ne l'identifie pas, mais Halley s'en charge, dans un calcul précis, où toutes les hypothèses sont démontrées[128] : cet astre est η Psc, au nord du nœud des Poissons.

---

[124] Souciet, 5ᵉ dissertation, p119
[125] Il détermine une distance angulaire de 45' en utilisant une autre remarque d'Hipparque (Souciet, 5ᵉ dissertation, p120), mais il se trompe car Hipparque parle d'une distance de 45' entre colure et première étoile, sans dire que l'étoile en question est celle de l'oreille : ça, c'est l'hypothèse de Souciet ! Il reprendra le texte un peu plus loin (p150) en affirmant cette fois qu'Hipparque dit que la première des trois étoiles brillantes de la tête, l'oreille, selon Souciet, est éloignée de la 20ᵉ partie d'une heure, soit 45', du colure. La Nauze (5ᵉ lettre, p429-431) y voit, lui, une distance en ascension droite et non en longitude écliptique...
[126] La Nauze, 1ᵉ lettre, p 343-345 ; 5ᵉ lettre p 381
[127] La Nauze, 5ᵉ lettre, p429-431
[128] Halley, Philosophical Transactions, vol 35

Se pose aussi la question de la tête de la Baleine, ou du monstre marin : le colure newtonien passe-t-il dessus, comme le précisent les textes anciens ? Comme Whiston (voir ci-dessus), Souciet assure que le colure de Newton passe trop loin de la tête[129]. La Nauze n'est pas du même avis[130] mais il se contredit à ce sujet[131] en assurant d'un côté que la tête est étroite, contrairement à ce qu'affirme Souciet, après avoir affirmé cinq pages plus tôt que la tête couvre le large intervalle situé entre 29.5° ♈ et 11° ♉ – l'oreille du Bélier se trouvant à 29° ♈ et le début de la queue en 14° ♉, La Nauze en déduit que ce qui passe au milieu de l'un passe au milieu de l'autre... Rien n'est plus faux ! La Nauze n'a en effet pas tenu compte du tout de l'effet d'inclinaison du colure équinoxial, d'autant plus fort que l'on s'éloigne de l'écliptique (Fig. 3, il finira par s'en rendre compte bien plus tard, comme le montrent les Mémoires de Trévoux, p2537). En outre, l'identification des étoiles pose problème : celle en 11° ♉ est probablement x ou λ Cet, celle en 29° ♈ l'étoile γ Ari et celle en 14° ♉ l'étoile ε Ari – mais où va-t-il donc chercher une étoile de la tête de la Baleine à 29° ♈ ? Il n'en existe tout simplement pas ! Les étoiles situées à la tête, chez Bayer (Uranometria) sont α, κ, et λ Cet, μ Cet étant déjà dans la nuque. Utilisant leurs coordonnées mesurées par Flamsteed, on trouve que le colure équinoxial passant près de ces astres coupe l'écliptique en 15.6°, 20.9°, 14.2°, et 10.0° ♉, respectivement. La solution de Newton (6°29' ♉) est donc assez peu en accord avec ces valeurs, alors que celle proposée par Whiston et Fréret (12°15' ♉) s'en accommode beaucoup mieux.

Souciet s'embarque également dans des discussions sur quelques étoiles supplémentaires[132] : la tête du Bélier, de nouveau, le pied du Bouvier et diverses parties du Centaure. Il prend à témoin les textes d'Hipparque... tout en utilisant les définitions que Newton attribue au temps de Chiron ! Il en résulte un méli-mélo absurde, qui ne prouve évidemment rien... La Nauze, cette fois, se rend compte du problème et le signale[133]. Il reprend aussi Souciet sur son affirmation péremptoire qu'une étoile située au 1$^{er}$ degré du signe du temps d'Hipparque se trouve en 1700 en 29° du même signe : c'est bien évidemment impossible, car cela placerait Hipparque au 4$^{e}$ siècle avant notre ère au lieu du 2$^{e}$ siècle avant notre ère (car 29×72=2088 ans, et 1700-2088 = 388 avant notre ère).

Souciet, toujours taraudé par le problème du début des signes, se plonge finalement sur le partage du zodiaque dans le système newtonien (soit avec un début à plus de 7° de l'oreille) et le sien (début sur l'oreille). En effet, constellations et signes diffèrent : les constellations couvrent de 20° à 50° de longitude écliptique, selon les cas, alors que les signes, par définition, occupent tous 30°. Il est donc inévitable qu'il existe un certain chevauchement entre signes et constellations. On peut cependant minimiser ce chevauchement en choisissant au mieux la position des signes. Souciet prend l'exemple du Bélier[134] : comme il ne couvre que 20°, il faut « caser » les 10° supplémentaires du signe quelque part. On peut soit mettre 5° de part et d'autre de la constellation, soit mettre les 10° au début, avant la constellation, soit les mettre après, soit... couper de manière plus étrange, en utilisant les 7.5° de Newton. Passant en revue les signes et constellations, Souciet montre que son système conduit à un recouvrement moindre que celui de Newton. La Nauze conteste ces conclusions. Non seulement, il place l'extension du Bélier à 25°[135], mais il propose de regarder les chevauchements de part et d'autre des signes : à la fin, comme Souciet, mais aussi au début, un point que ce dernier se garde d'aborder[136]. Il montre ainsi que l'arrangement de Newton semble plus équilibré que celui de Souciet. Au passage, La Nauze note plusieurs erreurs, dans le choix des

---

[129] Souciet, 5$^e$ dissertation, p144
[130] La Nauze, 1$^e$ lettre, p359
[131] La Nauze, 5$^e$ lettre – on compare ici p416 et p421
[132] Souciet, 5$^e$ dissertation, p148-150
[133] La Nauze, 5$^e$ lettre, p 422 et p426-429
[134] Souciet, 5$^e$ dissertation, p124-125 et p158-168
[135] La Nauze, 5$^e$ lettre, p394-396
[136] La Nauze, 1$^e$ lettre, p346-353 ; 5$^e$ lettre, p440-464

« dernières » étoiles associées aux constellations. La Table 1 et la Fig. 6 ci-dessous comparent les deux systèmes : la solution newtonienne est effectivement plus équilibrée que celle de Souciet.

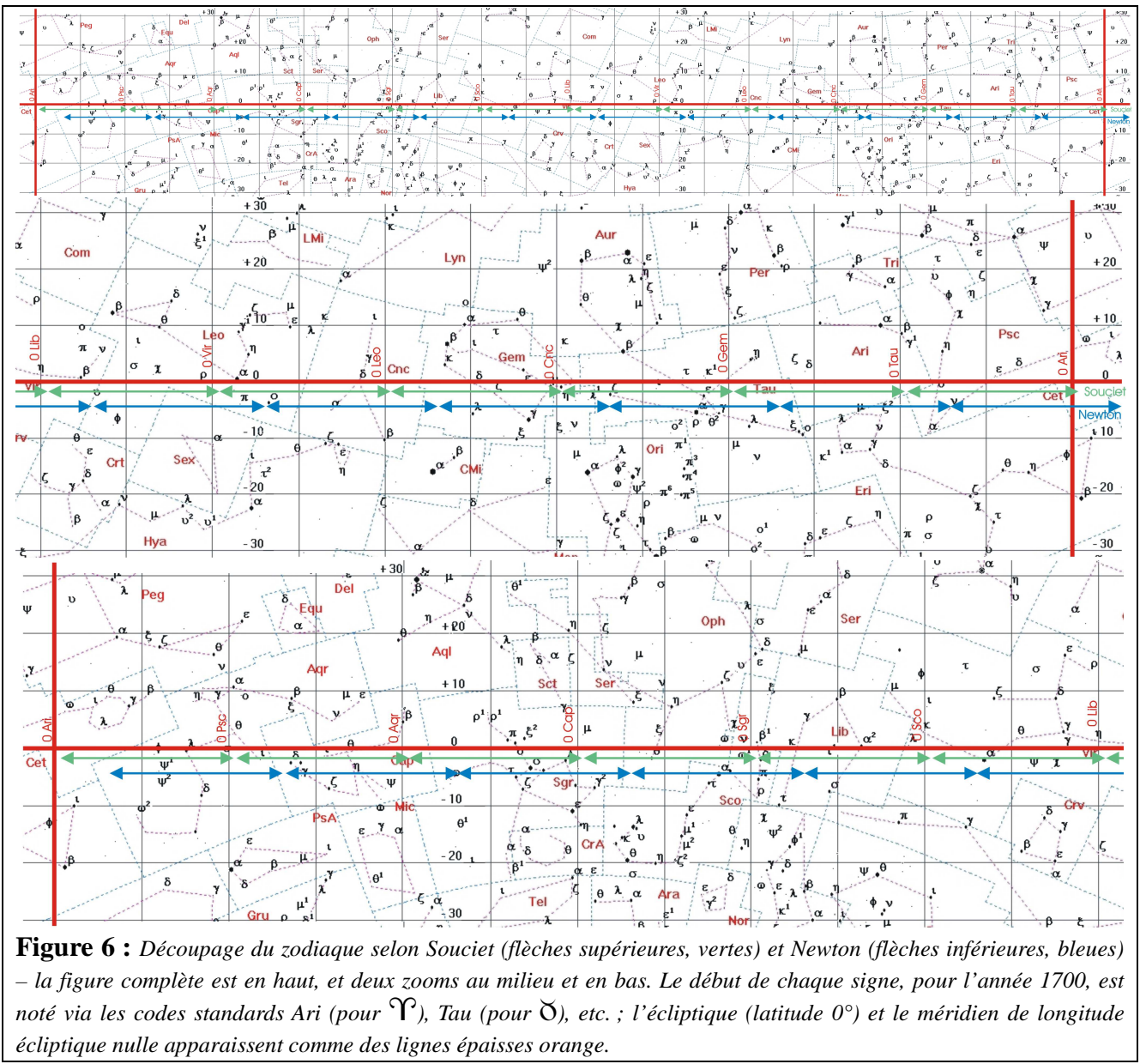

**Figure 6 :** *Découpage du zodiaque selon Souciet (flèches supérieures, vertes) et Newton (flèches inférieures, bleues) – la figure complète est en haut, et deux zooms au milieu et en bas. Le début de chaque signe, pour l'année 1700, est noté via les codes standards Ari (pour ♈), Tau (pour ♉), etc. ; l'écliptique (latitude 0°) et le méridien de longitude écliptique nulle apparaissent comme des lignes épaisses orange.*

**Table 1 :** *Longitudes écliptiques des limites des signes et constellations, en utilisant le système de Newton (début 7°24' à l'ouest de l'oreille du Bélier) et celui de Souciet (début à l'oreille du Bélier). Les valeurs fournies par La Nauze apparaissent en italiques, les incohérences entre les deux tables de Souciet en rouge souligné. Comme point de comparaison, la dernière colonne donne les longitudes extrêmes des étoiles de ces constellations, dans le catalogue de Flamsteed. Noter que Souciet, suivi par la Nauze, considère un début des signes pour Newton en 21°25', alors que 29°1'-7°24'=21°37' ; il s'agit de la confusion habituelle de Souciet entre 7°24' (début de signe-oreille) et 7°36' (oreille-colure). Les longitudes reproduites ici correspondent à l'équinoxe 1700 pour les deux premières colonnes et 1690 pour la dernière.*

| Signe et constellation | Newton | Souciet | Flamsteed |
|---|---|---|---|
| **ARIES ♈** | | | |
| Première étoile de la constellation | *29°1' ♈* | *29°1' ♈* | 26°36' ♈ |
| Dernière étoile de la constellation | 19°13' ♉ | 19°<u>1</u>' ♉ | 21°6' ♉ |
| Début du signe | 21°25' ♈ | 29°1' ♈ | |
| Fin du signe | 21°25' ♉ | 29°1' ♉ | |
| **TAURUS ♉** | | | |
| Première étoile de la constellation | *14°7' ♉* | *14°7' ♉* | 16°49' ♉ |
| Dernière étoile de la constellation | 20°16' ♊ | 20°16' ♊ | 26°3' ♊ |
| Début du signe | 21°25' ♉ | 29°1' ♉ | |
| Fin du signe | 21°25' ♊ | 29°1' ♊ | |
| **GEMINI ♊** | | | |
| Première étoile de la constellation | *29°14' ♊* | *29°14' ♊* | 26°37' ♊ |
| Dernière étoile de la constellation | 24°27' ♋ | <span style="color:red">27°</span>27' ♋ | 22°43' ♋ |
| Début du signe | 21°25' ♊ | 29°1' ♊ | |
| Fin du signe | 21°25' ♋ | 29°1' ♋ | |
| **CANCER ♋** | | | |
| Première étoile de la constellation | *18°0' ♋* | *18°0' ♋* | 22°49' ♋ |
| Dernière étoile de la constellation | <span style="color:red">25°39'</span> ♌ | 12°41' ♌ | 12°20' leo |
| Début du signe | 21°25' ♋ | 29°1' ♋ | |
| Fin du signe | 21°25' ♌ | 29°1' ♌ | |
| **LEO ♌** | | | |
| Première étoile de la constellation | *11°1' ♌* | *11°1' ♌* | 10°57' ♌ |
| Dernière étoile de la constellation | 20°50' ♍ | 20°50' ♍ | 20°42' ♍ |
| Début du signe | 21°25' ♌ | 29°1' ♌ | |
| Fin du signe | 21°25' ♍ | 29°1' ♍ | |
| **VIRGO ♍** | | | |
| Première étoile de la constellation | *14°45' ♍* | *14°45' ♍* | 17°30' ♍ |
| Dernière étoile de la constellation | 29°33' ♎ | 29°33' ♎ | 8°17' ♏ |
| | *5°54' sco* | *5°54' sco* | |
| Début du signe | 21°25' ♍ | 29°1' ♍ | |
| Fin du signe | 21°25' ♎ | 29°1' ♎ | |
| **LIBRA ♎** | | | |
| Première étoile de la constellation | *26°40' ♎* | *26°40' ♎* | 3°5' ♏ |
| Dernière étoile de la constellation | 29°23' ♏ | <span style="color:red">16°41'</span> ♏ | 27°4' ♏ |
| | *26°29' ♏* | *26°29' ♏* | |
| Début du signe | 21°25' ♎ | 29°1' ♎ | |
| Fin du signe | 21°25' ♏ | 29°1' ♏ | |
| **SCORPIUS ♏** | | | |
| Première étoile de la constellation | *16°28' ♏* | *16°28' ♏* | 26°48' ♏ |
| Dernière étoile de la constellation | <span style="color:red">29°7'</span> ♐ | 23°31' ♐ | 20°15' ♐ |
| Début du signe | 21°25' ♏ | 29°1' ♏ | |
| Fin du signe | 21°25' ♐ | 29°1' ♐ | |
| **SAGITTARIUS ♐** | | | |
| Première étoile de la constellation | *26°56' ♐* | *26°56' ♐* | 19°23' ♐ |
| Dernière étoile de la constellation | 24°18' ♑ | <span style="color:red">22°6'</span> ♑ | 26°38' ♑ |
| | *0°32' ♒* | *0°32' ♒* | |
| Début du signe | 21°25' ♐ | 29°1' ♐ | |
| Fin du signe | 21°25' ♑ | 29°1' ♑ | |
| **CAPRICORNUS ♑** | | | |
| Première étoile de la constellation | *28°34 ♑* | *28°34 ♑* | 28°6' ♑ |
| Dernière étoile de la constellation | 21°53' ♒ | 21°53' ♒ | 21°29' ♒ |
| Début du signe | 21°25' ♑ | 29°1' ♑ | |
| Fin du signe | 21°25' ♒ | 29°1' ♒ | |
| **AQUARIUS** ♒ | | | |
| Première étoile de la constellation | *7°35' ♒* | *7°35' ♒* | 7°24' ♒ |
| Dernière étoile de la constellation | 24°52' ♓ | 24°52' ♓ | 15°58' ♓ |
| Début du signe | 21°25' ♒ | 29°1' ♒ | |
| Fin du signe | 21°25' ♓ | 29°1' ♓ | |
| **PISCES** ♓ | | | |
| Première étoile de la constellation | *14°25' ♓* | *14°25' ♓* | 11°6' ♓ |
| Dernière étoile de la constellation | 25°33' ♈ | 25°33' ♈ | 28°22' ♈ |
| Début du signe | 21°25' ♓ | 29°1' ♓ | |
| Fin du signe | 21°25' ♈ | 29°1' ♈ | |

## *5. Résumé et conclusion*

Au 17e siècle et au début du 18e siècle, la chronologie est une « science » en plein essor. Les tentatives de datation dépendent surtout des analyses de textes, mais l'astronomie vient alors faire une apparition plus ou moins discrète. Il manque cependant un chronomètre global précis – la datation par isotope relevant encore de la science-fiction à l'époque. Newton propose un tel chronomètre par un concept inédit : l'utilisation de la précession des équinoxes.

Attribuant la description de la sphère céleste par Eudoxe à l'époque des Argonautes (en en particulier Chiron), choisissant avec soin les étoiles potentiellement décrites par ces textes, il date l'expédition du 10e siècle avant notre ère. En outre, il utilise la différence entre l'année solaire réelle et celle du calendrier de 365 jours pour relier le calendrier égyptien au calendrier chaldéen. Couplé au raccourcissement de la durée moyenne des règnes, cela conduit à une ligne du temps étonnante, raccourcie de cinq siècles. Une telle chronologie « démontre » l'ancienneté des hébreux et de leur religion, et postule la relative jeunesse des civilisations égyptiennes ou chinoises.

On relève cependant plusieurs problèmes. Pour les travaux sur les calendriers, il y a la mauvaise identification du pharaon concerné, l'héritage inversé, et le début mal placé à l'équinoxe. Quant à l'idée, originale et innovante, de l'utilisation de la précession, elle a l'apparence d'une démonstration… qui serait valable si les hypothèses étaient vérifiées (la sphère décrite par Eudoxe identique à celle de Chiron) et si les étoiles utilisées pour le calcul étaient bien choisies (càd suffisamment brillantes et correspondant à la description d'Eudoxe). Dans les deux cas, Newton donne souvent l'impression de choisir ce qui soutient sa thèse, qui peut donc très logiquement être combattue. Certes, le concept de départ est génial, mais pour être applicable, il faut une civilisation organisée, qui a déjà beaucoup observé (elle connaît écliptique, zodiaque, colures etc.), le fait précisément (mieux qu'un degré) et dans un système connu (pas d'ambiguïté sur l'identification des étoiles, le système de référence, etc.) – on ne trouve cela que dans des cas assez récents. Newton suppose donc présomptueusement en disposer, alors qu'il ne dispose que d'observations peu claires (« milieu ») dans un système mal connu (problème de délimitation des constellations) : son travail est donc logiquement voué à l'échec.

Même si la démarche de Newton s'inscrit dans la tradition de son époque, sa chronologie surprenante par sa durée fera débat jusqu'au début du 19e siècle. La plupart des discussions se focalisent sur la durée des règnes et l'interprétation des textes, mais quelques sources examinent de près la question astronomique, les écrits de Whiston, Souciet, Fréret (pour la critique) et La Nauze (pour le soutien). L'analyse la plus claire, la plus complète et la plus précise est celle de Fréret, bien que non astronome, qui reprend en partie le travail de Whiston. Le jésuite Souciet multiplie les erreurs de calcul et autres problèmes, pour lesquels il sera tancé par Halley ; il est aussi brouillon et part dans des directions parallèles, parfois intéressantes mais peu ou mal exploitées. La Nauze, virulent avec Souciet, ne s'avère pas bien meilleur, reprenant même les erreurs qu'il critique. Le débat semble ici ailleurs que sur la justesse des travaux : c'est la personne de Newton qui est visée. La Chronologie représente en effet son talon d'Achille, son œuvre « scientifique » la plus discutable. En l'attaquant, c'est l'œuvre entière du savant anglais qu'on espère remettre en question. Ce ne sera cependant pas le cas : si la Chronologie tombera bien sous les coups de butoir, les travaux mécaniques et optiques conserveront leur aura et s'imposeront partout.

Au final, la Chronologie sombrera simplement dans l'oubli, évacuée sous le tapis avec les travaux alchimiques pour ne pas écorner l'image du grand scientifique. Toutefois, négliger ces aspects pseudo-scientifiques a pour conséquence de donner une fausse image de la personnalité de Newton. Il a travaillé durant des années à ces choses aujourd'hui considérées comme mineures : elles étaient pourtant très importantes à ses yeux. Tout comme lui (dans la Chronologie), nous ne devons pas écarter des faits sous prétexte qu'ils ne conviennent pas à nos désirs…

## *6.   Annexes*

Pour mieux comprendre les discussions des arguments astronomiques de la Chronologie, il est nécessaire de connaître les bases du repérage sur la sphère céleste, que nous rappelons brièvement ci-dessous.

<u>a) Coordonnées célestes</u>

Dès les débuts de l'astronomie quantitative, mathématique, un repérage précis de n'importe quel objet du ciel, où qu'il soit sur la voûte céleste, fut nécessaire. On définit deux systèmes, encore utilisés aujourd'hui (Fig. A1).

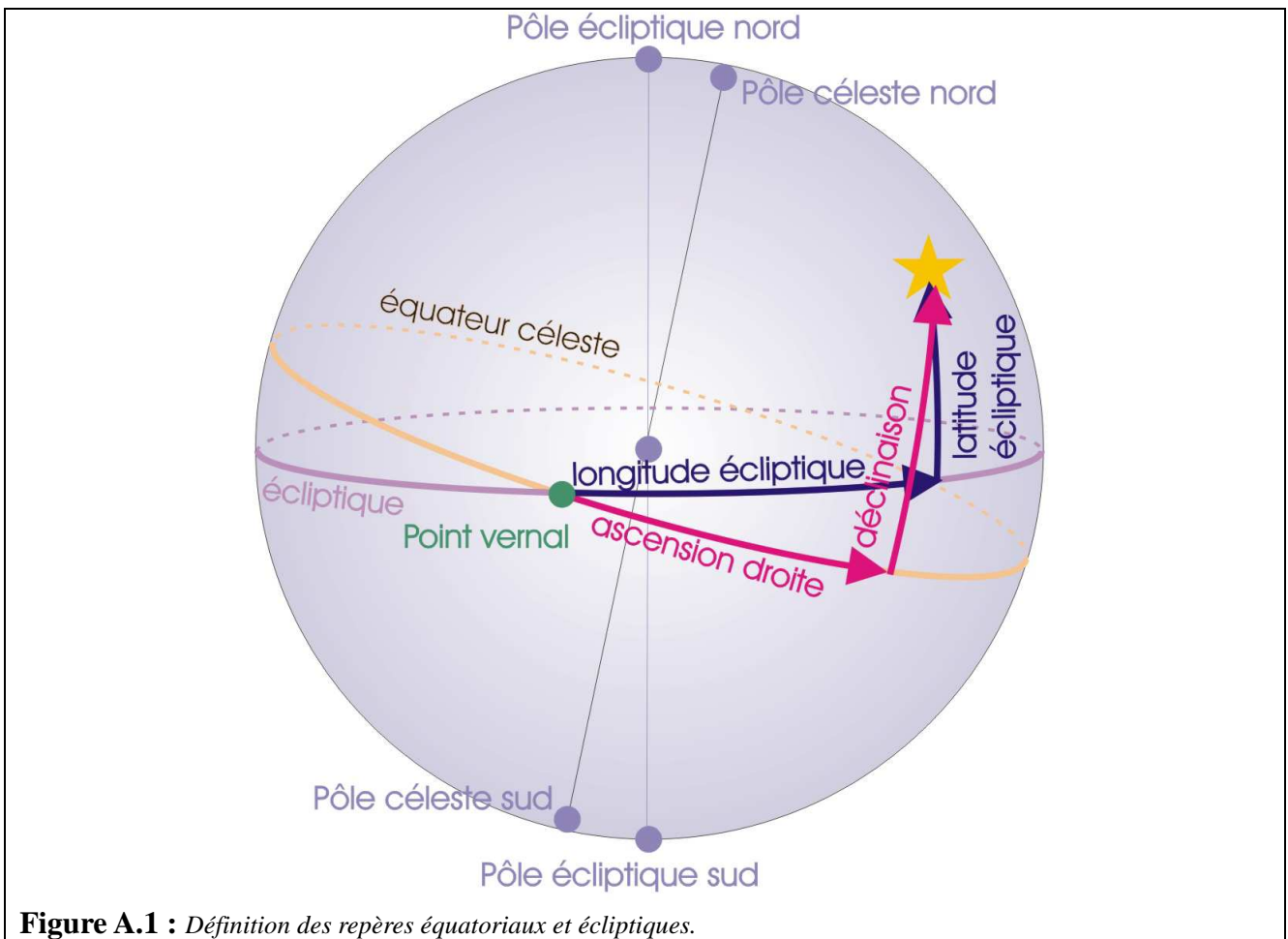

**Figure A.1 :** *Définition des repères équatoriaux et écliptiques.*

Le premier système de coordonnées fait appel à l'ascension droite et à la déclinaison, équivalents de la longitude et la latitude sur la sphère terrestre. Comme pour cette dernière, il faut préciser les repères de base ou points zéro. La **sphère céleste** est divisée en deux parties par l'**équateur céleste**, projection de l'équateur terrestre sur la voûte céleste. L'axe des pôles, perpendiculaire à l'équateur, coupe la voûte céleste aux deux **pôles célestes**, nord et sud. La **déclinaison** est comptée (de 0 à 90°, positif côté nord, négatif au sud) depuis l'équateur le long d'un méridien céleste, grand cercle passant par les pôles. L'**ascension droite** est mesurée (de 0 à 24 heures ou de 0 à 360°) le long de l'équateur, depuis le point vernal (voir ci-dessous) – à noter que la direction des ascensions droites croissante indique l'Est et celle des décroissantes l'Ouest. Au cours d'une journée, un observateur terrestre voit la sphère céleste tourner – c'est le reflet de la rotation de notre planète sur elle-même. Les coordonnées des astres ne sont évidemment pas affectées par ce mouvement.

Comme la Terre tourne autour du Soleil, cet astre change de position par rapport aux étoiles de la voûte céleste. Ce trajet apparent du Soleil se fait le long d'un cercle appelé **écliptique**, qui n'est autre que l'intersection entre la sphère céleste et le plan de l'orbite terrestre. L'écliptique fait donc un angle de 23.5° par rapport à l'équateur céleste. Quatre positions sur l'écliptique sont particulièrement remarquables : celles où se trouve le Soleil aux équinoxes et solstices (Fig. A.2). Les deux équinoxes se produisent lorsque le Soleil se trouve aux intersections entre équateur et écliptique, tandis que les solstices se produisent quand le Soleil culmine, extrêmes nord et sud, par rapport à l'équateur. Le **point vernal** correspond à la position du Soleil lors de l'équinoxe de printemps ; et l'appellation abusive « points cardinaux » utilisée dans les sources évoquées dans l'article correspond à l'ensemble des deux points solsticiaux et des deux points équinoxiaux.

Tout comme l'équateur céleste, l'écliptique peut servir de base à un système de coordonnées, longitude écliptique et latitude écliptique (voir Fig. A.1). La latitude est comptée en degrés depuis l'écliptique, et la longitude est mesurée à partir du point vernal, soit sur une échelle de 360° ou en utilisant une échelle de 30° et 12 signes. Ces signes sont les signes du zodiaque, chacun occupant 30° de longitude et notés traditionnellement (par ordre de sucession depuis le point vernal) ♈ (Aries-Bélier), ♉ (Taurus-Taureau), ♊ (Gemini-Gémeaux), ♋ (Cancer-Cancer), ♌ (Leo-Lion), ♍ (Virgo-Vierge), ♎ (Libra-Balance), ♏ (Scorpio-Scorpion), ♐ (Sagittarius-Sagittaire), ♑ (Capricornus ou Caper-Capricorne), ♒ (Aquarius-Verseau), ♓ (Pisces-Poissons). Pour passer de la coordonnée en degrés+signe en angles dans l'échelle habituelle de 360°, il faut ajouter 30° par signe traversé entièrement depuis le point vernal : 5° ♊ vaut donc 30° (largeur du signe du Bélier) +30° (largeur du signe du Taureau) + 5° = 65° de longitude écliptique.

Considérant la sphère céleste avec ces deux systèmes de repérage, on peut définir deux grands cercles particuliers, les **colures** (Fig. A2). Le colure équinoxial passe par les deux points d'équinoxe et les pôles célestes : il est donc perpendiculaire à l'équateur et fait un angle de 66.5° avec l'écliptique. Le colure solsticial comprend les deux points de solstice, les pôles célestes, et les pôles écliptiques : il est donc perpendiculaire à la fois à l'équateur et à l'écliptique. Ces deux colures se trouvent dans deux plans perpendiculaires.

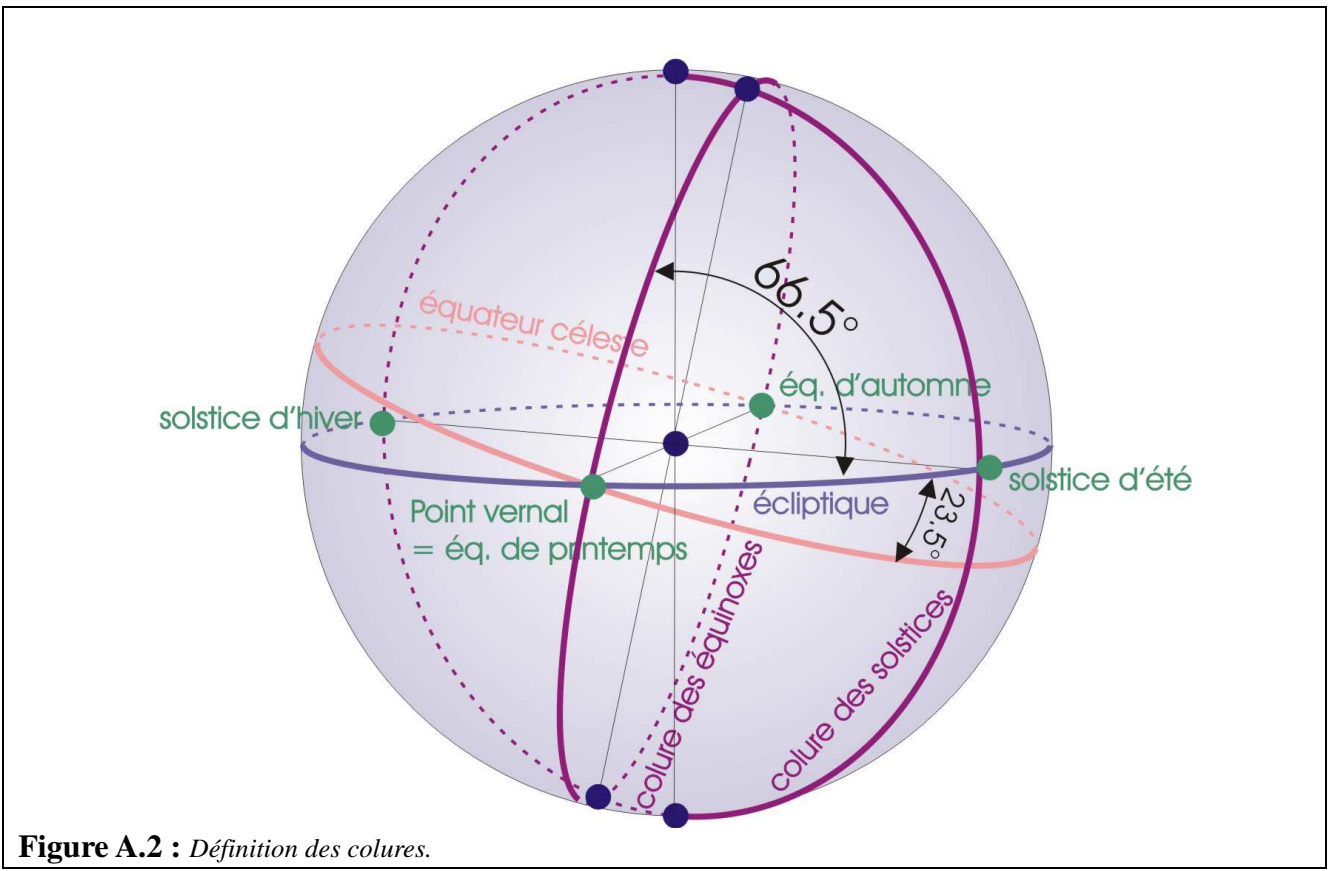

**Figure A.2 :** *Définition des colures.*

A noter que l'on passe d'un système de coordonnées à l'autre par les formules :

$$\sin(\lambda) = \sin(\delta)\cos(\varepsilon) - \cos(\delta)\sin(\varepsilon)\sin(\alpha) \quad ; \quad tg(\beta) = \frac{tg(\delta)\sin(\varepsilon) + \sin(\alpha)\cos(\varepsilon)}{\cos(\alpha)}$$

$$\sin(\delta) = \sin(\lambda)\cos(\varepsilon) + \cos(\lambda)\sin(\varepsilon)\sin(\beta) \quad ; \quad tg(\alpha) = \frac{\sin(\beta)\cos(\varepsilon) - tg(\lambda)\sin(\varepsilon)}{\cos(\beta)}$$

Avec α l'ascension droite, δ la déclinaison, β la longitude écliptique, la λ latitude écliptique, et ε l'obliquité de la Terre (soit 23.5°).

b) Précession

L'axe de rotation de la Terre n'est pas fixe par rapport aux étoiles. Il tourne autour de l'axe des pôles écliptiques avec une période de 25 800 ans. Ce phénomène est dû à la forme de notre planète, qui n'est pas une sphère parfaite : sous l'action gravifique du Soleil, le bourrelet équatorial tend à rejoindre le plan orbital, créant un couple de force qui fait tourner l'axe principal.

Ce phénomène porte le nom de **précession**. Il a pu être expliqué théoriquement par Isaac Newton dans ses Principia, quoique de nombreuses erreurs parsèment son raisonnement[137], qui fut d'ailleurs corrigé par d'Alembert en 1749.

Puisque l'équateur céleste est perpendiculaire à l'axe de rotation de la Terre, toute modification de l'un entraîne un changement de position de l'autre. Les systèmes de coordonnées célestes sont alors affectés, puisqu'ils ont pour référence le point vernal, intersection entre écliptique et équateur[138]. Les variations diffèrent toutefois d'un système à l'autre (Fig. A3) : d'un côté, la longitude écliptique varie tandis que la latitude écliptique reste inchangée ; de l'autre, ascension droite et déclinaison changent tous deux.

Ce changement dans les coordonnées a été repéré en premier par Hipparque de Nicée, lorsqu'il compara ses mesures à celles de ses collègues Aristille et Timocharis. Cela l'encouragea à produire un catalogue complet de la voûte céleste, pouvant servir de référence aux générations futures. Avec les mesures imparfaites de l'époque, il évalua la précession à un degré par siècle (contre 1° par 72 ans en réalité). A noter qu'il voyait alors la précession comme un changement d'ascension droite à déclinaison constante, alors qu'il s'agit en fait d'un changement de longitude écliptique à latitude écliptique constante.

La précession possède au moins deux conséquences :
- L'étoile polaire change avec le temps : celle des Egyptiens et des premiers Grecs était Thuban, dans la constellation du Dragon, tandis qu'il s'agit de Polaris, dans la Petite Ourse, à l'heure actuelle. A l'époque d'Hipparque, aucune étoile n'était proche du pôle, d'où son étonnement à la mention d'un tel astre par Eudoxe[139].
- Signes et constellations du zodiaque s'éloignent l'un de l'autre. En effet, depuis l'époque d'Hipparque, on compte les signes de 30° en 30° depuis le point vernal, qui reste donc en permanence à 0° ♈. Comme la position céleste du point vernal change avec le temps, les signes se décalent progressivement par rapport à leurs constellations d'origine : le décalage (1° en 72 ans, soit 30° en 2160 ans) atteint environ un signe à l'heure actuelle. Une autre convention était possible : garder le lien entre signe et constellation, en faisant voyager le point vernal parmi les signes – cette possibilité est centrale dans les débats liés à la Chronologie[140].

---

[137] G.J. Dobson, Arch. Hist. Exact Sci. Vol 53, 125-145, 1998
[138] Le point vernal étant relatif à l'équinoxe, on parle donc généralement de "précession des équinoxes" pour cet effet.
[139] Voir par exemple Fréret, Défense, p448
[140] Voir par exemple Whiston cité par Fréret, Défense, p 424-427

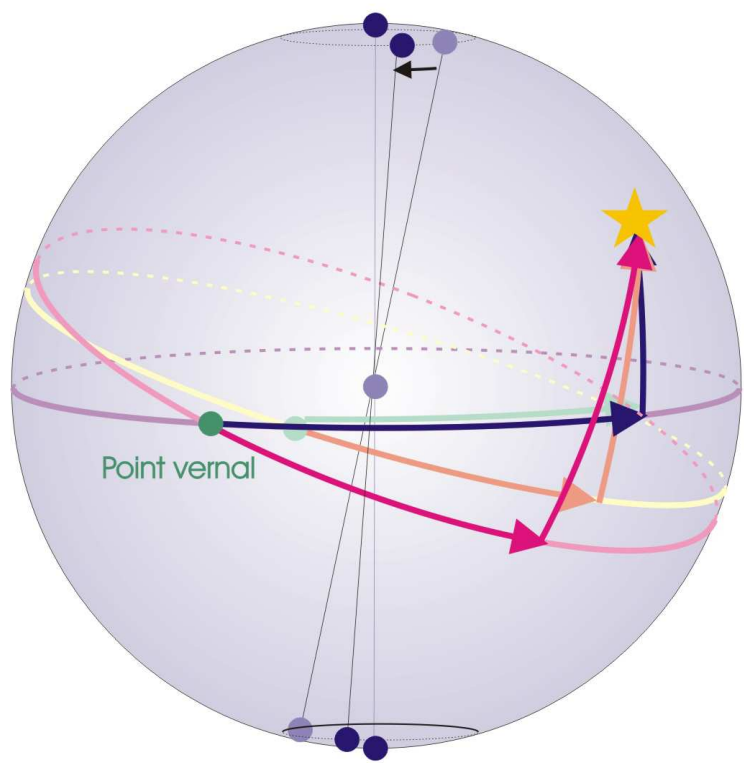

**Figure A3 :** *Changement des coordonnées au cours du temps à cause de la précession.*

c) Cartes du ciel

      Cette annexe fournit les identifications des étoiles discutées sur les cartes de Bayer (Uranometria), puisque leur notation est utilisée par tous, ainsi que les cartes des colures aux deux époques discutées (939 et 1353 avant notre ère).

bayer*.jpg

**Figure A4 :** *Extraits de l'Uranometria de Bayer avec mise en évidence des étoiles évoquées par Newton (en jaune) et celles discutées par Fréret, Whiston, Halley et Souciet (en rouge). Les scans de l'Uranometria viennent de http://www.lindahall.org/services/digital/ebooks/bayer/*

sky_939.jpg + sky_1353.jpg

**Figure A5 :** *Position des étoiles discutées pour leur appartenance aux colures des équinoxes et des solstices (lignes orange foncé) en 939 (étoiles à problème en jaune foncé, autres en vert) et 1353 avant notre ère (couples d'étoiles de Whiston en bleu). Les coordonnées sont ascension droite et déclinaison, et les étoiles choisies par Fréret pour indiquer tropiques et équateur sont notées en vert pour l'époque 1353. Les astres appartenant au Navire n'ont pas été indiquées.*